\documentclass[letterpaper,11pt]{article}

\usepackage{jheppub}
\usepackage{graphicx}
\usepackage{amsmath}
\usepackage{amssymb}
\usepackage{mathrsfs}
\usepackage{xspace}
\usepackage[dvipsnames]{xcolor}
\usepackage[utf8]{inputenc}
\usepackage[normalem]{ulem}

\newcommand{\eq}[1]{eq.~\eqref{eq:#1}}
\newcommand{\eqs}[2]{eqs.~\eqref{eq:#1} and \eqref{eq:#2}}
\renewcommand{\sec}[1]{sec.~\ref{sec:#1}}
\newcommand{\secs}[2]{secs.~\ref{sec:#1} and \ref{sec:#2}}

\newcommand{\app}[1]{app.~\ref{app:#1}} 
\newcommand{\fig}[1]{fig.~\ref{fig:#1}}

\newcommand{\ord}[1]{{\mathcal O}(#1)}

\newcommand{\nn}{\nonumber}

\newcommand{\df}{\mathrm{d}}

\newcommand{\lra}{\leftrightarrow}

\newcommand{\al}{\alpha}
\newcommand{\bt}{\beta}
\newcommand{\ga}{\gamma}
\newcommand{\Ga}{\Gamma}
\newcommand{\de}{\delta}

\newcommand{\la}{\lambda}
\newcommand{\si}{\sigma}

\newcommand{\cL}{{\mathcal L}}

\newcommand{\cS}{{\mathscr S}}

\newcommand{\bn}{{\bar{n}}}

\newcommand{\lqcd}{\Lambda_\mathrm{QCD}}
\newcommand{\cusp}{\mathrm{cusp}}

\newcommand{\SCETa}{\ensuremath{{\rm SCET}_{\rm I}}\xspace}

\newcommand{\SCETab}{\ensuremath{{\rm SCET}_+}\xspace}

\newcommand{\Pythia}{\textsc{Pythia}\xspace}

\newcommand{\Eventtwo}{\textsc{Event2}\xspace}

\newcommand{\ea}{e_\alpha}
\newcommand{\eb}{e_\beta}
\newcommand{\Lb}{L_\beta}
\newcommand{\La}{L_\alpha}

\allowdisplaybreaks[2]

\preprint{\vbox{\hbox{Nikhef 2018-013}\hbox{UWThPh 2018-19}\hbox{MITP/18-051}}}

\title{Joint resummation of two angularities \\ at next-to-next-to-leading logarithmic order}

\author[a,b]{Massimiliano Procura,}

\author[c,d]{Wouter J.~Waalewijn,}

\author[d,e]{Lisa Zeune}

\affiliation[a]{Theoretical Physics Department, CERN, 1 Esplanade des Particules, Geneva 23, Switzerland}
\affiliation[b]{Fakult\"at f\"ur Physik, Universit\"at Wien, Boltzmanngasse 5, 1090 Wien, Austria}
\affiliation[c]{Institute for Theoretical Physics Amsterdam and Delta Institute for Theoretical Physics, University of Amsterdam, Science Park 904, 1098 XH Amsterdam, The Netherlands}
\affiliation[d]{Nikhef, Theory Group, Science Park 105, 1098 XG, Amsterdam, The Netherlands}
\affiliation[e]{PRISMA Cluster of Excellence \& Mainz Institute for Theoretical Physics, Johannes Gutenberg University, 55099 Mainz, Germany}

\abstract{Multivariate analyses are emerging as important tools to understand properties of hadronic jets, which play a key role in the LHC experimental program. 
We take a first step towards precise \emph{and} differential theory predictions, by calculating the cross section for $e^+e^- \to 2$ jets differential in the angularities $e_\al$ and $e_\bt$. The logarithms of $e_\al$ and $e_\bt$ in the cross section are jointly resummed to next-to-next-to-leading logarithmic accuracy, using the \SCETab framework we developed, and are matched to the next-to-leading order cross section. We perform analytic one-loop calculations that serve as input for our numerical analysis, provide controlled theory uncertainties, and compare our results to \Pythia. We also  obtain predictions for the cross section differential in the ratio $e_\al/e_\bt$, which cannot be determined from a fixed-order calculation. The effect of nonperturbative corrections is also investigated. Using \Eventtwo, we validate the logarithmic structure of the single angularity cross section predicted by factorization theorems at $\ord{\al_s^2}$, highlighting the importance of recoil for specific angularities when using the thrust axis as compared to the winner-take-all axis.}
\begin{document}
\maketitle

%%%%%%%%%%%%%%%%%%%%%%%%%%%%%%%%%%%%%%%%%%%%%%%%%%%%%%%%%%%%%%%%%%%%%%%%%%%%%%%%
\section{Introduction}
\label{sec:intro}
%%%%%%%%%%%%%%%%%%%%%%%%%%%%%%%%%%%%%%%%%%%%%%%%%%%%%%%%%%%%%%%%%%%%%%%%%%%%%%%%

Hadronic jets play a central role in collider physics as proxies of the hard quarks and gluons produced in short-distance interactions. A precise theoretical understanding of jet properties is often key to establishing measurements which can either further confirm the predictions of the Standard Model at higher accuracy, or identify deviations that could hint at New Physics. Progress in QCD calculations involving jets has been impressive in the last years, boosted by the demands of the LHC experimental program. Jet properties are typically studied using two complementary tools: analytic resummation and Monte Carlo parton showers. The latter offer a fully exclusive description of the final state, enabling the user to perform any measurement, but their formal accuracy is currently limited to leading-logarithmic order.\footnote{See ref.~\cite{Dasgupta:2018nvj} for a recent discussion of the logarithmic accuracy of parton showers.} Corrections to the traditional parton shower method have been considered lately, {\it e.g.}~by incorporating additional information about interference effects~\cite{Nagy:2007ty,Nagy:2012bt,Nagy:2015hwa} and higher-order splitting functions~\cite{Jadach:2016zgk,Hoche:2017iem,Hoche:2017hno,Li:2016yez}. On the other hand, with an analytic approach one can often achieve higher logarithmic resummations, and obtain uncertainty estimates that can be validated by comparing different orders. However, this approach is traditionally limited to single differential measurements. 

Inspired by ref.~\cite{Larkoski:2014tva}, we recently took the first step towards a precise {\it and} more differential characterization of jets by constructing an effective field theory (called \SCETab) and using it to derive a factorization formula~\cite{Procura:2014cba}, which enables the simultaneous resummation of two independent observables to higher logarithmic accuracy. This opens up the possibility of performing multivariate analyses, including correlations with controlled theory uncertainties. Applications of our framework are particularly relevant in the context of jet substructure studies (see {\it e.g.}~ref.~\cite{Larkoski:2017jix} for a recent review), where a more detailed characterization of the QCD radiation pattern within a jet is exploited to obtain crucial information about the hard scattering process, thereby providing innovative ways to search for New Physics. This generally involves multi-differential cross sections, with several independent measurements performed on a single jet and the possibility to exploit shared information content among these observables. Furthermore, several of the most powerful discriminants of quark- vs.~gluon-initiated jets or of QCD jets vs.~boosted hadronically decaying heavy particles are formed by taking ratios of two observables, as is done for $N$-subjettinesses~\cite{Thaler:2010tr,Thaler:2011gf}, energy-energy-correlation functions~\cite{Banfi:2004yd,Larkoski:2013eya} and planar flow~\cite{Almeida:2008yp}. These are typically not infrared- and collinear-safe~\cite{Soyez:2012hv} but still Sudakov safe, meaning that they can nevertheless be properly defined and calculated by marginalizing the corresponding \emph{resummed} double differential cross section~\cite{Larkoski:2013paa}.

As a first step in understanding multi-differential cross sections beyond next-to-leading logarithmic (NLL) accuracy, we demonstrate in this paper how to exploit our theoretical framework for the case of the simultaneous measurement of two event shapes for $e^+ e^-$ collisions in the dijet limit, where all-order resummations are essential to obtain reliable theory predictions. In order to avoid complications related to non-global logarithms~\cite{Dasgupta:2001sh}, we restrict ourselves to event-based observables and postpone the case of jet-based measurements at NNLL to a future publication. Our focus here is on the family of infrared and collinear safe angularities~\cite{Berger:2003iw}, which generalize the classic event-shape variables thrust and broadening, and characterize the energy distribution of final-state particles as function of the angle with respect to some axis. 
Calculations of single angularities with respect to the thrust axis were carried out at NLL accuracy in refs.~\cite{Berger:2003iw,Berger:2003pk,Hornig:2009vb,Almeida:2014uva}. Recently these observables have been analyzed up to NNLL, including next-to-next-leading fixed-order (NNLO) corrections~\cite{Bell:2018gce}, for the purpose of a precision determination of $\al_s(m_Z)$ from LEP data, which will provide complementary information to analogous precision fits based on event shapes like thrust~\cite{Becher:2008cf,Abbate:2010xh} and $C$-parameter~\cite{Hoang:2014wka}. NNLL+NLO accuracy for angularities has also been reached using the {\tt ARES} method~\cite{Banfi:2014sua} in~\cite{Banfi:2018mcq}, while the NNLL resummation of jet broadening in the framework of SCET was achieved in~\cite{Becher:2012qc}. Furthermore, (generalized) angularities measured on individual jets are useful tools to investigate jet substructure~\cite{Almeida:2008yp,Ellis:2010rwa,Larkoski:2014pca}. 

In this paper we go beyond state-of-the-art NLL accuracy for the jointly resummed cross section of two angularities, and use \SCETab to achieve NNLL precision throughout the phase space. We match to \SCETa theories that describe the phase-space boundaries to maintain NNLL accuracy there, and to the fixed-order QCD result at NLO to obtain a reliable description of the cross section beyond the dijet limit. We also correct typos in expressions for the necessary one-loop ingredients that have been derived elsewhere. In our numerical analysis, theoretical uncertainties are provided by suitable ``profile functions" which we design to produce scale variations that smoothly interpolate between the distinct kinematic regions where resummations must be handled differently. We also investigate nonperturbative corrections, and compare the results of our numerical analysis to the parton shower of \Pythia 8.2~\cite{Sjostrand:2014zea}.

By projecting our double differential cross section, we obtain predictions for the cross section differential in the ratio of two angularities, which cannot be determined from a fixed-order calculation.
Furthermore, we analyze single angularity distributions up to NNLL+NLO and investigate their logarithmic structure by comparing the fixed-order expansions from our resummed distributions against numerical results from \Eventtwo, for angularities calculated with respect to the thrust axis and winner-take-all (WTA) axis~\cite{Salam:WTAUnpublished,Larkoski:2014uqa}. Our analysis demonstrates that for the WTA axis the same factorization formulae can be used for the whole range of angularities, even for those measurements that would be sensitive to recoil effects from soft radiation if the thrust axis was used. 

The paper is organized as follows: \sec{framework} describes in detail our theoretical framework and the analytic input for our numerical analysis. After reviewing the factorization for the double differential angularity distribution, we collect all relevant fixed-order ingredients (correcting typos in the literature) and anomalous dimensions. We describe in detail our scale choices and procedure to estimate the perturbative uncertainty. In \sec{results} we show numerical results for single and double differential angularity distributions at NNLL+NLO accuracy, as well as for the ratio of two angularities. We conclude in \sec{conc}.

%%%%%%%%%%%%%%%%%%%%%%%%%%%%%%%%%%%%%%%%%%%%%%%%%%%%%%%%%%%%%%%%%%%%%%%%%%%%%%%%
\section{Framework}
\label{sec:framework}
%%%%%%%%%%%%%%%%%%%%%%%%%%%%%%%%%%%%%%%%%%%%%%%%%%%%%%%%%%%%%%%%%%%%%%%%%%%%%%%%

%===============================================================================
\subsection{Angularities}
%===============================================================================

The angularities $e_\al$ are a one-parameter family of global $e^+e^-$ event shapes, defined as~\cite{Berger:2003iw}
%%%
\begin{align} \label{eq:ea_def}
  e_\al &= \frac{1}{Q} \sum_i E_i (\sin \theta_i)^{2-\al} (1-|\cos \theta_i|)^{\al-1} 
\,,\end{align}
%%%
where $Q$ is the center-of-mass energy.
The sum runs over all particles $i$, where $E_i$ denotes its energy and $\theta_i$ its angle with respect to an appropriately chosen axis. Smaller values of angularities correspond to more collimated radiation, where the parameter $\al$ determines the weight of the angle. Our convention for $\al$ is such that for small angles, 
%%%
\begin{align}
(\sin \theta_i)^{2-\al} (1-|\cos \theta_i|)^{\al-1} \approx 2^{1-\al} \,\theta_i^\al
\,,\end{align}
%%%
{\it i.e.}~$\al=2$ corresponds to thrust~\cite{Farhi:1977sg} and $\al=1$ to (total) broadening~\cite{Rakow:1981qn,Catani:1992jc} when calculated with respect to the thrust axis.

For angularities with $\al \gtrsim 2$, the direction of the thrust axis is insensitive to (recoil by) soft radiation, but as $\al \to 1$, and certainly for $\al \leq 1$, this effect cannot be ignored~\cite{Dokshitzer:1998kz}. Thus we find it convenient to use an axis that is recoil-insensitive~\cite{Larkoski:2014uqa}. This is accomplished by clustering the event with exclusive $k_T$~\cite{Catani:1991hj}, which splits the event into two jets, using the WTA recombination scheme~\cite{Salam:WTAUnpublished,Bertolini:2013iqa}.\footnote{One can also run e.g.~anti-$k_T$~\cite{Cacciari:2008gp} in exclusive mode with WTA recombination scheme. At the accuracy we are working, there is no difference, and this is corroborated by both \Pythia and \Eventtwo. Alternatively, the broadening axis~\cite{Larkoski:2014uqa} can be used but this is more complicated to implement.} The angle $\theta_i$ in \eq{ea_def} will be taken with respect to the axis of the jet the particle belongs to, so there is no global axis for the event. 

In our previous publication~\cite{Procura:2014cba}, we focussed on jet-based angularities~\cite{Almeida:2008yp,Ellis:2010rwa}. 
However, since the correlation between soft radiation inside and outside the jet makes these observables theoretically more complicated, introducing non-global logarithms~\cite{Dasgupta:2001sh}, we shall limit ourselves here to event-based angularities.

%===============================================================================
\subsection{Power counting and modes for double angularity measurements}
%===============================================================================

\begin{figure}[t]
  \centering
   \includegraphics[width=0.5\textwidth]{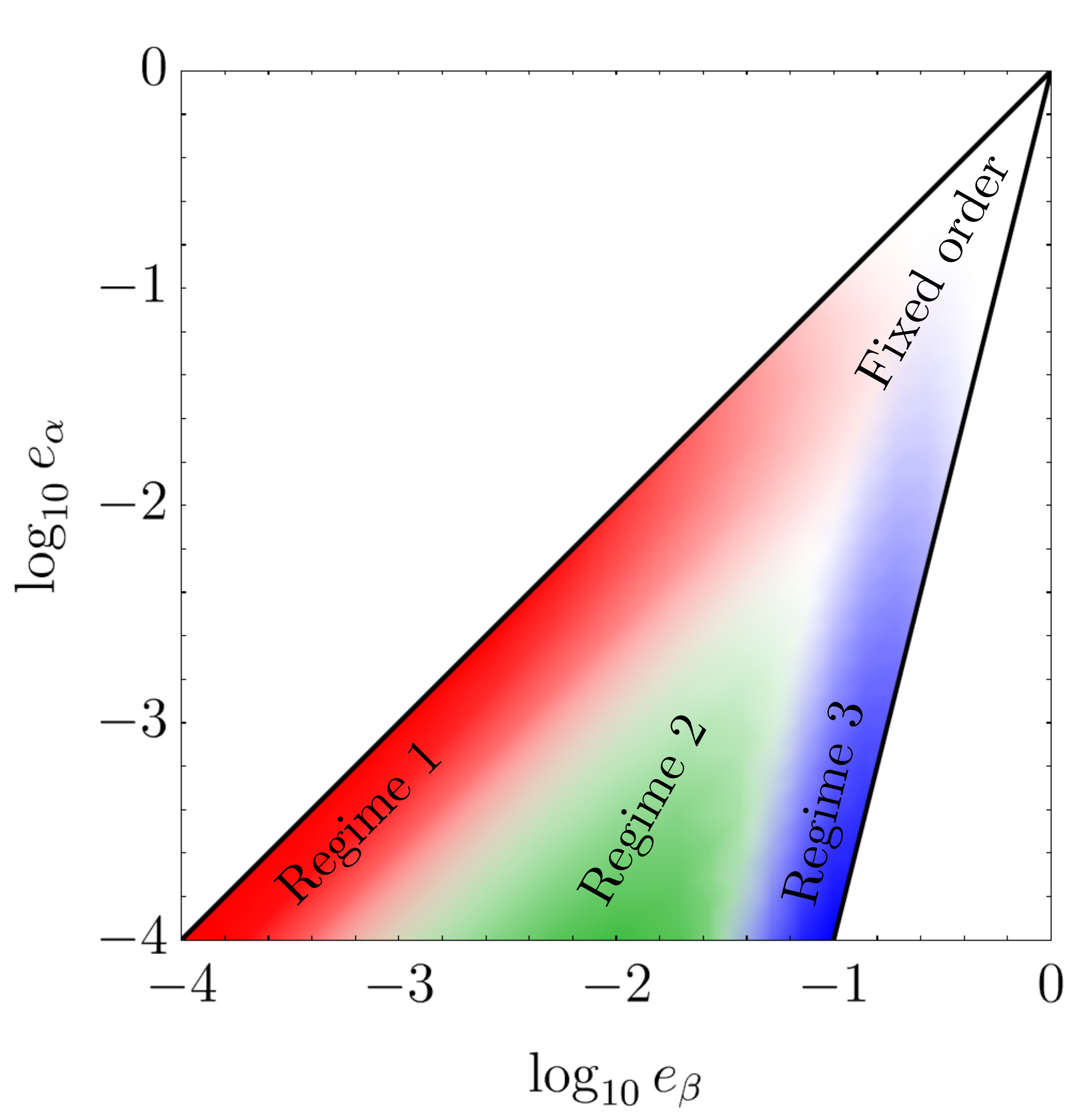}
   \caption{The phase space for the simultaneous measurement of $\ea$ and $\eb$ and the various regimes of Soft-Collinear Effective Theory.}
   \label{fig:regions}
  \end{figure}

We calculate the $e^+ e^- \to 2$ jets cross section differential in two angularities $e_\al$ and $e_\bt$ taking into account the fact that the phase space is characterized by three different regions
(\fig{regions}), corresponding to
%%%
\begin{align} \label{eq:regimes}
 {\rm Regime}\ 1:~e_\beta \sim e_\alpha
 \,, \qquad
 {\rm Regime}\ 2:~e_\beta \gg e_\alpha \gg e_\beta^{\alpha/\beta}
 \,, \qquad
 {\rm Regime}\ 3:~e_\alpha \sim  e_\beta^{\alpha/\beta}
\,,\end{align}
%%%
each one with its own factorization theorem that enables the resummation of logarithms of $e_\al$ and $e_\bt$. Regime 1 and 3 correspond to the boundaries and were discussed in ref.~\cite{Larkoski:2014tva}, while we obtained the factorization theorem for regime 2 describing the bulk of the phase space in ref.~\cite{Procura:2014cba}. 
%---------------------------------------
%---------------------------------------

   \begin{table}
   \centering
   \begin{tabular}{l|ccc}
     \hline \hline
      &  Regime 1  & \hspace{15ex} Regime 2 \hspace{15ex} & Regime 3 \\
     Mode &   $e_\beta \sim e_\alpha$  & \hspace{15ex} $e_\beta \gg e_\alpha \gg e_\beta^{\alpha/\beta}$ \hspace{13ex} & $e_\alpha \sim  e_\beta^{\alpha/\beta}$ \\ \hline
     $n$-coll. & $(1,e_\bt^{2/\bt},e_\bt^{1/\bt})$ & \hspace{15ex}$(1,e_\bt^{2/\bt},e_\bt^{1/\bt})$ \hspace{15ex} & $(1,e_\al^{2/\al},e_\al^{1/\al})$ \\     
     $\bn$-coll. & $(e_\bt^{2/\bt},1,e_\bt^{1/\bt})$ & \hspace{15ex} $(e_\bt^{2/\bt},1,e_\bt^{1/\bt})$ \hspace{15ex} & $(e_\al^{2/\al},1,e_\al^{1/\al})$  \\
     $n$-csoft & \multicolumn{3}{c}{\hspace{13ex}$\big((e_\al^{-\bt} e_\bt^\al)^{1/(\al-\bt)}, ( e_\al^{2-\bt} e_\bt^{\al-2})^{1/(\al-\bt)}, 
     (e_\al^{1-\bt} e_\bt^{\al-1})^{1/(\al-\bt)}\big)$\hspace{13ex}\phantom{.}} \\ 
     $\bn$-csoft & \multicolumn{3}{c}{\hspace{13ex}$\big(( e_\al^{2-\bt} e_\bt^{\al-2})^{1/(\al-\bt)}, (e_\al^{-\bt} e_\bt^\al)^{1/(\al-\bt)}, 
     (e_\al^{1-\bt}  e_\bt^{\al-1})^{1/(\al-\bt)}\big)$\hspace{13ex}\phantom{.}} \\ 
     soft & $(e_\bt, e_\bt, e_\bt)$ & \hspace{15ex} $(e_\al, e_\al, e_\al)$ \hspace{15ex} & $(e_\al, e_\al, e_\al)$  \\
     \hline \hline
   \end{tabular}
   \caption{The parametric size of the light-cone components of the momenta $(p^-,p^+,p_\perp^\mu)/Q$ of the various degrees of freedom in SCET.}
   \label{tab:modes}
   \end{table}

We will briefly review how the regimes in \eq{regimes} arise, and present the factorization theorems in the next section. The relevant modes (degrees of freedom) in the framework of Soft-Collinear Effective Theory (SCET)~\cite{Bauer:2000ew, Bauer:2000yr, Bauer:2001ct, Bauer:2001yt} are summarized in table~\ref{tab:modes}. In SCET, the real radiation in the two-jet region is either collinear or soft. The corresponding momenta have the following parametric scaling
%%%
\begin{align} \label{eq:pc}
  p_n^\mu  \sim Q(1, \la_c^2, \la_c)
  \,, \qquad
  p_{\bar n}^\mu  \sim Q(\la_c^2, 1, \la_c)
  \,, \qquad
  p_s^\mu \sim Q(\la_s, \la_s, \la_s)
\,,\end{align}
%%%
in terms of light-cone coordinates
%%%
\begin{align}
 p^\mu =  (p^-, p^+, p_\perp^\mu) = p^- \frac{n^\mu}{2} + p^+ \frac{\bn^\mu}{2} + p_\perp^\mu
\,.\end{align}
%%%
Here $n^\mu$ and $\bn^\mu$ are light-like vectors along the axes used to define the angularities in \eq{ea_def}, and $n \cdot \bn = 2$. 

The scaling of $\la_c$ and $\la_s$ in \eq{pc} is fixed by the measurement: the parametric size of the contribution of collinear or soft radiation to the angularities simplifies to
%%%
\begin{align} \label{eq:e_pc}
   e_\al \sim \la_c^\al + \la_s 
   \,, \qquad
   e_\bt \sim \la_c^\bt + \la_s
\,.\end{align}
%%%
Assuming $\al>\bt$ for definiteness, this implies $\la_c \sim e_\bt^{1/\bt}$ and $\la_s \sim e_\al$. A consistent theory is obtained if $\la_c^\al \sim \la_s$ or $\la_c^\bt \sim \la_s$, which correspond to regime 3 and 1 in \eq{regimes}. In regime 2 there is an additional collinear-soft mode, whose power counting 
%%%
\begin{align}
  p_{n,cs}^\mu \sim Q \big(\la_{cs}^-, \la_{cs}^+, (\la_{cs}^- \la_{cs}^+)^{1/2}\big)
  \,, \qquad
  p_{\bar n,cs}^\mu \sim Q \big(\la_{cs}^+, \la_{cs}^-, (\la_{cs}^- \la_{cs}^+)^{1/2}\big)
\,,\end{align}
%%%
is uniquely fixed by requiring that it contributes to \emph{both} $e_\al$ and $e_\bt$,
%%%
\begin{align}
  e_\al \,Q \sim \la_{cs}^- (\la_{cs}^+/\la_{cs}^- )^{\al/2}
  \,, \qquad
  e_\bt \,Q \sim \la_{cs}^- (\la_{cs}^+/\la_{cs}^- )^{\bt/2}
\end{align}
%%%
when $\la_{cs}^+<\la_{cs}^-$. This leads to
%%%
\begin{align}  
  \la_{cs}^- \sim (e_\al^{-\bt} \,e_\bt^\al)^{1/(\al-\bt)}
  \,, \qquad
  \la_{cs}^+ \sim ( e_\al^{2-\bt} \,e_\bt^{\al-2})^{1/(\al-\bt)}
\,,\end{align}
%%%
and $\la_{cs}^- \lra \la_{cs}^+$ for the collinear-soft mode in the other direction.
These extensions of SCET have been named \SCETab\cite{Bauer:2011uc,Procura:2014cba,Larkoski:2015zka,Pietrulewicz:2016nwo}. As one approaches regime 1 and 3 from regime 2, the collinear-soft mode merges with the soft mode or the collinear mode, respectively. 

%===============================================================================
\subsection{Factorization}
%===============================================================================

Before presenting the factorization theorems for the various regimes, we want to point out that these all describe the full cross section up to power corrections, 
%%%
\begin{align} \label{eq:si_pc}
\frac{\df^2 \si}{\df e_\al \, \df e_\bt} &= \frac{\df^2 \si_{1}}{\df e_\al \, \df e_\bt}\, 
\big[1+ \mathcal{O}\big(e_\bt^{\min(2/\bt,1)} \big)\big]
\,, \nn \\
\frac{\df^2 \si}{\df e_\al \, \df e_\bt} &= \frac{\df^2 \si_{2}}{\df e_\al \, \df e_\bt}\, 
\bigg\{1+ \mathcal{O}\bigg[\bigg(\frac{e_\bt}{e_\al^{\bt/\al}}\bigg)^{\frac{\al \min(2/\bt,1)}{\al-\bt}}, 
\bigg(\frac{e_\al}{e_\bt}\bigg)^{\frac{\al \min(2/\al,1)}{\al-\bt}} \bigg] \bigg\}
\,, \nn \\
\frac{\df^2 \si}{\df e_\al \, \df e_\bt} &= \frac{\df^2 \si_{3}}{\df e_\al \, \df e_\bt}\, 
\big[1+ \mathcal{O}\big(e_\al^{\min(2/\al,1)}\big)\big]
\,.\end{align}
%%%
Regime 2 resums the most logarithms but also involves two expansions. Starting from regime 2 and approaching either of the phase space boundaries, one of the power corrections becomes of order one and the other smoothly matches onto the power correction for regime 1 or 3, respectively. We will discuss how to combine these formulae to obtain predictions throughout phase space in \sec{matching}.

In regime 1, the power counting in \eqs{regimes}{e_pc} implies that collinear and soft radiation both contribute to $e_\bt$ but only soft radiation contributes to $e_\al$. This leads to the following factorization theorem~\cite{Larkoski:2014tva}, 
%%%
\begin{align} \label{eq:si_1}
\frac{\df^2 \si_{1}}{\df e_\al \, \df e_\bt} &= \hat{\sigma}_0 H (Q^2,\mu) \int\! \df (Q^\bt e_\bt^n)\, J (Q^\bt e_\bt^n,\mu)  \int\! \df (Q^\bt e_\bt^\bn)\,  J (Q^\bt e_\bt^\bn,\mu)\,
\nn \\ & \quad \times 
\int\! \df (Q e_\al^s)\, \df (Q^\bt e_\bt^s)\, 
 S (Q e_\al^s,Q^\bt e_\bt^s,\mu)\, \de(e_\al-e_\al^s)\, \de(e_\bt-e_\bt^n-e_\bt^\bn-e_\bt^s)
\,.\end{align}
%%%
Here, $\hat{\sigma}_0$ denotes the Born cross section, the hard function $H$ contains hard virtual corrections, the jet functions $J$ describe the contribution of collinear radiation to $e_\bt$, and the soft function $S$ accounts for the contribution of soft radiation to $e_\al$ and $e_\bt$. Their expressions at one loop are collected in \sec{fixed_order}. The delta functions simply sum the various contributions, since angularities are additive. This is basically the factorization theorem for a single angularity $e_\bt$~\cite{Hornig:2009vb,Bauer:2008dt}, with a soft function that is differential in $e_\al$ too. 

Similarly, in regime 3 only collinear radiation contributes to $e_\bt$, but collinear and soft radiation contributes to $e_\al$, leading to~\cite{Larkoski:2014tva}\footnote{This setup was already considered in ref.~\cite{Jain:2011iu} for initial-state radiation in $pp \to 0$ jets with $\al=2$ and $\bt=1$.}
%%%
\begin{align} \label{eq:si_3}
\frac{\df^2 \si_{3}}{\df e_\al \, \df e_\bt} &= 
\hat{\sigma}_0 H (Q^2,\mu)  \int\! \df (Q e_\al^n)\,\df (Q^\bt e_\bt^n)\,  J(Q e_\al^n,Q^\bt e_\bt^n,\mu) 
\int\! \df (Q e_\al^\bn)\, \df (Q^\bt e_\bt^\bn)\, J(Q e_\al^\bn,Q^\bt e_\bt^\bn,\mu) 
\nn \\ & \quad \times
\int\! \df (Q e_\al^s)\, S (Q e_\al^s,\mu) \,
\de(e_\al-e_\al^n-e_\al^\bn-e_\al^s)\, \de(e_\bt-e_\bt^n-e_\bt^\bn)
\,.\end{align}
%%%
This is the factorization theorem for $e_\al$ but with double differential jet functions.

Finally, the factorization theorem for regime 2 is given by~\cite{Procura:2014cba}
%%%
\begin{align} \label{eq:si_2}
 \frac{\df^2 \si_{2}}{\df e_\al \, \df e_\bt} &= 
\hat{\sigma}_0 H (Q^2,\mu) 
\int \! \df (Q^\bt e_\bt^n)\,J(Q^\bt e_\bt^n,\mu)
\int \! \df  (Q e_\al^{ns})\, \df (Q^\bt e_\bt^{ns})\,  \cS(Q e_\al^{ns}, Q^\bt e_\bt^{ns},  \mu) 
\nn \\ & \quad \times
\int \! \df (Q^\bt e_\bt^\bn)\,J(Q^\bt e_\bt^\bn,\mu)
\int \! \df  (Q e_\al^{\bn s})\, \df (Q^\bt e_\bt^{\bn s})\, \cS(Q e_\al^{\bn s}, Q^\bt e_\bt^{\bn s}, \mu) 
\nn \\ & \quad \times 
\int\! \df (Q e_\al^s)\, S (Q e_\al^s,\mu)\,
\de(e_\al - e_\al^{ns} - e_\al^{\bn s} -e_\al^s )\, \de(e_\bt - e_\bt^n -e_\bt^\bn - e_\bt^{ns}- e_\bt^{\bn s})
\,,\end{align}
%%%
where the collinear-soft function $\cS$ accounts for the contribution of collinear-soft radiation to $e_\al$ and $e_\bt$. The jet functions $J$ are the same as in \eq{si_1} and the soft function $S$ is the same as in \eq{si_3}.

Expanding $\si_1$ and $\si_3$ in the \SCETab regime described by $\si_2$ (when resummation is turned off), we obtain the following consistency relations\footnote{The power corrections to this equation are smaller than in \eq{si_pc}, because the former describes the power corrections of regime 2 with respect to 1 and 3, whereas the latter also contains the power corrections with respect to the full cross section.}
%%%
\begin{align}\label{eq:consistency}
    J(Q e_\al,Q^\bt e_\bt,\mu) &= 
    \int \! \df (Q^\bt e_\bt^n)\,J(Q^\bt e_\bt^n,\mu)
    \int \! \df  (Q e_\al^{ns})\, \df (Q^\bt e_\bt^{ns})\,  \cS(Q e_\al^{ns}, Q^\bt e_\bt^{ns},  \mu) 
     \\ & \quad \times
     \de(e_\al - e_\al^{ns})\,  \de(e_\bt - e_\bt^n -e_\bt^\bn)
     \bigg\{1+ \mathcal{O}\bigg[\bigg(\frac{e_\bt}{e_\al^{\bt/\al}}\bigg)^{\frac{\al}{\al-\bt}} \bigg] \bigg\}
    \,, \nn \\
    S (Q e_\al,Q^\bt e_\bt,\mu) &=
    \int \! \df  (Q e_\al^{ns})\, \df (Q^\bt e_\bt^{ns})\,  \cS(Q e_\al^{ns}, Q^\bt e_\bt^{ns},  \mu) 
    \int \! \df  (Q e_\al^{\bn s})\, \df (Q^\bt e_\bt^{\bn s})
    \nn \\ & \quad \times
     \cS(Q e_\al^{\bn s}, Q^\bt e_\bt^{\bn s}, \mu)  \int\! \df (Q e_\al^s)\, S (Q e_\al^s,\mu)    
    \nn \\ & \quad \times
    \de(e_\bt - e_\bt^{ns}- e_\bt^{\bn s})\, \de(e_\al - e_\al^{ns} - e_\al^{\bn s} -e_\al^s )
    \bigg\{1+ \mathcal{O}\bigg[\bigg(\frac{e_\al}{e_\bt}\bigg)^{p\,} \bigg] \bigg\}    
\,.\nn\end{align}
%%%
We have verified these relation at one-loop order using the expressions in \sec{fixed_order}. In the second case, the power corrections turn out to vanish at this order, so we could not determine the exponent $p>0$.

%===============================================================================
\subsection{Fixed-order ingredients}
\label{sec:fixed_order}
%===============================================================================

In this section we collect all fixed-order ingredients needed for our numerical analysis, some of which we calculated ourselves. We use the perturbative expansion
%%%
\begin{align}
F &=\sum_n \Big(\frac{\alpha_s(\mu)}{4 \pi}\Big)^n\, F^{(n)}
\,,\end{align}
%%%
where $F=H, J, S, \cS$, and give $F^{(0)}$ and $F^{(1)}$. The following shorthand notation for plus distributions is used
%%%
\begin{align}
  \cL_n(x) \equiv \Big[\frac{\ln^n x}{x}\Big]_+
\,.\end{align}
%%%
These functions have been computed before. In our independent calculations, however, we found some typos in the literature concerning the double differential jet and soft functions, which we correct here. All one-loop ingredients are presented in the form we implemented in our numerical analysis, and are written in such a way to make it straightforward to carry out the convolutions appearing in formulae for the factorized cross section.

%~~~~~~~~~~~~~~~~~~~~~~~~~~~~~~~~~~~~~~~~~~~~~~~~~~~~~~~~~~~~~~~
\subsubsection{Hard function}
%~~~~~~~~~~~~~~~~~~~~~~~~~~~~~~~~~~~~~~~~~~~~~~~~~~~~~~~~~~~~~~~

The hard function entering all aforementioned factorization theorems encodes virtual corrections in the $q \bar{q}$- production at the hard scale $Q$, and is given by the square of the Wilson coefficient in the matching of QCD onto SCET currents~\cite{Manohar:2003vb,Bauer:2003di},
%%%
\begin{align}
H^{(0)} (Q^2,\mu) &= 1
\,, \nn \\
H^{(1)} (Q^2,\mu) &= 2 C_F \Big( - \ln^2 \frac{Q^2}{\mu^2} + 3 \ln \frac{Q^2}{\mu^2}  -8 +\frac{7 \pi^2}{6} \Big)
\,.
\end{align}
%%%

%~~~~~~~~~~~~~~~~~~~~~~~~~~~~~~~~~~~~~~~~~~~~~~~~~~~~~~~~~~~~~~~
\subsubsection{Jet functions}
%~~~~~~~~~~~~~~~~~~~~~~~~~~~~~~~~~~~~~~~~~~~~~~~~~~~~~~~~~~~~~~~

The single differential jet function in \eqs{si_1}{si_2} is \cite{Larkoski:2014uqa}
%%%
\begin{align} 
J^{(0)} (Q^\bt e_\bt,\mu) &= \delta (Q^\bt e_\bt) 
\,,\nn \\
J^{(1)} (Q^\bt e_\bt,\mu) &=   \frac{6 C_F}{\bt(\beta-1)}\, \bigg\{\frac{4}{3}\frac{1}{ \mu^\beta}\cL_{1}\Big( \frac{Q^\bt e_\bt}{\mu^\bt}\Big) +
(1-\bt) \frac{1}{\mu^\beta}\cL_{0}\Big( \frac{Q^\bt e_\bt}{\mu^\bt}\Big) +
\bigg[ 1-\frac{19}{6}\bt + \frac{13}{6} \bt^2 
\nn \\ & \quad
+ \pi^2 \Big( -\frac{1}{9} + \frac{\bt}{3} -  \frac{\bt^2}{4} \Big) + ( 1- \bt )  \ln 2 \bigg] \delta (Q^\bt e_\bt)  \bigg\}
\,.\end{align}
%%%
Note that the constant terms differ from those obtained in ref.~\cite{Hornig:2009vb} because we employ the WTA axis. For angularities with $\al \gtrsim 2$, this is the only difference between using the thrust axis or the WTA axis in the factorization formula.\footnote{This is a consequence of consistency: Since the hard and soft functions are insensitive to such axis choice, this cannot affect the anomalous dimensions and thus logarithmic terms.}

The double differential jet function in \eq{si_3} is given by~\cite{Larkoski:2014tva}\footnote{\label{footnote:AID}We have calculated this independently, as eqs.~(A.6) through (A.9) and (A.14) of ref.~\cite{Larkoski:2014tva} contain typos.}
%%%
\begin{align} \label{eq:J_double}
J^{(0)} (Q e_\al,Q^\bt e_\bt,\mu)&= \delta(Q e_\al)\delta(Q^\bt e_\bt)
\,,\nn \\ 
J^{(1)} (Q e_\al,Q^\bt e_\bt,\mu)&=  \frac{4 C_F}{Q^{\bt+1}} \frac{\df}{\df e_\bt} \frac{\df}{\df e_\al}
\Bigg(
\frac{1}{\al(\al\!-\!1)} \ln^2 e_\al
\!+\! \frac{1}{24 \al (\al\!-\!1)}  \bigg[ 
6 \al \ln \frac{Q^2}{\mu^2} \Big(\al \ln \frac{Q^2}{\mu^2} \!+\! 3(1\!-\!\al)\Big)
\nn \\ & \quad 
 +6 (\al-1) (13 \al -6 -6 \ln 2) - (2-3\al)^2 \pi^2
\bigg]
\nn \\ & \quad 
+ \theta \Big( 2^{\frac{\al-\bt}{\bt}} e_\bt^{\frac{\al}{\bt}} - e_\al \Big) \frac{1}{2 \al (\al-1)} \Big(2 \al \ln \frac{Q^2}{\mu^2} + 3(1-\al) \Big) \ln e_\al
\nn \\ & \quad 
+ \theta \Big( e_\al - 2^{\frac{\al-\bt}{\bt}} e_\bt^{\frac{\al}{\bt}}  \Big) \frac{1}{6 \al \bt (\al-\bt) (\al-1)}\, e_\al^{\frac{-\bt}{\al-\bt}}
\bigg\{
- 6 (\al-1) \bt^2\,e_\al^{\frac{\bt}{\al-\bt}} \ln^2 e_\al
\nn \\ & \quad 
+ (\al\!-\!1) \bigg[ (\al\!-\!\bt)^2 \Big(18\, e_\bt^{\frac{\al}{\al-\bt}}\!+(\pi^2 \!-\!9 \!-\!9 \ln 2)e_\al^{\frac{\bt}{\al-\bt}} \Big)
-9 \al (\al\!-\!\bt)\,  e_\al^{\frac{\bt}{\al-\bt}} \ln e_\bt 
\nn \\ & \quad 
- 6 \al^2\,  e_\al^{\frac{\bt}{\al-\bt}} \ln^2 e_\bt \bigg]
+ 12 \al \bt  \bigg[ (\al-1) \ln e_\bt + \frac{\al-\bt}{2} \ln \frac{Q^2}{\mu^2}\bigg]  e_\al^{\frac{\bt}{\al-\bt}} \ln e_\al
\nn \\ & \quad 
- 12\, (\al-1) (\al-\bt)^2\, e_\al^{\frac{\bt}{\al-\bt}}\,{\rm Li}_2 \Big( e_\al^{\frac{-\bt}{\al-\bt}} e_\bt^{\frac{\al}{\al-\bt}} \Big)
\bigg\}
\Bigg)
\,.\end{align}
%%%
In principle one can perform the derivatives, but we find it more convenient to work the cumulative distributions to avoid complicated plus distributions. Note that we can perform the necessary convolutions using cumulative distributions, as discussed in \sec{conv}.

Integrating the double differential jet function over $e_\bt$ yields the single differential jet function of $e_\al$~\cite{Jain:2011iu}
%%%
\begin{align}\label{eq:intdoublejet}
\int \df (Q^\bt e_\bt) \, J^{(1)} (Q^\bt e_\bt,Q e_\al,\mu) = Q^{\al-1}\, J^{(1)}(Q^\al e_\al,\mu)\,.
\end{align}
%%%
This is obvious from comparing \eq{si_1} with the factorization theorem for a single angularity $e_\al$, as the only difference is that the double differential jet function is replaced by this single differential jet function.

%~~~~~~~~~~~~~~~~~~~~~~~~~~~~~~~~~~~~~~~~~~~~~~~~~~~~~~~~~~~~~~~
\subsubsection{Soft functions}
%~~~~~~~~~~~~~~~~~~~~~~~~~~~~~~~~~~~~~~~~~~~~~~~~~~~~~~~~~~~~~~~

The soft function encodes the effects of soft radiation, which is described by a matrix element of eikonal Wilson lines (along the two outgoing quarks) on which the appropriate measurement is performed. The soft function for a single angularity in \eqs{si_3}{si_3} is given by~\cite{Hornig:2009vb}\footnote{Note that our convention for $\al$ differs from ref.~\cite{Hornig:2009vb} by $\al \to 2 - \al$.} 
%%%
\begin{align} \label{eq:S_ea}
S^{(0)} (Q e_\al,\mu) &= \delta (Q e_\al)
\,, \nn \\
S^{(1)} (Q e_\al,\mu) &= \frac{C_F}{\alpha-1}\bigg[ - \frac{16}{\mu}\cL_{1}\Big( \frac{Q e_\al}{\mu}\Big) +\frac{\pi^2}{3}\delta (Q e_\al)  \bigg]
\,.\end{align}
%%%
For the double differential soft function in \eq{si_1} one obtains~\cite{Larkoski:2014tva}\textsuperscript{\ref{footnote:AID}}
%%%
\begin{align} \label{eq:S_double}
S^{(0)}(Q e_\al,Q^\bt e_\bt,\mu) &= \delta (Q e_\al) \delta (Q^\bt e_\bt)
\,, \nn \\ 
S^{(1)}(Q e_\al,Q^\bt e_\bt,\mu) &=   C_F \bigg\{
\bigg[- \frac{16 }{(\bt-1)}\, \frac{1}{\mu^\bt} \cL_{1}\Big( \frac{Q^\bt e_\bt}{\mu^\bt}\Big)
+ 8 \ln \frac{Q^2}{\mu^2}\, \frac{1}{\mu^\bt} \cL_{0}\Big( \frac{Q^\bt e_\bt}{\mu^\bt}\Big)
\nn \\ & \quad
+ \left(- 2(\bt-1) \ln^2 \frac{Q^2}{\mu^2} + \frac{\pi^2}{3 (\bt-1)} \right) \delta (Q^\bt e_\bt) \bigg] \delta (Q e_\al)
\nn \\ & \quad
-\frac{8}{\al-\bt}\, \frac{\df}{\df (Q e_\al)}\, \frac{\df}{\df (Q^\bt e_\bt)}\,
\theta(e_\al)\theta(e_\bt-e_\al)
\bigg[
\ln \frac{Q e_\al}{\mu} - \ln \frac{Q^\bt e_\bt}{\mu^\bt}
\nn \\ & \quad
+\frac{1}{2}(\bt-1)\ln \frac{Q^2}{\mu^2}
\bigg]^2
\bigg\}
\,.\end{align}
%%%
Note that this expression is more complicated because we chose to write it in terms of $Q^\bt e_\bt$ instead of $Q e_\bt$. In particular, the first two lines of the one-loop expression correspond directly to the single differential soft function in \eq{S_ea}, but for $e_\bt$.

Integrating the double differential soft function over $e_\al$ produces the single differential soft function of $e_\bt$~\cite{Procura:2014cba},
%%%
\begin{align}\label{eq:intdoublesoft}
\int \df (Q\, e_\al) \, S^{(1)} (Q^\bt e_\bt,Q e_\al,\mu) = Q^{1-\bt} S^{(1)} (Q e_\bt,\mu)
\,.\end{align}
%%%
This is obvious from comparing \eq{si_1} with the factorization theorem for a single angularity $e_\bt$, as the only difference is that the double differential soft function is replaced by this single differential soft function.

%~~~~~~~~~~~~~~~~~~~~~~~~~~~~~~~~~~~~~~~~~~~~~~~~~~~~~~~~~~~~~~~
\subsubsection{Collinear-soft function}
%~~~~~~~~~~~~~~~~~~~~~~~~~~~~~~~~~~~~~~~~~~~~~~~~~~~~~~~~~~~~~~~
Finally, the collinear-soft function that enters in \eq{si_2} is given by~\cite{Kasemets:2015uus}
%%%
\begin{align}
\cS^{(0)} (Q e_\al,Q^\bt e_\bt,\mu)&= \delta(Q e_\al)\delta(Q^\bt e_\bt)
\,, \\ 
\cS^{(1)} (Q e_\al,Q^\bt e_\bt,\mu)&=  C_F \bigg\{ 
\Big( - \frac{8}{\bt-1} - \frac{8}{\al-\bt} \Big)\delta(Q e_\al) \frac{1}{\mu^\bt} \cL_{1}\Big( \frac{Q^\bt e_\bt}{\mu^\bt}\Big)
\nn \\ & \quad
+\! \Big( \frac{8}{\al \!-\! 1} - \frac{8}{\al-\bt} \Big) \delta(Q^\bt e_\bt) \frac{1}{\mu} \cL_{1}\Big( \frac{Q e_\al}{\mu}\Big)
\!+\!  \frac{8}{\al \!-\! \bt} \frac{1}{\mu} \cL_{0}\Big( \frac{Q e_\al}{\mu}\Big) \frac{1}{\mu^\bt} \cL_{0}\Big( \frac{Q^\bt e_\bt}{\mu^\bt}\Big)
\nn \\ & \quad 
+\! \frac{4 (\al\!-\!1)}{\al\!-\!\bt} \ln \frac{Q^2}{\mu^2}
\delta(Q e_\al) \frac{1}{\mu^\bt} \cL_{0}\Big( \frac{Q^\bt e_\bt}{\mu^\bt}\Big)
\!-\! \frac{4 (\bt\!-\!1)}{\al-\bt}  \ln \frac{Q^2}{\mu^2}
\delta(Q^\bt e_\bt) \frac{1}{\mu} \cL_{0}\Big( \frac{Q e_\al}{\mu}\Big)
\nn \\ & \quad 
+\bigg[ \frac{\pi^2 (\al-\bt)}{6 (\al-1) (\bt-1)} -\frac{(\al-1)(\bt-1)}{ \al-\bt } \ln^2 \frac{Q^2}{\mu^2} \bigg]
\delta(Q e_\al)\delta(Q^\bt e_\bt)
\bigg\}
\,. \nn\end{align}
%%%
This is the simplest double differential function, as it contains pure logarithms. To see that it involves a single scale, it is the easiest to consider the double cumulative distribution, which only involves logarithms of $e_\al^{\bt-1} e_\bt^{1-\al} (Q/\mu)^{\bt-\al}$. 

%===============================================================================
\subsection{Resummation}
%===============================================================================

The factorization theorems enable the resummation of large logarithms of $e_\al$ and $e_\bt$ through renormalization group (RG) evolution, since each ingredient is only sensitive to a single scale. By evaluating the ingredients at their natural scale, where they contain no large logarithms, and evolving them to a common scale, these logarithms get exponentiated.
In this section we give the form of anomalous dimensions and evolution kernels, with explicit expressions provided in \app{RGE}. We also discuss how to perform convolutions with cumulative distributions.

%~~~~~~~~~~~~~~~~~~~~~~~~~~~~~~~~~~~~~~~~~~~~~~~~~~~~~~~~~~~~~~~
\subsubsection{Anomalous dimensions}
%~~~~~~~~~~~~~~~~~~~~~~~~~~~~~~~~~~~~~~~~~~~~~~~~~~~~~~~~~~~~~~~

The RG equation of the hard function is
%%%
\begin{align} \label{eq:H_evo}
\mu \frac{\df}{\df\mu}\, H(Q^2, \mu) &= \gamma_H(Q^2, \mu)\, H(Q^2, \mu) 
\,, \nn \\
 \ga_H(Q^2,\mu) &= 2\Ga_\text{cusp}(\al_s) \ln \frac{Q^2}{\mu^2} + \ga_H(\al_s)
\,.
\end{align}
%%%
Here $\Ga_\text{cusp}$ is the cusp anomalous dimension~\cite{Korchemsky:1987wg}, and $\ga_H(\al_s)$ the non-cusp contribution.
Similarly, for the jet functions
%%%
\begin{align}
\mu \frac{\df}{\df\mu}\, J(Q^{\bt} e_\bt , \mu) &= \int_0^{Q^{\bt} e_\bt}\! \df (Q^{\bt}{e_\bt'})\,  \ga_J(Q^{\bt} e_\bt - Q^{\bt} e_\bt',\mu)\, J(Q^{\bt} e_\bt', \mu) 
\,, \nn \\
\mu \frac{\df}{\df\mu}\, J(Q e_\al,Q^{\bt} e_\bt , \mu) &= \int_0^{Q^{\bt} e_\bt}\! \df (Q^{\bt}{e_\bt'})\,  \ga_J(Q^{\bt} e_\bt - Q^{\bt} e_\bt',\mu)\, J(Q e_\al, Q^{\bt} e_\bt', \mu) 
\,, \nn \\
\ga_J(Q^{\bt} e_\bt,\mu) &= -\frac{2}{\bt-1}\, \Ga_\text{cusp}(\al_s)\, \frac{1}{\mu^{\bt}} \cL_0\Big(\frac{Q^{\bt} e_\bt}{\mu^{\bt}}\Big) + \ga_J(\al_s, \bt)\,\de(Q^{\bt} e_\bt)
\,,
\end{align}
for the soft functions
%%%
\begin{align}
\mu \frac{\df}{\df\mu}\, S(Q e_\al, \mu) &= \int_0^{Q e_\al}\! \df (Q{e_\al'})\,  \ga_S(Q e_\al - Q e_\al',\mu)\, S(Q e_\al', \mu) 
\,, \nn \\
\mu \frac{\df}{\df\mu}\, S(Q e_\al, Q^{\bt} e_\bt, \mu) &= \int_0^{Q e_\al}\! \df (Q{e_\al'})\,  \ga_S(Q e_\al - Q e_\al', \mu)\, S(Q e_\al', Q^{\bt} e_\bt, \mu) 
\,, \nn \\
\ga_S(Q e_\al,\mu) &= \frac{4}{\al-1}\, \Ga_\text{cusp}(\al_s)\, \frac{1}{\mu} \cL_0\Big(\frac{Q e_\al}{\mu}\Big) + \ga_S(\al_s, \al)\,\de(Q e_\al)
\,,\end{align}
%%%
and for the collinear-soft function
%%%
\begin{align}
\mu \frac{\df}{\df\mu}\, \cS(Q e_\al, Q^{\bt} e_\bt, \mu) &= \!\int_0^{Q e_\al}\!\!\! \df (Q{e_\al'})\, \!\int_0^{Q^{\bt} e_\bt}\!\!\!  \df (Q^{\bt}{e_\bt'})\, \ga_\cS(Q e_\al \!\!-\! Q e_\al', Q^{\bt} e_\bt \!-\! Q^{\bt} e_\bt', \mu)\, \cS(Q e_\al', Q^{\bt} e_\bt,  \mu) 
, \nn \\
\ga_\cS(Q e_\al, Q^{\bt} e_\bt, \mu) &= \Ga_\text{cusp}(\al_s) \bigg[- \frac{2}{\al\!-\!1}\, \frac{1}{\mu} \cL_0\Big(\frac{Q e_\al}{\mu}\Big) \de(Q^{\bt} e_\bt) 
\!+\! \frac{2}{\bt\!-\!1}\, \de(Q e_\al) \frac{1}{\mu^\bt} \cL_0\Big(\frac{Q^\bt e_\bt}{\mu^\bt}\Big) 
\nn \\ & \quad
 - \ln \frac{Q^2}{\mu^2} \de(Q e_\al)\,\de(Q^\bt e_\bt) \bigg] + \ga_\cS(\al_s,\al,\bt)\,\de(Q e_\al)\,\de(Q^\bt e_\bt)
\,.\end{align}
%%%

Using these expressions, one can verify that the cross sections in eq.~\eqref{eq:si_1}, \eqref{eq:si_3} and \eqref{eq:si_2} are $\mu$-independent up to the order that we are working, if the following relations hold
%%%
\begin{align} \label{eq:noncusp_rel}
  \ga_H(\al_s) + 2 \ga_J(\al_s, \al)  + \ga_S(\al_s, \al) &= 0
  \,,\nn \\
  \ga_H(\al_s) + 2 \ga_J(\al_s, \bt)  + 2 \ga_\cS(\al_s, \al,\bt) + \ga_S(\al_s, \al)&= 0  
\,.\end{align}
%%%
We have checked this equation at one-loop order, and use it to extract the two-loop non-cusp anomalous dimensions, taking the known results for the hard function and soft function to fix all the others. We stress that an essential ingredient to achieve NNLL accuracy in our analysis is provided by the novel calculation of the two-loop soft anomalous dimension in~\cite{Bell:2018vaa}. Cusp and non-cusp contributions to the anomalous dimensions are collected in \app{RGE}.

%~~~~~~~~~~~~~~~~~~~~~~~~~~~~~~~~~~~~~~~~~~~~~~~~~~~~~~~~~~~~~~~
\subsubsection{Evolution equations}
%~~~~~~~~~~~~~~~~~~~~~~~~~~~~~~~~~~~~~~~~~~~~~~~~~~~~~~~~~~~~~~~

For the hard function, the solution to RG equation in \eq{H_evo} is given by
%%%
\begin{align}
H(\mu) = H(\mu_0) \, \exp \bigl[ K_H (\mu, \mu_0) \bigr] \Bigl(\frac{\mu_0}{Q} \Bigr)^{\omega_H (\mu, \mu_0)} 
\,.\end{align}
%%%
Here $K_H$ and $\omega_H$ are given by
%%%
\begin{align}
K (\mu, \mu_0) &= - 4 K_{\Gamma} (\mu, \mu_0) + K_{\gamma_H} (\mu, \mu_0) 
\,, \qquad
\omega_H (\mu, \mu_0) = -4\, \eta_\Gamma (\mu, \mu_0) \,,
\end{align}
%%%
where $\ga_H$ in the subscript denotes the non-cusp anomalous dimension and
%%%
\begin{align} \label{eq:Keta_def}
K_\Gamma(\mu, \mu_0)
& = \int_{\alpha_s(\mu_0)}^{\alpha_s(\mu)}\!\frac{\df\alpha_s}{\beta(\alpha_s)}\,
\Gamma_\cusp(\alpha_s) \int_{\alpha_s(\mu_0)}^{\alpha_s} \frac{\df \alpha_s'}{\beta(\alpha_s')}
\,,\qquad
\eta_\Gamma(\mu, \mu_0)
= \int_{\alpha_s(\mu_0)}^{\alpha_s(\mu)}\!\frac{\df\alpha_s}{\beta(\alpha_s)}\, \Gamma_\cusp(\alpha_s)
\,,\nn \\
K_{\gamma_F}(\mu, \mu_0)
& = \int_{\alpha_s(\mu_0)}^{\alpha_s(\mu)}\!\frac{\df\alpha_s}{\beta(\alpha_s)}\, \gamma_F(\alpha_s)
\,.\end{align}
%%%
These integrals can be performed analytically in a perturbative expansion, see \app{RGE}.

Similarly for the jet and soft function ($F=J,\,S$), 
%%%
\begin{align} 
F(t_F,\mu) = \int\! \df t_F' U_F(t_F - t_F',\mu,\mu_0)\, F(t_F', \mu_0) \, ,
\end{align}
%%%
where $t_J = Q^\bt e_\bt$\,, $t_S = Q e_\al$. The evolution kernel $U_F$ is given by
%%%
\begin{align} \label{eq:Uevol}
U_F(t_F,\mu,\mu_0) = \frac{\exp\left[K_F(\mu,\mu_0) + \gamma_E \omega_F(\mu,\mu_0)\right]}{\Gamma[1-\omega_F(\mu,\mu_0)]} \biggl[ -\frac{\omega_F(\mu,\mu_0)}{\mu_0^{j_F}} \cL^{-\omega_F(\mu,\mu_0)}\Bigl(\frac{t_F}{\mu_0^{j_F}}\Bigr) + \delta(t_F)\biggr] \, .
\end{align}
%%%
where $j_J = \bt$ and $j_S = 1$. We use the plus distribution
%%%
\begin{align}
  \cL^\eta (x) \equiv \bigg[\frac{\theta(x)}{x^{1-\eta}}\bigg]_+
\end{align}
%%%
and
%%%
\begin{align}
K_J (\mu, \mu_0) &= \frac{2\bt}{\bt - 1} K_{\Gamma} (\mu, \mu_0) + K_{\gamma_J} (\mu, \mu_0) 
\,,&
\omega_J (\mu, \mu_0) &= \frac{2}{\bt-1} \, \eta_\Gamma (\mu, \mu_0) 
\,, \nn  \\
K_S (\mu, \mu_0) &= \frac{4}{1-\al} K_{\Gamma} (\mu, \mu_0) + K_{\gamma_S} (\mu, \mu_0) 
\,,&
\omega_S (\mu, \mu_0) &= \frac{4}{1-\al} \, \eta_\Gamma (\mu, \mu_0) 
\,.\end{align}
%%%
We do not need the evolution kernel for the collinear-soft function, as we choose the collinear-soft scale as the endpoint of our evolution.

%~~~~~~~~~~~~~~~~~~~~~~~~~~~~~~~~~~~~~~~~~~~~~~~~~~~~~~~~~~~~~~~
\subsubsection{Convolutions with cumulative distributions}
\label{sec:conv}
%~~~~~~~~~~~~~~~~~~~~~~~~~~~~~~~~~~~~~~~~~~~~~~~~~~~~~~~~~~~~~~~

The most complicated step in our numerical evaluations is the convolution of the evolution kernel in \eq{Uevol} with the one-loop double differential jet and soft functions in \eqs{J_double}{S_double}. To avoid subtleties with plus functions, we perform these convolutions using cumulants, as follows. For the cumulative distributions  $F$ and $G$, if we want to perform the convolution of $F'$ and $G'$ and take the cumulant of the result, we can rewrite
%%%
\begin{align}
  \int_0^{y_c}\! \df y \int_0^y\! \df x\, F'(x) G'(y-x)
  &= \int_0^{y_c}\! \df x\, \int_0^{y_c-x}\! \df y\, F'(x) G'(y)
  = \int_0^{y_c}\! \df x\, F'(x) G(y_c - x)
  \nn \\
  &= F(y_c) G(y_c) + \int_0^{y_c}\! \df x\, F'(x) [G(y_c - x) - G(y_c)]~.
\end{align}
%%%
Note that since $G(y_c - x) - G(y_c)$ vanishes for $x \to 0$, the final integral does not require a plus prescription for $F'(x)$. In our case it is convenient to take $G$ to be the cumulant of the double differential jet or soft function, so its derivative is never needed.

%===============================================================================
\subsection{The next-to-leading order cross section}
\label{sec:NLO}
%===============================================================================

In this section we first present the calculation of the double angularity cross section at NLO. We subsequently decompose this result into a singular and nonsingular component. By adding the latter to our resummed cross section, the matching at NLO is achieved. We also give the nonsingular contribution for single angularity measurements.

%~~~~~~~~~~~~~~~~~~~~~~~~~~~~~~~~~~~~~~~~~~~~~~~~~~~~~~~~~~~~~~~
\subsubsection{Calculation}
%~~~~~~~~~~~~~~~~~~~~~~~~~~~~~~~~~~~~~~~~~~~~~~~~~~~~~~~~~~~~~~~

Since the virtual corrections are already included in our factorization theorems, we only need the real contribution to calculate the double differential cross section at $\ord{\al_s}$. The final state consists of three massless partons, which can be characterized by their energy fractions $x_i$ in the center of mass frame, normalized to  $x_1+x_2+x_3=2$. Assuming $x_1 \geq x_2 \geq x_3$, partons 2 and 3 get clustered together into one jet by exclusive $k_T$ algorithm~\cite{Catani:1993hr}, because the angle $\theta_{23}$ is smaller than $\theta_{12}$ and $\theta_{13}$. The other jet then only consists of parton 1. Due to the WTA recombination scheme, the jet axes are along the momenta of particles 1 and 2, so that the angularity is determined by the energy fraction $x_3$ and the angle $\theta_{23}$,
%%%
\begin{align} \label{eq:ea_nlo}
 e_\al = \frac12 x_3\, (1-\cos^2 \theta_{23})^{1-\al/2} (1-|\cos \theta_{23}|)^{\al-1}
 \,, \qquad
 \cos \theta_{23} = \frac{2(x_1-1)}{x_2 x_3} +1
\,.\end{align}
%%%

The cross section is then calculated numerically using
%%%
\begin{align}
  \frac{\df^2 \si}{\df x_q\, \df x_{\bar q}} = \hat{\sigma}_0\, \frac{\al_s C_F}{2\pi}\, \frac{x_q^2 + x_{\bar q}^2}{(1-x_q)(1-x_{\bar q})} + \ord{\al_s^2}
\,,\end{align}
%%%
for $x_q, \bar x_q <1$. Specifically, we sample logarithmically in $1-x_q$ and $1-x_{\bar q}$, using a cutoff that is outside our plot ranges.  

Our result is shown in \fig{fullnloandns} for three pairs of angularities with exponents $(\al,\bt)$. From the double differential jet and soft function in \eqs{J_double}{S_double}, we see that in the resummation regime the phase-space boundaries are
%%%
\begin{align} \label{eq:PS_NLO}
  e_\bt \geq e_\al \geq 2^{\frac{\al-\bt}{\bt}} e_\bt^{\frac{\al}{\bt}} 
\,,\end{align}
%%%
at one-loop order.
Note that the lower boundary is slightly shifted compared to the canonical expression in \eq{regimes}. The upper boundary $e_\bt \geq e_\al$ corresponds to $\cos \theta_{23}=0$ and not to one of the phase-space boundaries ($x_1 = x_2$ or $x_1 = 2-2x_2$).

At NNLO, the phase-space boundaries in the resummation regime are 
%%%
\begin{align} \label{eq:PS_NNLO}
  e_\bt \geq e_\al \geq e_\bt^{\frac{\al}{\bt}} 
\,.\end{align}
%%%
The lower boundary follows from considering two one-loop jet functions, whose contribution $e_{\al_i}$, $e_{\bt_i}$ to the respective hemispheres each satisfies \eq{PS_NLO}.\footnote{Contributions from a tree-level jet function combined with a two-loop jet function also satisfy \eq{PS_NNLO}.} 

%~~~~~~~~~~~~~~~~~~~~~~~~~~~~~~~~~~~~~~~~~~~~~~~~~~~~~~~~~~~~~~~
\subsubsection{Fixed-order nonsingular}
\label{sec:FOns}
%~~~~~~~~~~~~~~~~~~~~~~~~~~~~~~~~~~~~~~~~~~~~~~~~~~~~~~~~~~~~~~~

%---------------------------------------
 \begin{figure}[t]
  \centering
    \includegraphics[width=0.49\textwidth]{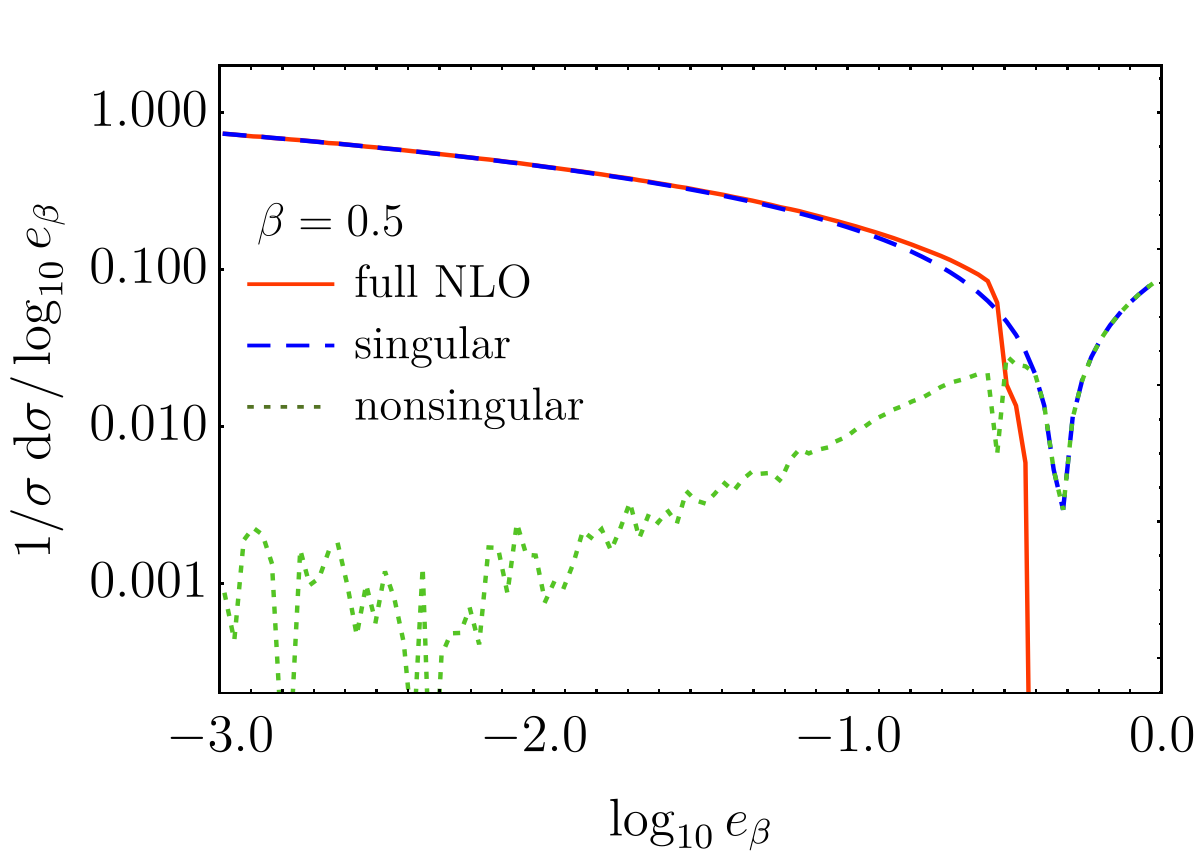}
   \includegraphics[width=0.49\textwidth]{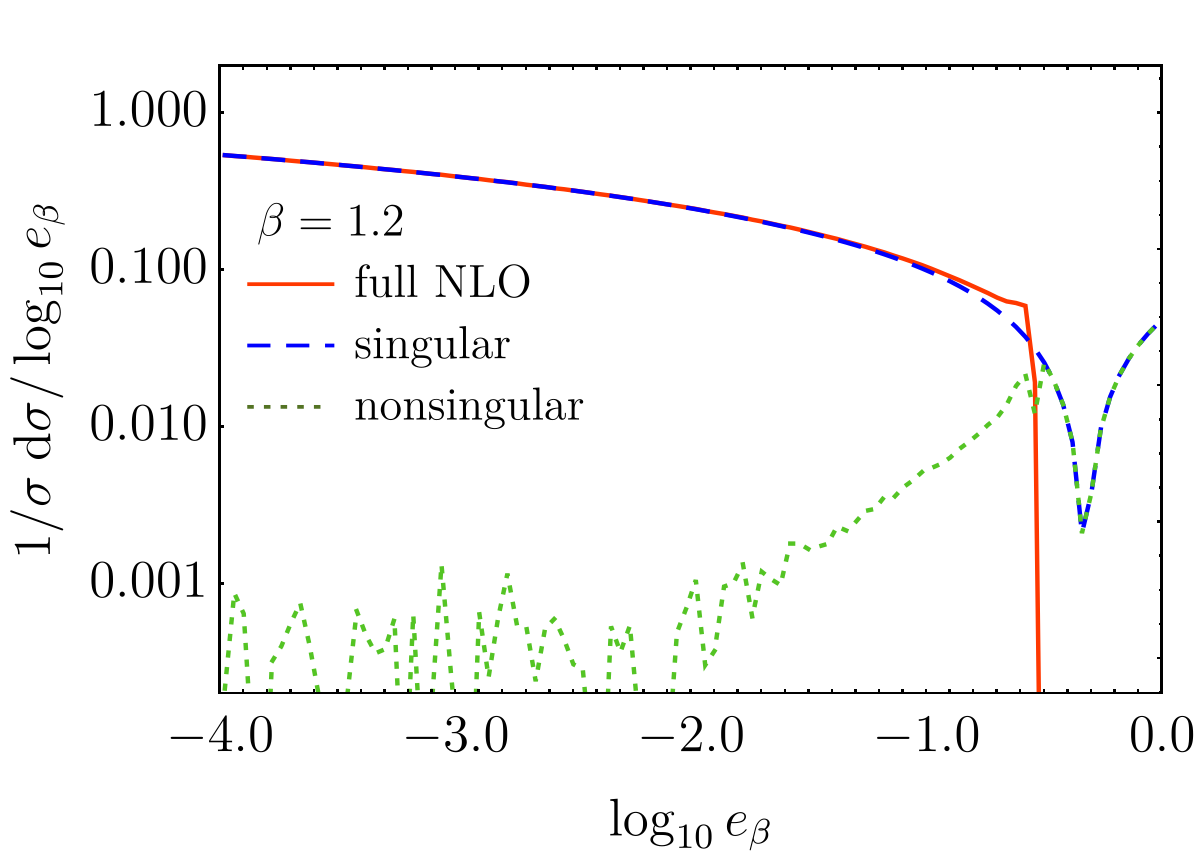}\\
   \includegraphics[width=0.49\textwidth]{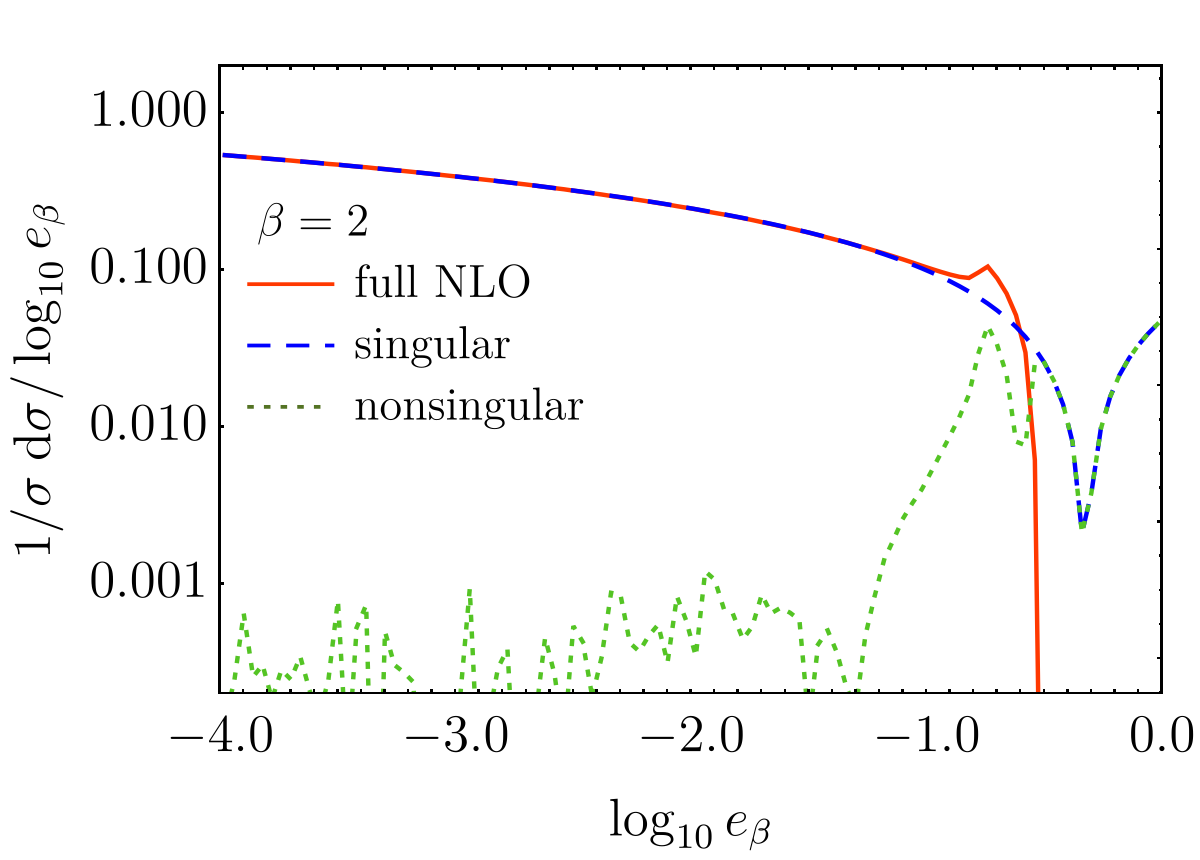}  
   \includegraphics[width=0.49\textwidth]{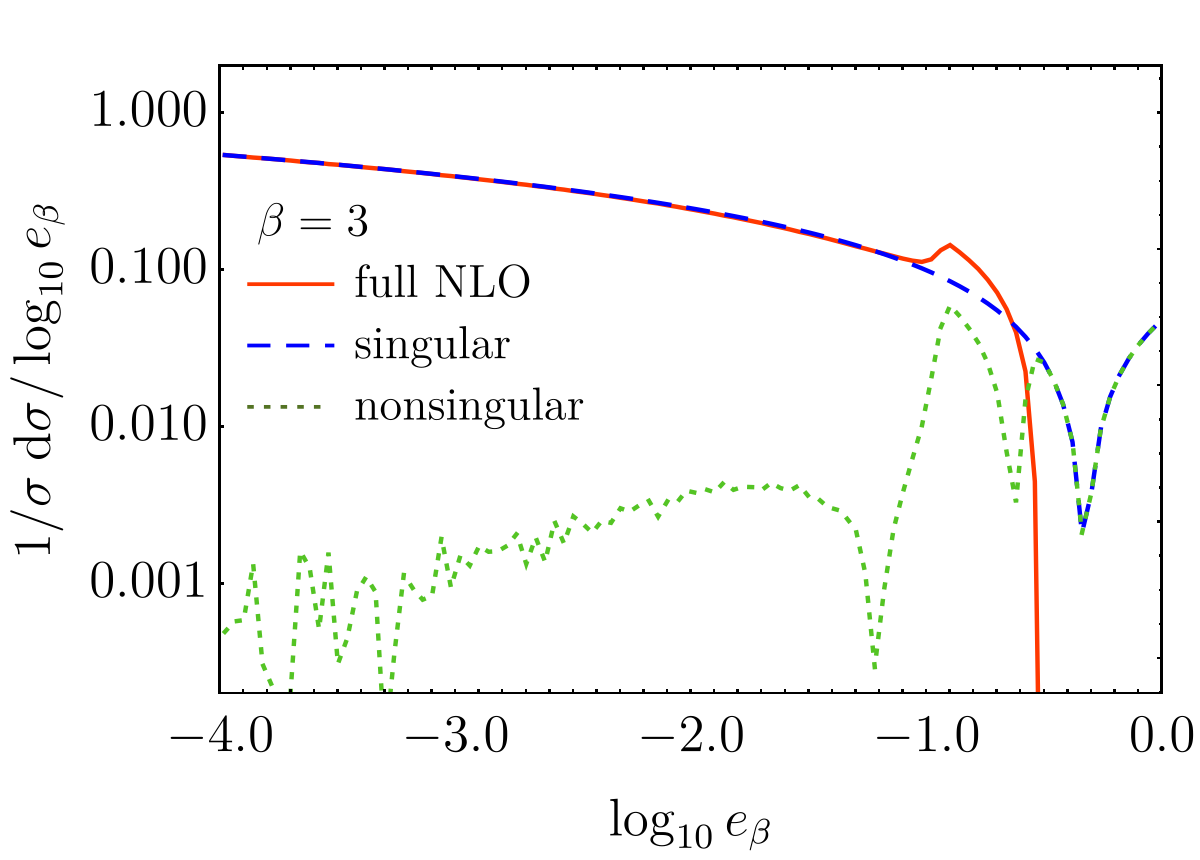}
   \caption{The full NLO (red solid), the NLO singular (blue dashed) and nonsingular (green dotted) cross sections, for four angularities $\bt = 0.5, 1.2, 2$ and $3$.}
   \label{fig:nlons1D}
  \end{figure}
%---------------------------------------

%---------------------------------------
 \begin{figure}[t]
  \centering
    \includegraphics[width=0.325\textwidth]{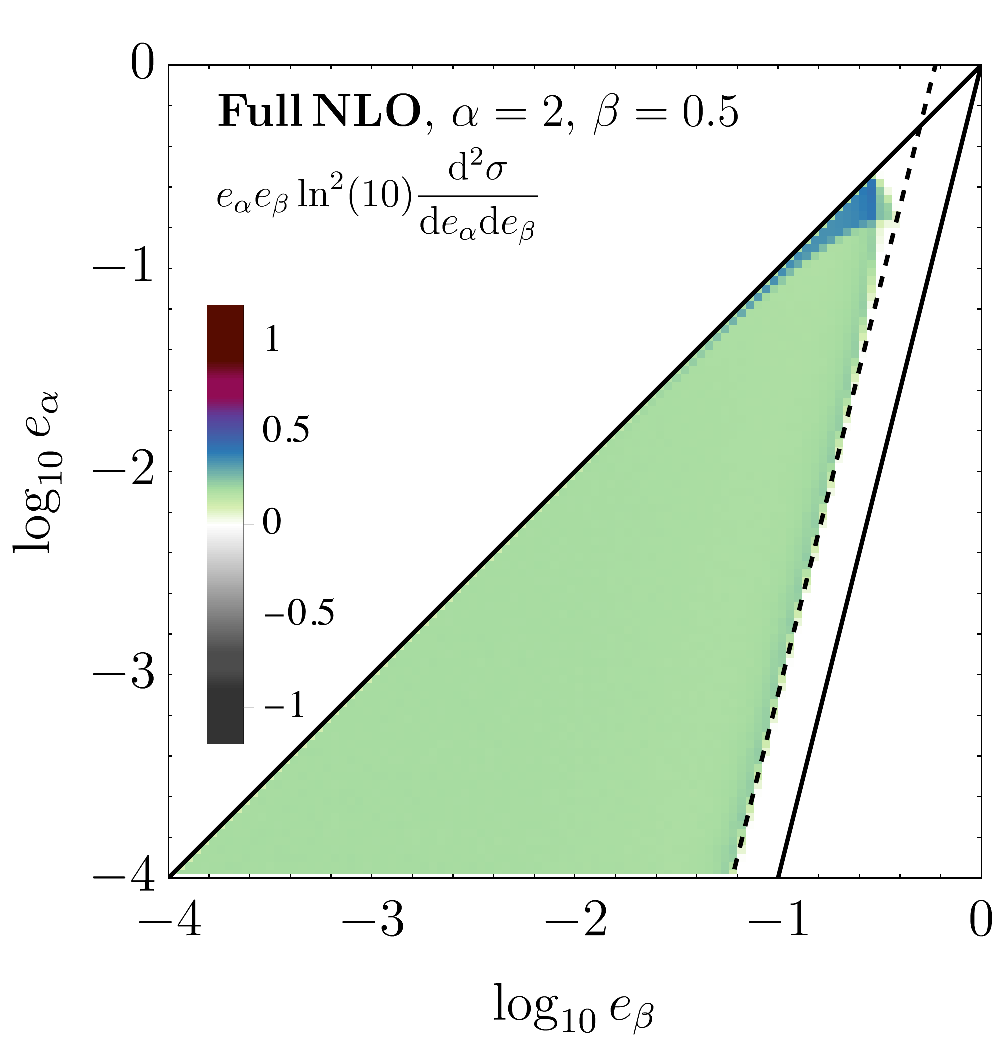}
   \includegraphics[width=0.325\textwidth]{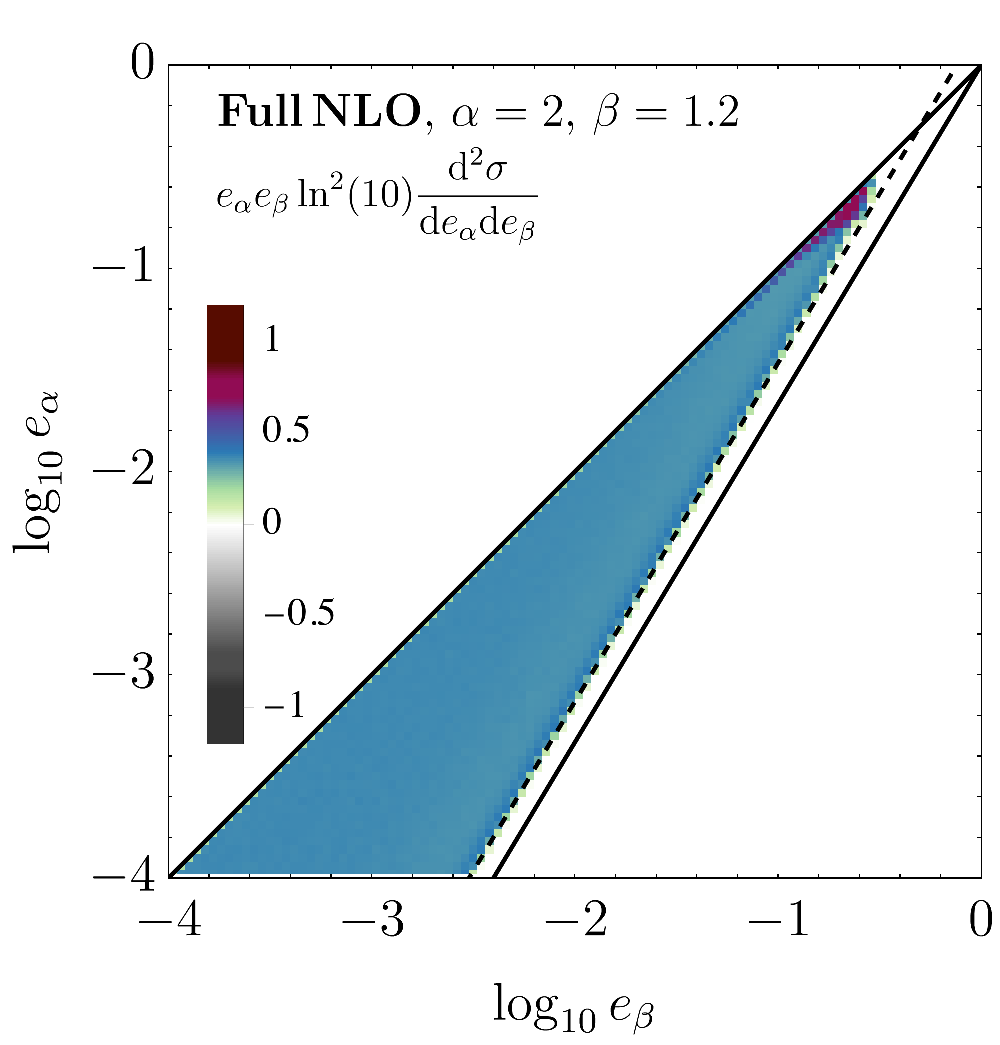}
   \includegraphics[width=0.325\textwidth]{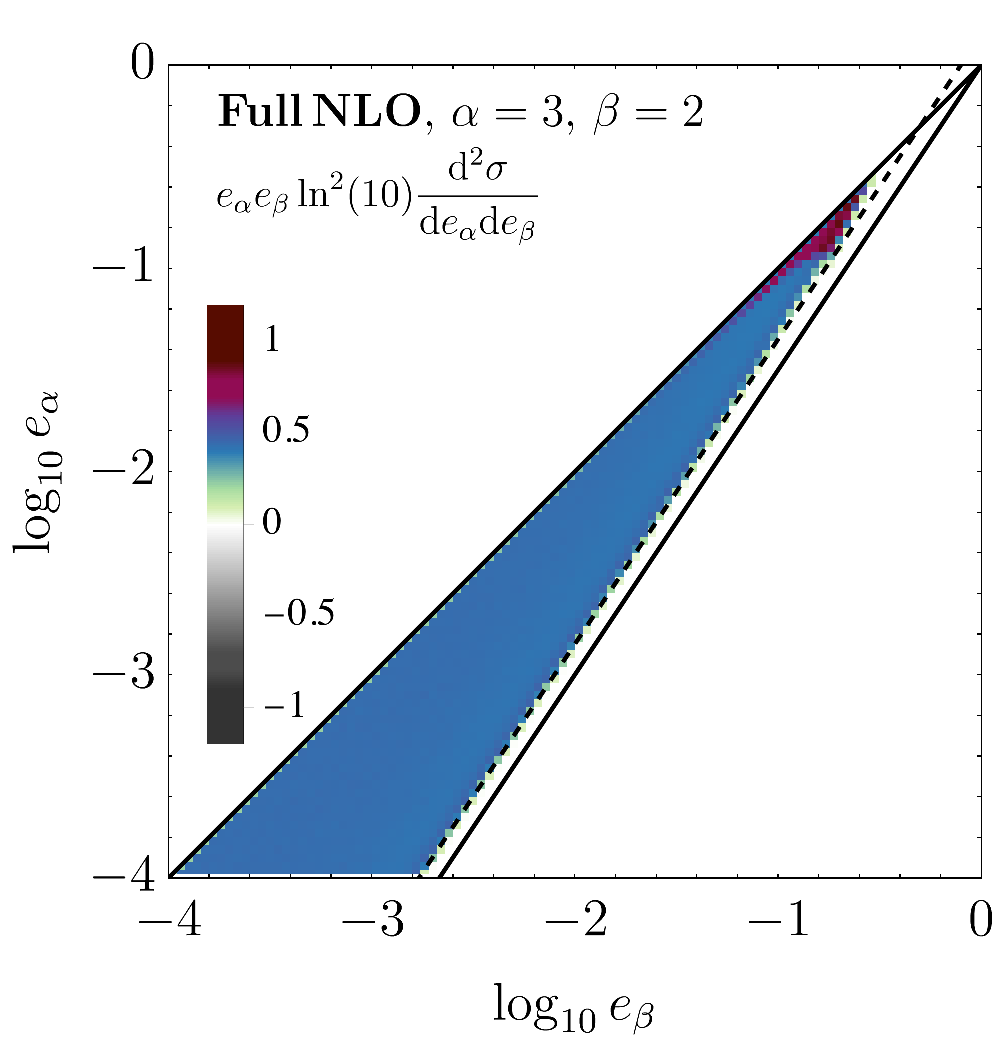}  \\
   \includegraphics[width=0.325\textwidth]{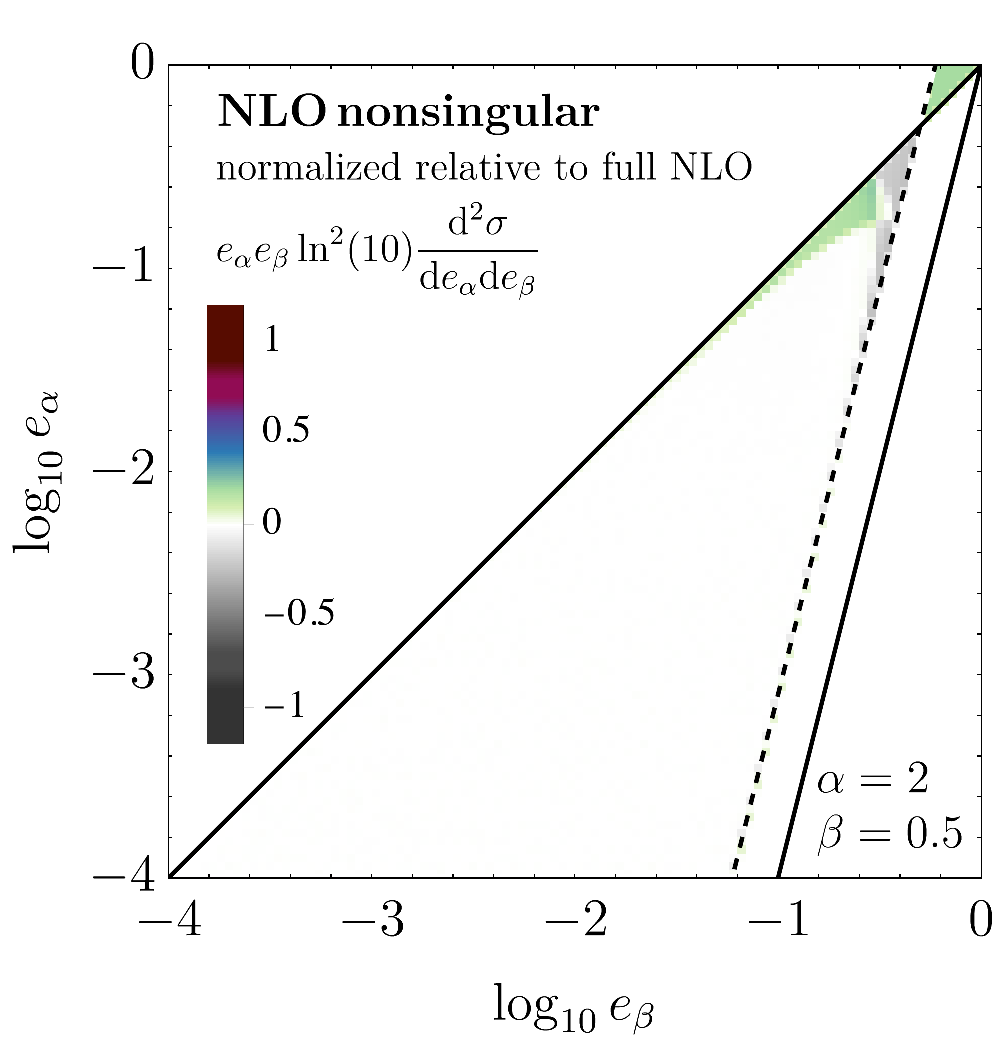}
   \includegraphics[width=0.325\textwidth]{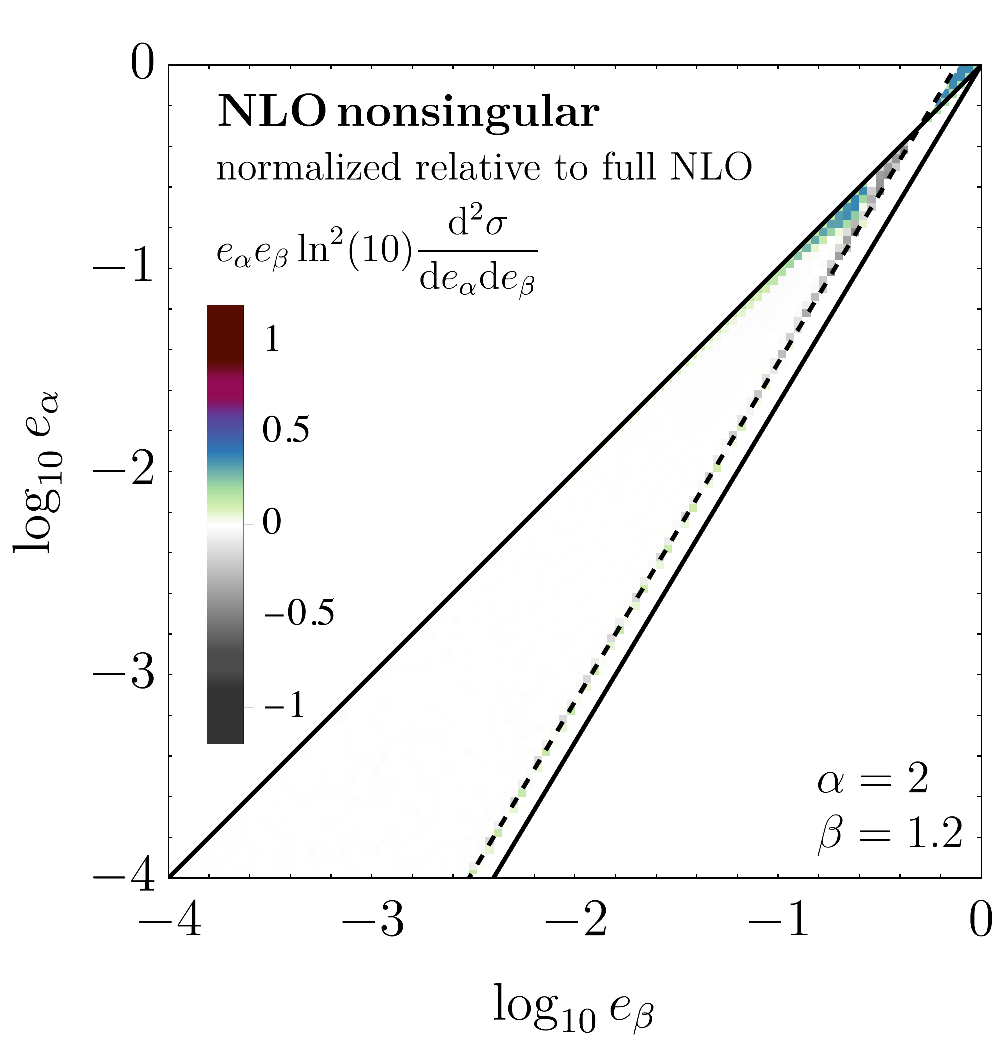}
   \includegraphics[width=0.325\textwidth]{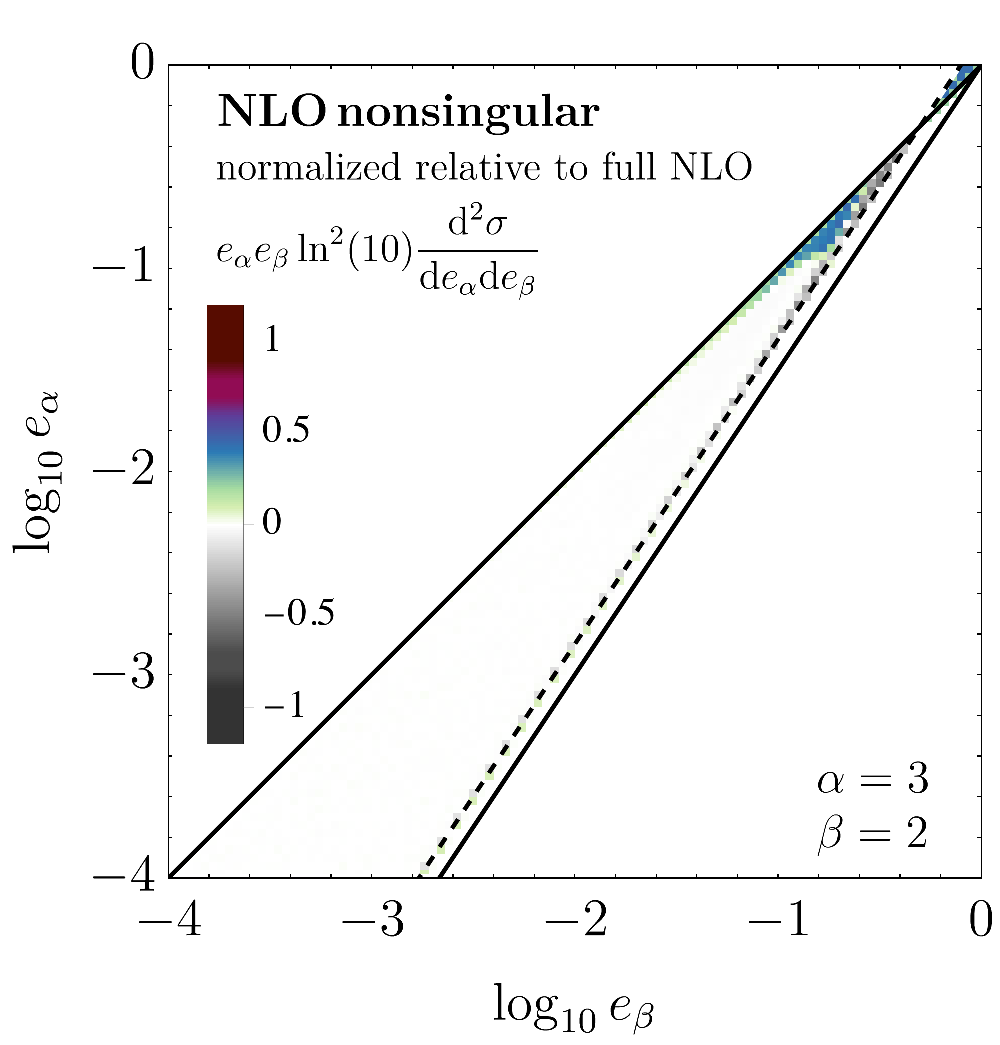}  
   \caption{The full NLO and NLO nonsingular cross sections, for three pairs of angularities
   $(\al,\bt) = (2, 0.5), (2, 1.2) \text{ and } (3,2)$.
}
   \label{fig:fullnloandns}
  \end{figure}
%---------------------------------------
We start by showing the nonsingular cross section for a single angularity in \fig{nlons1D}, for four representative angularity exponents $\bt = 0.5, 1.2, 2$ and $3$. Here $Q=1000$ GeV.\footnote{All plots in this paper are for $Q=1000$ GeV, unless another $Q$ value is explicitly specified.}
The full NLO cross section is normalized to 1. The singular and nonsingular cross sections are rescaled by the same amount as the full NLO.
For small values of the angularities, the singular contribution to the cross section dominates and the nonsingular cross section is power suppressed, demonstrating the validity of the factorization theorem at this order.\footnote{The noise in the nonsingular at small values of the angularity is due to the limited statistics of Monte Carlo integration, and is irrelevant for our final results.} On the other hand, at large values of the angularity, the singular and nonsingular contributions become equal in size, and matching to the NLO is important to correctly describe the cross section in this region. Indeed beyond the endpoint of the distribution, where the cross section vanishes, the singular and nonsingular are exactly equal and opposite in sign. Thus the resummation must be turned off in this region through an appropriate scale choice (see \sec{prof_1D}) to maintain this cancellation. The bump in the cross section in the fixed order region arises because we use the WTA axis rather than the thrust axis. (This bump corresponds to the Sudakov shoulder observed in ref.~\cite{Banfi:2018mcq}.)

Moving on to two angularities, \fig{fullnloandns} shows the total NLO cross sections and the corresponding NLO nonsingular for $(\al,\bt) = (2,0.5)$, $(2, 1.2)$ and $(3,2)$. 
The singular cross section is not shown separately as its shape can be derived by comparing the total NLO with the NLO nonsingular cross section. It happens to be constant in the  \SCETab region in this double-logarithmic plot.
As expected, the nonsingular is relevant in the fixed-order region of phase-space, where the angularities are large.
The feature which we observe in the fixed-order region is the two-dimensional analogue of the bump we saw for one angularity.
It occurs for the WTA axis, where regions in phase space in which two particles carry the same energy fractions ($x_1=x_2$ or $x_2=x_3$) lead to those sharp edges.

%===============================================================================
\subsection{Scales and uncertainties}
\label{sec:prof}
%===============================================================================

In this section we specify the central scale choices used to achieve resummation, and describe in detail the scale variations used to estimate the perturbative uncertainty. Since the spectra we are dealing with have distinct kinematic regions where resummations must be handled differently, we use angularity-dependent soft and jet scales given by ``profile functions", a method previously applied to e.g.~the thrust event shape~\cite{Abbate:2010xh} and the $B \to X_s \gamma$ spectrum~\cite{Ligeti:2008ac}. We start with the single angularity distribution and then extend our discussion to the case of two angularities, introducing profile functions depending simultaneously on both $e_\al$ and $e_\bt$. The hard scale is $\mu_H = Q$ and will not be varied to estimate the resummation uncertainties.

%~~~~~~~~~~~~~~~~~~~~~~~~~~~~~~~~~~~~~~~~~~~~~~~~~~~~~~~~~~~~~~~
\subsubsection{Single angularity}
\label{sec:prof_1D}
%~~~~~~~~~~~~~~~~~~~~~~~~~~~~~~~~~~~~~~~~~~~~~~~~~~~~~~~~~~~~~~~

The canonical scales for the single-$\eb$ resummation are 
\begin{align}
&\mu_J^{\rm can} = Q \eb^{1/\bt}, \quad  &&\log_{10} (\mu_J^{\rm can} / Q) = 1/\bt\, \Lb
\nn \\
&\mu_S^{\rm can} = Q \eb, \quad  &&\log_{10} (\mu_S^{\rm can} / Q) = \Lb
\end{align}
with $\Lb \equiv \log_{10} e_\beta$. Given these expressions, we find it convenient  -- in particular in view of the case of two angularities discussed later --  to construct profile functions in terms of the logarithms of the angularities, rather than the angularities themselves.

For the central value of our predictions we take
\begin{align} \label{eq:central}
&\log_{10} (\mu_J / Q) = 1/\bt\, \Lb \times
h\bigl(\Lb,  t_1,  t_3\bigr)
\,,\nn \\
&\log_{10} (\mu_S / Q) = \Lb \times
h\bigl(\Lb,  t_1,  t_3\bigr)
\,.\end{align}
%%%
The function
%%%
\begin{align}
\label{eq:hfunc}
  h( t,  t_1,  t_3) &= 
  \begin{cases}
  1 &  t \leq  t_1\,, \\
  - \dfrac{( t- t_3)^2}{( t_1-  t_3)^3}   (2 t - 3 t_1+ t_3)
   &  t_1 \leq  t \leq  t_3\,, \\
  0 & t \geq  t_3\,.
 \end{cases}
\end{align}
smoothly connects the canonical region $t\leq t_1$ to the fixed-order region $t \geq t_3$ using a cubic polynomial.

For $t_1$ we take the value of $e_\bt$ where the NLO nonsingular cross section is 10\% of the NLO singular cross section (see \fig{nlons1D}). This leads to
%%%
\begin{align} \label{eq:t1central}
(\beta, t_1) = (0.5, -0.795)\,, \quad (1.2,-0.82)\,, \quad (2,-0.90)\,, \quad (3,-0.98)
\,.\end{align}
%%%
For the $t_3$ parameter, we take the point where the NLO singular vanishes, which is 
%%%
\begin{align}
t_3 = -0.33
\end{align}
%%%
for all angularities. To simplify our scale choices in the double differential case, we do not introduce a profile to handle the transition to the nonperturbative regime but instead freeze $\al_s$ below 2 GeV to avoid the Landau pole.\footnote{Alternatively we could have constructed the profiles 
as a direct extension of those for thrust in ref.~\cite{Mo:2017gzp} to other values of $\beta$, i.e.~$\mu_J = Q f_{\rm run}(t)^{1/\bt}\,, \,\mu_S = Q f_{\rm run}(t)$, with $f_{\rm run}$ in ref.~\cite{Mo:2017gzp}. 
In this case the transition to fixed order would be done in $\eb$ instead of $\log_{10} \eb$, two quadratics instead of a cubic are used in the transition region, and the transition to the nonperturbative regime is also handled by $f_{\rm run}$.
We have checked that the difference between these profile choices is very small (compared to our uncertainties).}

We now consider a range of scale variations to estimate the perturbative uncertainty.

\begin{itemize}
\item[(i)] Fixed-order uncertainty: We \emph{simultaneously} vary all scales $\mu_i$, including $\mu_H$, by a factor of 2 or 1/2.
This variation smoothly transitions into the fixed-order uncertainty in the region where the resummation is turned off, since there only a single scale remains.

\item[(ii)]  Resummation uncertainty: Following refs.~\cite{Gangal:2014qda,Mo:2017gzp}, we vary the jet and soft scale according to 
%%%
\begin{align} \label{eq:resum_unc}
\log_{10} (\mu_J^{\rm vary} /Q)&= (1/\bt - b)  \Big[  \bt \log_{10} (\mu_J / Q) + a \, f_{\rm vary}(\Lb,  t_1,  t_3) \Big]
\,, \nn \\
\log_{10} (\mu_S^{\rm vary} /Q) &= \log_{10} (\mu_S / Q) + a\,  f_{\rm vary}(\Lb,  t_1,  t_3)
\end{align}
%%%
where we take the following values for the parameters $a$ and $b$
%%%
\begin{align} \label{eq:ab_vary}
(a,b) =  (\min(\bt,1),0)\,, \quad (-\min(\bt,1),0)\,, \quad \big(0,\tfrac{(\bt-1)}{3\bt}\big)\,, \quad \big(0,-\tfrac{(\bt-1)}{3\bt}\big)
\,,\end{align}
%%%
and 
%%%
\begin{align}
f_{\rm vary}(t,t_1,t_3) &= \log_{10} 2 \times h(t,t_1,t_3)~.
\end{align}
%%%
This form of $f_{\rm vary}$ corresponds to a factor 2 variation in the canonical region but no variation in the fixed-order region, where we are not allowed to vary $\mu_H$, $\mu_J$ and $\mu_S$ independently of each other, and a smooth transition in between. 

For $a=b=0$, \eq{resum_unc} reproduces our central scale choice. Scale variations with $a\neq 0$ and $b=0$ preserve the canonical relation
%%%
\begin{align} \label{eq:can_rel}
  \Big(\frac{\mu_J}{\mu_H}\Big)^\bt = \frac{\mu_S}{\mu_H}
\,.\end{align}
%%%
We choose the size of these variations in \eq{ab_vary} such that the smallest scale varies by a factor of 2 or 1/2 in the canonical region.\footnote{For $\beta<1$ the smallest scale is the jet scale.} Setting $a=0$ and $b \neq 0$ does not preserve \eq{can_rel}. We also impose that these variations vanish for $\bt \to 1$, since $\mu_S$ and $\mu_J$ coincide in this limit, and agree with the choice for thrust in ref.~\cite{Mo:2017gzp}. Furthermore, the deviations from \eq{can_rel} for $\beta>1$ are required to be of the same size as the deviations from
%%%
\begin{align} 
  \Big(\frac{\mu_S}{\mu_H}\Big)^{1/\bt} = \frac{\mu_J}{\mu_H}
\end{align}
%%%
for $\beta<1$. 

\item[(iii)] Variations of the transition points: Since there is also a certain amount of arbitrariness in choosing the transition points $t_1$ and $t_3$, these get varied as well (but only one at a time).
For $t_1$ we consider the following alternatives to \eq{t1central}
%%%
\begin{align}
(\beta, t_1) &= (0.5, -0.645)\,,\quad (1.2, -0.700)\,,\quad (2, -0.820)\,,\quad (3, -0.98) \,,\nn \\
(\beta, t_1) &= (0.5, -0.975)\,,\quad (1.2, -0.98)\,,\quad (2, -0.98)\,,\quad (3, -1.06)\,,
\end{align}
%%%
depending on $\beta$. These values correspond to the points where the NLO nonsingular is $20\%$ and $5\%$ of NLO singular, respectively. The parameter $t_3$ is alternately set 
%%%
\begin{align}
(\beta, t_3) &= (0.5, -0.405)\,,\quad (1.2, -0.5)\,,\quad (2, -0.5)\,,\quad (3, -0.5) \,,
\end{align}
%%%
This corresponds to the point where the total NLO cross section vanishes.

\end{itemize}
Our final uncertainty band is obtained by adding the fixed-order uncertainty in quadrature to the resummation uncertainty, which is obtained by taking the envelope of the $a$ and $b$ variations above, and the variations of the transition points.
The resummation uncertainty is dominated by the $a$ and $b$ variations, whereas the uncertainty from the variations of the transition points is rather small (but still contributes to the envelope in parts of the transition region).

%~~~~~~~~~~~~~~~~~~~~~~~~~~~~~~~~~~~~~~~~~~~~~~~~~~~~~~~~~~~~~~~
\subsubsection{Two angularities}
\label{sec:prof_2D}
%~~~~~~~~~~~~~~~~~~~~~~~~~~~~~~~~~~~~~~~~~~~~~~~~~~~~~~~~~~~~~~~

Next we consider the case where two angularities are measured. We will first construct running scales for the central value of our prediction, using the canonical values of the scales in the three regimes listed in table \ref{tab:can2d} as a starting point. Although in regime 1 $\mu_S \sim e_\al Q \sim e_\bt Q$, we choose $\mu_S \sim e_\bt Q$ as our canonical scale because the resummation in this regime is governed by $e_\bt$ (and conversely the jet scale in regime 3 involves $e_\al$). The collinear-soft scale merges with the soft scale in region 1 and with the jet scale in region 3. 
%%%
\begin{table}
  \centering
  \begin{tabular}{c | c c c c c c}
  \hline \hline
     & $\log_{10} (\mu_J^{\rm can} / Q)$ & $\log_{10} (\mu_\cS^{\rm can} / Q)$ & $\log_{10} (\mu_S^{\rm can} / Q)$  \\ \hline
  1 
  & $1/\bt \Lb$ &  & $\Lb$  \\ 
  2 
  &  $1/\bt \Lb$ &  $(1-\bt)/(\al-\bt) \La + (\al -1)/(\al-\bt)  \Lb$ & $\La$  \\
  3 
  &  $1/\al \La$ &  &  $\La$ \\
  \hline\hline
  \end{tabular}
  \caption{Canonical scales for the measurement of two angularities in the three regions, with $\La \equiv \log_{10} e_\al$ and $\Lb \equiv \log_{10} e_\beta$.}
\label{tab:can2d}
\end{table}
%%%

We take for the jet and soft scale 
%%%
\begin{align} \label{eq:2D_profiles}
  \log_{10} \frac{\mu_J}{Q}
  &= g \biggl[ \Lb,\, \frac{1}{\bt},\, \frac{1}{\al} \La,\,
\La + \tilde t_{\rm R3} 
\Big[ \log_{10}\big(2^{(\bt-\al)/\al} e_\al^{\bt/\al}\big) - \La \Big],  
\log_{10}\big(2^{(\bt \!-\! \al)/\al} e_\al^{\bt/\al}\big) 
 \!\biggr]
\nn \\ & \quad \times
h\bigl[\min(\La, \Lb), \tilde t_1,
\tilde t_3\bigr]
\,,\nn \\ 
  \log_{10} \frac{\mu_S}{Q}
  &= g \biggl[
\La,\, 1,\,
\Lb,\,
\log_{10}\big(2^{(\al-\bt)/\bt} e_\beta^{\al/\bt} \big) \!+\! \tilde t_{\rm R1} \Big[\Lb - \log_{10}\big(2^{(\al-\bt)/\bt} e_\beta^{\al/\bt} \big)\Big], 
\Lb \biggr]
\nn \\ & \quad \times
h\bigl[\min(\La, \Lb), \tilde t_1,
\tilde t_3\bigr]
\,,
\end{align}
%%%
and we fix the collinear-soft scale using the canonical relation
%%%
\begin{align} \label{eq:CS_can}
  \log_{10} \frac{\mu_\cS}{Q}
  &= \frac{1-\bt}{\al-\bt}\,  \log_{10} \frac{\mu_S}{Q}  + \frac{(\al -1)\bt}{\al-\bt}\, \log_{10}\frac{\mu_J}{Q}
\,.\end{align}
%%%

The transition between the SCET regions is handled by the function $g$, 
%%%
\begin{align}
  g(x, a, b, x_1, x_2) &= 
  \begin{cases}
  a\, x & x \leq x_1\,, \\
  b - \dfrac{(x-x_2)^2}{(x_1-x_2)^3} [a(x \,x_1+x \,x_2 -2x_1^2)+b(-2x+3x_1-x_2)]& x_1 \leq x \leq x_2\,, \\
 b & x \geq x_2\,.
 \end{cases}
\end{align}
%%%
In the intermediate region, $x_1 \leq x \leq x_2$, $g$ is given by the cubic polynomial that is continuous and has a continuous derivative. The first three arguments of $g$ in \eq{2D_profiles} directly follow from the canonical scales in table~\ref{tab:can2d}. 
The transition points $x_1$ were chosen as a fraction $\tilde t_{\rm \,R1 / R3}$ of the total distance (in logarithmic space) between the two phase-space boundaries in \eq{PS_NLO}.
For the central profiles we choose $\tilde t_{\rm \,R1} = 0.8$ and $\tilde t_{\rm \,R3} = 0.95$, which corresponds roughly to the region where the nonsingular terms (from the boundary regimes) are $10\%$ of the singular one.
The transition points $x_2$ were chosen at the phase-space boundary. 
For example, for $\mu_S$ we start the transition at 
%%%
\begin{align}
  \ln e_\al -
  \log_{10}\big(2^{(\al-\bt)/\bt} e_\beta^{\al/\bt} \big) = \tilde t_{\rm R1} \Big[\log_{10} e_\beta - \log_{10}\big(2^{(\al-\bt)/\bt} e_\beta^{\al/\bt} \big)\Big]
\,,\end{align}
%%%
and end at $e_\al = e_\bt$.
With the definition above, the profile scales remain constant at their canonical regime 3 values, beyond the NLO phase-space boundary, such that the $\ea$ resummation is turned off here.
This implies in particular that also the NLL results know about the NLO phase-space boundary. 
We will add a comment below, how our results change if the canonical (instead of NLO) phase-space boundary would have been used in the profile scales.

The transition to the fixed-order region is controlled through the function $h$ in \eq{hfunc}.
Due to the argument of $h$ in \eq{2D_profiles}, the fixed-order region 
%%%
\begin{align} \label{eq:FO_box}
  \min(\La, \Lb) \geq \tilde{t}_3
\end{align}
%%%
has a square shape. We have checked that other choices have minimal impact on the result. The transition points are taken from the single angularity case: for $\tilde t_1$ we take the minimum (which corresponds to a larger transition region) of $t_1$ for the single angularities $e_\al$ and $e_\bt$ from \eq{t1central}, and $\tilde t_3 = t_3 = -0.33$.

As for the single angularity spectrum, several scale variations are taken into account.
\begin{itemize}
\item[(i)]  Fixed-order uncertainty: We \emph{simultaneously} vary all scales $\mu_i$ by a factor of 2 or 1/2.

\item[(ii)] Resummation uncertainty: Extending the one-dimensional case, jet, collinear-soft and soft scales are varied according to 
%%%%
\begin{align} \label{eq:2d_vary}
\log_{10}\frac{\mu_J^{\rm vary}}{Q} &= 
(1-h_J) (1/\alpha-b(\alpha)) \big[ 
 \alpha \log_{10} \frac{\mu_J}{Q}  + a(\alpha) f_{\rm vary}(\La,\tilde t_1,\tilde t_3) 
\big]
\nn \\ & \quad +
h_J (1/\beta-b(\beta)) \big[ 
 \beta \log_{10} \frac{\mu_J}{Q} + a(\beta)  f_{\rm vary}(\Lb,\tilde t_1,\tilde t_3) 
\big]
\,, \nn \\
\log_{10}\frac{\mu_\cS^{\rm vary}}{Q} &= 
(1-h_S) \Big[ \log_{10} \frac{\mu_S}{Q} + a(\beta)  f_{\rm vary}(\Lb,\tilde t_1,\tilde t_3)\Big]
\nn \\ & \quad + 
(1-h_J) (1/\alpha-b(\alpha)) \big[ 
\alpha \log_{10} \frac{\mu_J}{Q}  + a(\alpha)   f_{\rm vary}(\La,\tilde t_1,\tilde t_3) \big]
\nn \\ & \quad + 
h_S\, h_J \bigg\{ 
\log_{10} \frac{\mu_\cS}{Q} + \frac{(1-\bt)}{\alpha-\beta} a(\alpha)   f_{\rm vary}(\La,\tilde t_1,\tilde t_3)
\nn \\ & \quad +
\frac{(\alpha-1)}{\alpha-\beta} 
a(\beta)  f_{\rm vary}(\Lb,\tilde t_1,\tilde t_3) 
\bigg\}
\,, \nn \\
\log_{10}\frac{\mu_S^{\rm vary}}{Q} &= \log_{10} \frac{\mu_S}{Q}  + 
h_S a(\alpha) f_{\rm vary}(\La,\tilde t_1,\tilde t_3) +
(1-h_S) a(\beta) f_{\rm vary}( \Lb,\tilde t_1,\tilde t_3)
\,,\end{align}
%%%
with $a(\alpha)=\pm \min(\alpha,1)$, $b(\alpha)=\pm (\alpha-1)/(3 \alpha)$, as in the single angularity case in \eq{ab_vary}. The transitions are governed by the functions 
%%%
\begin{align}
&h_S = h\biggl[
\La,
\log_{10}\big(2^{(\al-\bt)/\bt} e_\beta^{\al/\bt} \big) \!+\! \tilde t_{\rm R1} \Big[\Lb - \log_{10}\big(2^{(\al-\bt)/\bt} e_\beta^{\al/\bt} \big)\Big], 
\Lb \biggr]\,,
\nn \\ 
&h_J = h\biggl[
\Lb,
\La + \tilde t_{\rm R3} 
\Big[ \log_{10}\big(2^{(\bt-\al)/\al} e_\al^{\bt/\al}\big) - \La\Big],  
\log_{10}\big(2^{(\bt-\al)/\al} e_\al^{\bt/\al}\big) 
\biggr]\,,
\end{align}
%%%
which have the property that $h_J = 1$ in regime 1 and 2, and 0 in regime 3, and $h_S = 1$ in regime 2 and 3, and 0 in regime 1. 
Thus the scale variations in regime 1 and 3 in \eq{2d_vary} are the usual single angularity ones. In these regimes the collinear-soft scale is not independent and thus needs to be varied in tandem with the soft or jet scale it has merged with. In the intermediate regime we have used \eq{CS_can} to determine the collinear-soft scale, but setting $b=0$ there. This is necessary, because otherwise $\mu_\cS$ is varied by much more than a factor of 2 when the angularity exponents $\al$ and $\bt$ are close to each other.

\item[(iii)] Variations of the transition points: 
We vary $\tilde t_1$, using the maximal and the minimal value of the $t_1$ variations of the two single angularities considered in \sec{prof_1D}. 
Similarly, for $\tilde t_3$ we use the variation of each of the single angularities. To vary the transition between the boundary theories, we vary $\tilde t_{\rm R1/R3}$ by taking
$\tilde t_{\rm R1} = 0.7, 0.9$ or $\tilde t_{\rm R3} = 0.9$.
These values are motivated by looking at the contour where the nonsingular terms are $10\%$ of the singular one (focusing on the resummation region).

\end{itemize}
As in the one-dimensional case, the total uncertainty is obtained by adding the fixed-order uncertainty in quadrature to the envelope of the resummation variations and the variations of the transition points.

%~~~~~~~~~~~~~~~~~~~~~~~~~~~~~~~~~~~~~~~~~~~~~~~~~~~~~~~~~~~~~~~
\subsubsection{Differential vs.~cumulant scale setting}
%~~~~~~~~~~~~~~~~~~~~~~~~~~~~~~~~~~~~~~~~~~~~~~~~~~~~~~~~~~~~~~~

We implement our choice of scales at the differential level, {\it i.e.}~we calculate $\df^2 \si/(\df e_\al\, \df e_\bt)$ using scales evaluated at $e_\al$ and $e_\bt$. An alternative is to use cumulant scale setting, 
%%%
\begin{align} \label{eq:cum}
   \Sigma(e_\al^c, e_\bt^c) = \int_0^{e_\al^c}\! \df e_\al \int_0^{e_\bt^c}\! \df e_\bt\, \frac{\df^2 \si}{\df e_\al\, \df e_\bt}
\,,\end{align}
%%%
evaluating the scales at $e_\al^c$ and $e_\bt^c$. Differentiating this introduces derivatives of the scales, %%%
\begin{align}
   \frac{\df^2 \Sigma}{\df e_\al\,\df e_\bt} &= \frac{\df^2 \si}{\df e_\al\, \df e_\bt}
   + \sum_{i}  \frac{\df^2 \Sigma}{\df \ln \mu_i\,\df e_\bt}\, \frac{\df \ln \mu_i}{\df e_\al}\, 
   \nn \\ & \quad
   + \sum_{j}  \frac{\df^2 \Sigma}{\df e_\al\,\df \ln \mu_j}\, \frac{\df \ln \mu_j}{\df e_\bt}\, 
   + \sum_{i,j}  \frac{\df^2 \Sigma}{\df \ln \mu_i\,\df \ln \mu_j}\, \frac{\df \ln \mu_i}{\df e_\al}\, \frac{\df \ln \mu_j}{\df e_\bt}   
    \label{eq:cum_diff}
\,.\end{align}
%%%
For the unprimed orders in table~\ref{tab:orders}, such as NNLL, differential scale setting does not capture all the logarithms, as discussed in detail in {\it e.g.}~ref.~\cite{Almeida:2014uva}. However, our scales in \secs{prof_1D}{prof_2D} undergo fairly rapid changes in transition regions, leading to artefacts from the terms involving the derivatives of scales, when using the cumulant scale setting.

We investigate this issue by supplementing our cross section with differential scale setting with the additional terms on the right-hand side of \eq{cum_diff}. By using the canonical scales to determine the scale derivates $\df \ln \mu_{i,j}/\df e_{\al,\bt}$ in these terms, we maintain the required formal accuracy while avoiding artefacts from derivatives of our profile scales encountered with cumulative scale setting. For example, in region 2
%%%
\begin{align}
   \frac{\df \ln \mu_J}{\df e_\al} &= 0\,, &
   \frac{\df \ln \mu_J}{\df e_\bt} &= \frac{1}{\bt e_\bt}\,,
   \nn \\
   \frac{\df \ln \mu_\cS}{\df e_\al} &= \frac{1-\bt}{(\al-\bt) e_\al}\,, &
   \frac{\df \ln \mu_\cS}{\df e_\bt} &= \frac{\al-1}{(\al-\bt) e_\bt}\,,
   \nn \\
   \frac{\df \ln \mu_S}{\df e_\al} &= \frac{1}{e_\al}\,, &
   \frac{\df \ln \mu_S}{\df e_\bt} &= 0
\,.\end{align}
%%%

\begin{figure}[t]
  \centering
   \includegraphics[width=0.325\textwidth]{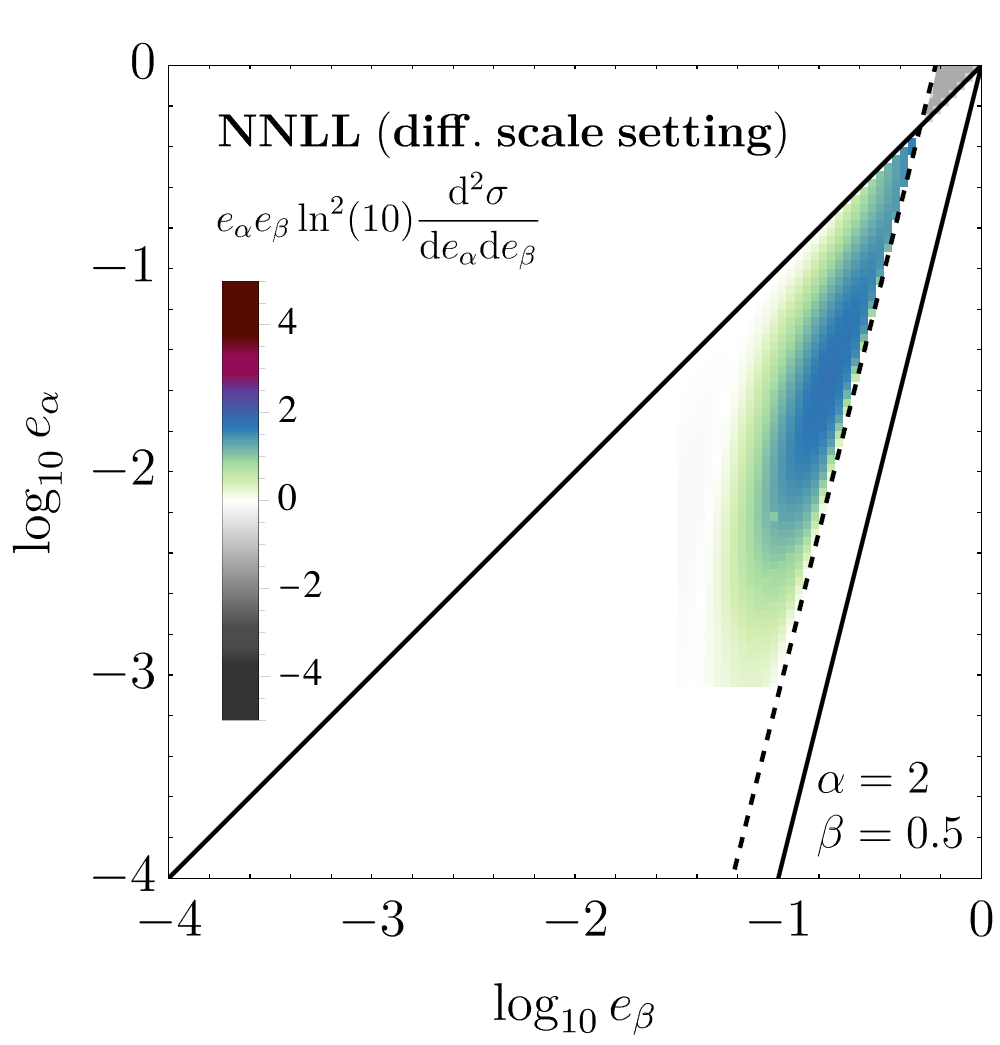}
   \includegraphics[width=0.325\textwidth]{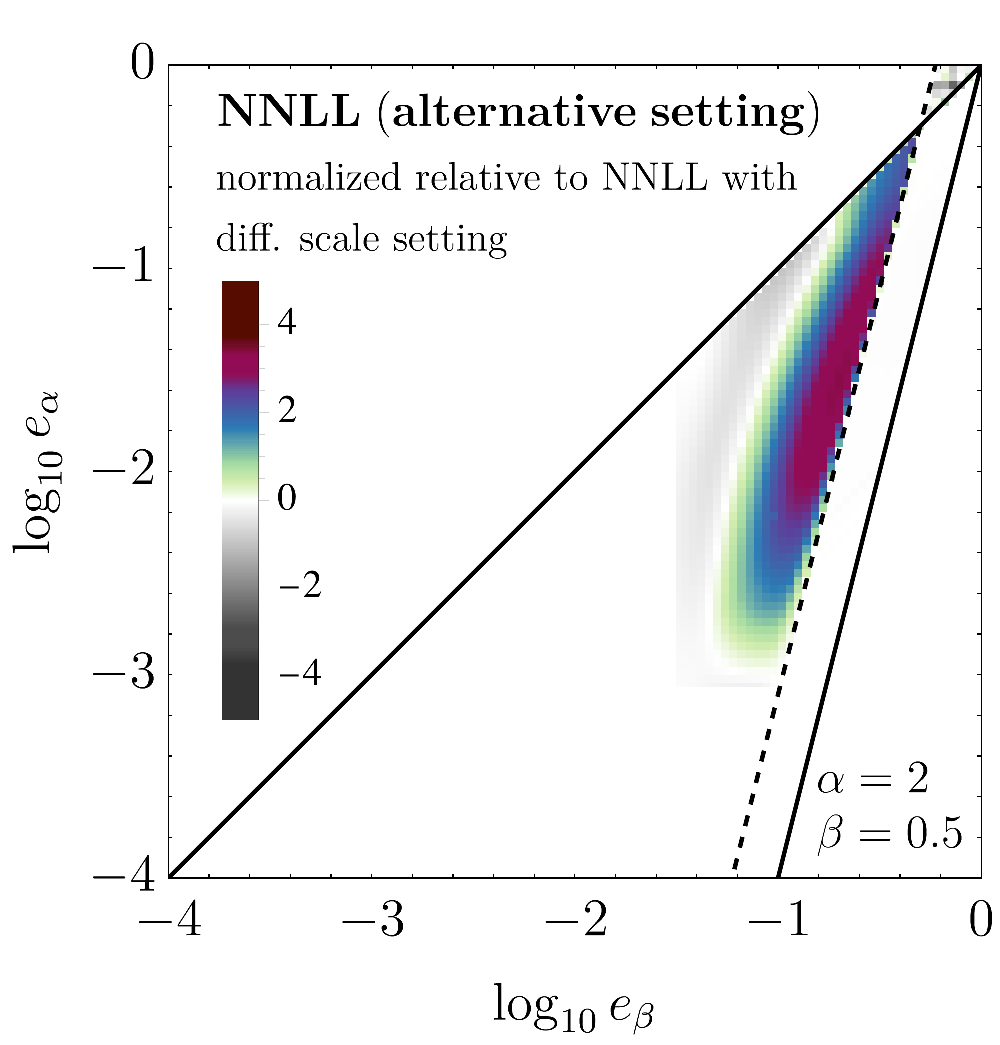}  
   \includegraphics[width=0.325\textwidth]{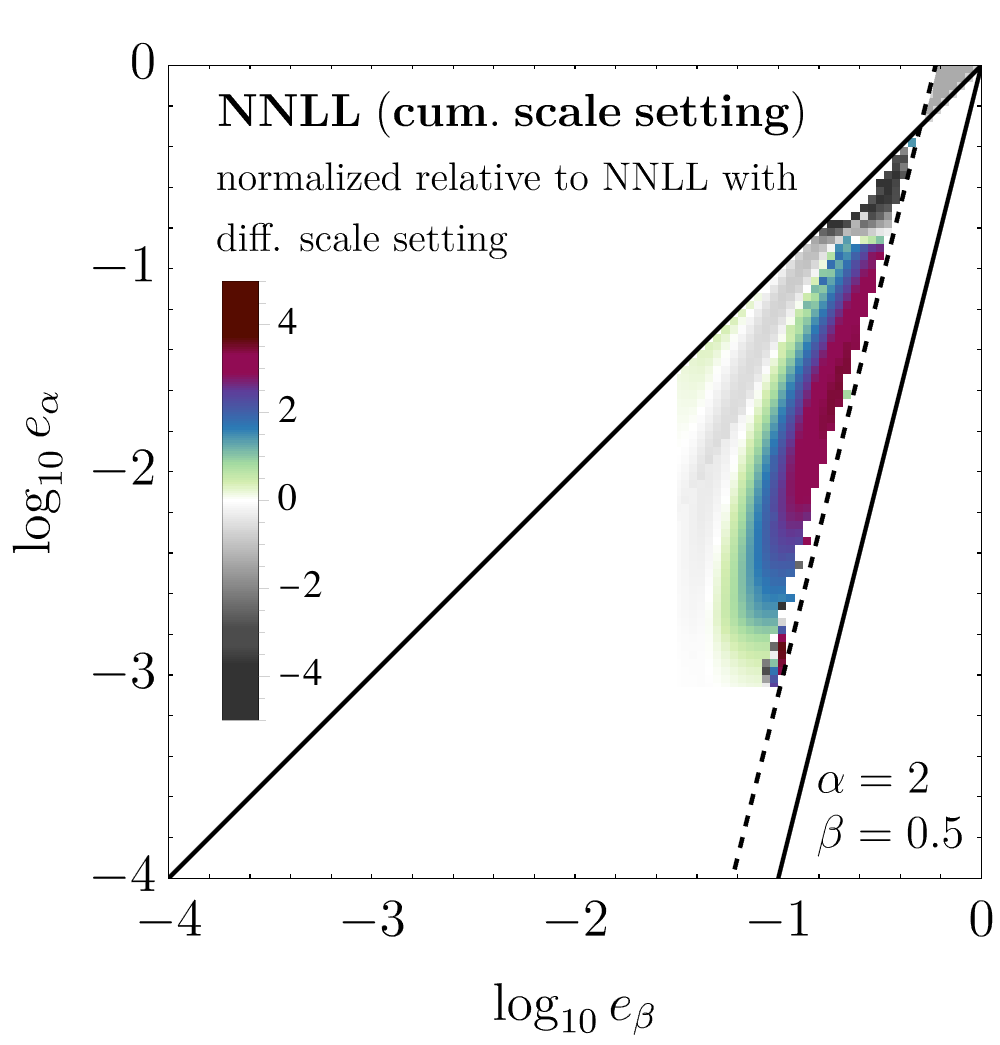}
   \caption{
   NNLL cross section with differential scale setting (left), 
   with differential scale setting plus extra NNLL terms in \eq{cum_diff} evaluated using canonical scales (center)
   and with cumulative scale setting (right). 
   }
   \label{fig:nnllextraterms}
  \end{figure}

In \fig{nnllextraterms} we compare the standard differential scale setting (left panel) and cumulative scale setting in \eq{cum} (right panel) to the alternative procedure we just described (middle panel). The cumulative scale setting leads to clear artefacts in the transition to fixed-order and to the boundaries of phase space, which are due to the derivatives of profiles scales in \eq{cum_diff}, which undergo a rapid transition. For example, the boundary of the box in \eq{FO_box} is clearly visible. Our alternative approach avoid these artefacts, by using canonical scales in the derivatives of scales. 
However, the alternative approach has a major disadvantage: 
Since the canonical scales do not turn off properly in the fixed-order region, the singular-nonsingular cancellation is spoiled there. Thus we are left with using standard differential scale setting in the results presented in \sec{results}, even though not all logarithms are captured.

%%%
\begin{table}[t]
  \centering
  \begin{tabular}{l | c c c c c c c}
  \hline \hline
  & Fixed-order & Non-cusp & Cusp & Beta \\ \hline
  LL & tree & -  & $1$-loop & $1$-loop \\
  NLL & tree & $1$-loop & $2$-loop & $2$-loop \\
  NLL$'$ & $1$-loop & $1$-loop & $2$-loop & $2$-loop\\
  NNLL & $1$-loop & $2$-loop & $3$-loop & $3$-loop\\
  NNLL$'$ & $2$-loop & $2$-loop & $3$-loop & $3$-loop\\
  \hline\hline
  \end{tabular}
  \caption{Perturbative ingredients needed at different orders in resummed perturbation theory. The columns correspond to 
  the loop order of the fixed-order ingredients, the non-cusp and cusp anomalous dimensions, and the QCD beta function.}
\label{tab:orders}
\end{table}
%%%

%---------------------------------------
%---------------------------------------

%===============================================================================
\subsection{Matching}
\label{sec:matching}
%===============================================================================

Given that we have a different factorization theorem for each of the regions of phase space, we would like to obtain an expression for the cross section which is valid everywhere. This is achieved by matching the cross section predictions from the various regions~\cite{Lustermans:2019plv}
%%%
\begin{align}
\sigma & =\sigma_2(\mu_J^{\rm R2}, \mu_S^{\rm R2}, \mu_{\cS}^{\rm R2}) 
\nn \\& \quad
+ \Big[\sigma_1(\mu_J^{\rm R2}, \mu_S^{\rm R1})-\sigma_2(\mu_J^{\rm R2},\mu_S^{\rm R1},\mu_S^{\rm R1} )\Big]
\nn \\& \quad
+ \Big[\sigma_3(\mu_J^{\rm R3}, \mu_S^{\rm R2})-\sigma_2(\mu_J^{\rm R3},\mu_S^{\rm R2},\mu_J^{\rm R3} )\Big]
\nn \\& \quad
+ \Big[\sigma_{\rm FO}(\mu_{\rm FO})-\sigma_1(\mu_{\rm FO},\mu_{\rm FO})-\sigma_3(\mu_{\rm FO},\mu_{\rm FO})+\sigma_2(\mu_{\rm FO},\mu_{\rm FO},\mu_{\rm FO})\Big] \,,
\label{eq:matching}
\end{align}
%%%
where each of the cross sections is differential in $e_\al$ and $e_\bt$. The first line describes the cross section for regime 2, using the scales which are appropriate for this regime. The second line ensures that in regime 1 we reproduce the correct cross section. This is achieved by including the nonsingular contribution obtained by adding the cross section of regime 1 and subtracting the one of regime 2 evaluated at the scales of regime 1 ({\it i.e.}~the overlap). Note that in regime 1, the R2 scales merge into the R1 scales, such that the second term on the second line cancels against the first line. This procedure is similar to the construction of the fixed-order nonsingular when a single type of logarithm is resummed. Similarly, the third line describes the nonsingular correction from the regime 3 boundary of phase space. The last line corresponds to the fixed-order nonsingular, shown in \fig{fullnloandns} above.
A smooth transition between the regimes is achieved by the profile scales discussed in \sec{prof}. 

\begin{figure}
  \centering
   \includegraphics[width=0.325\textwidth]{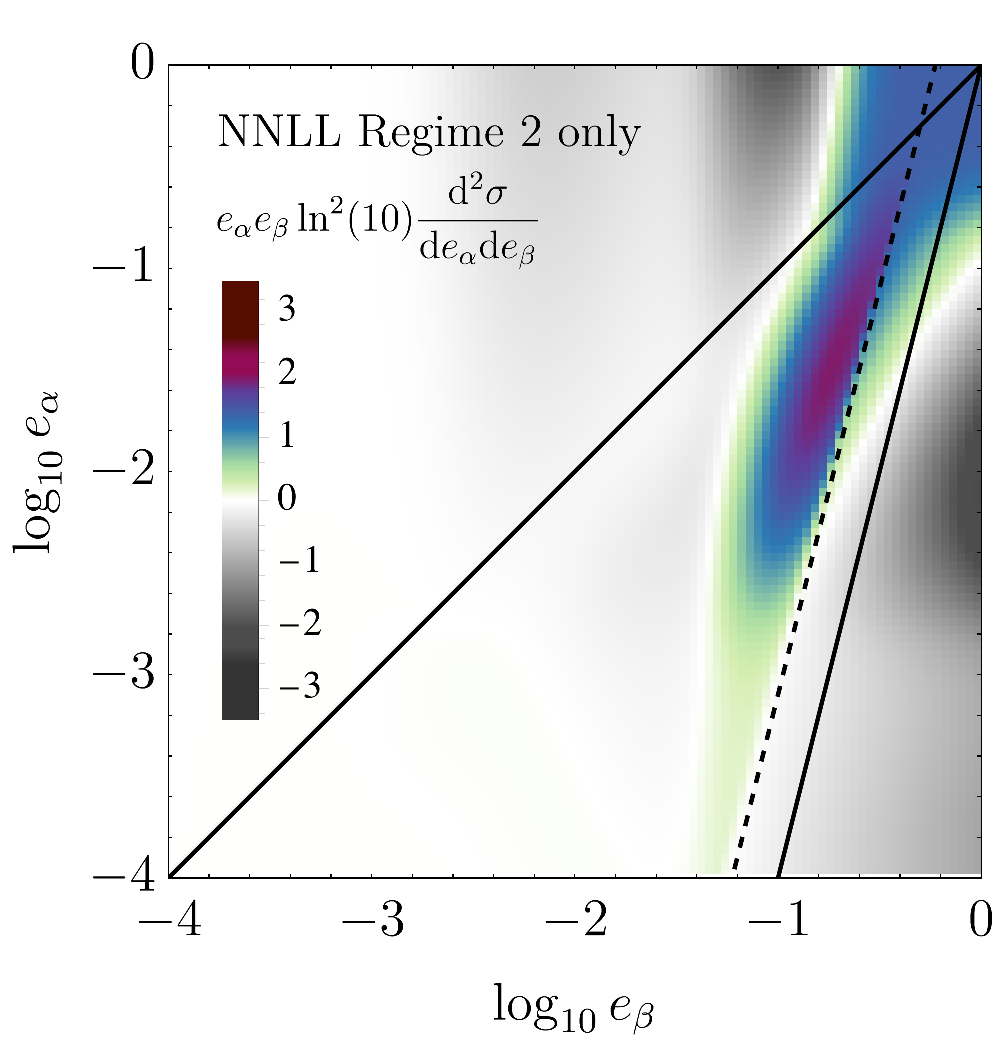}
   \includegraphics[width=0.325\textwidth]{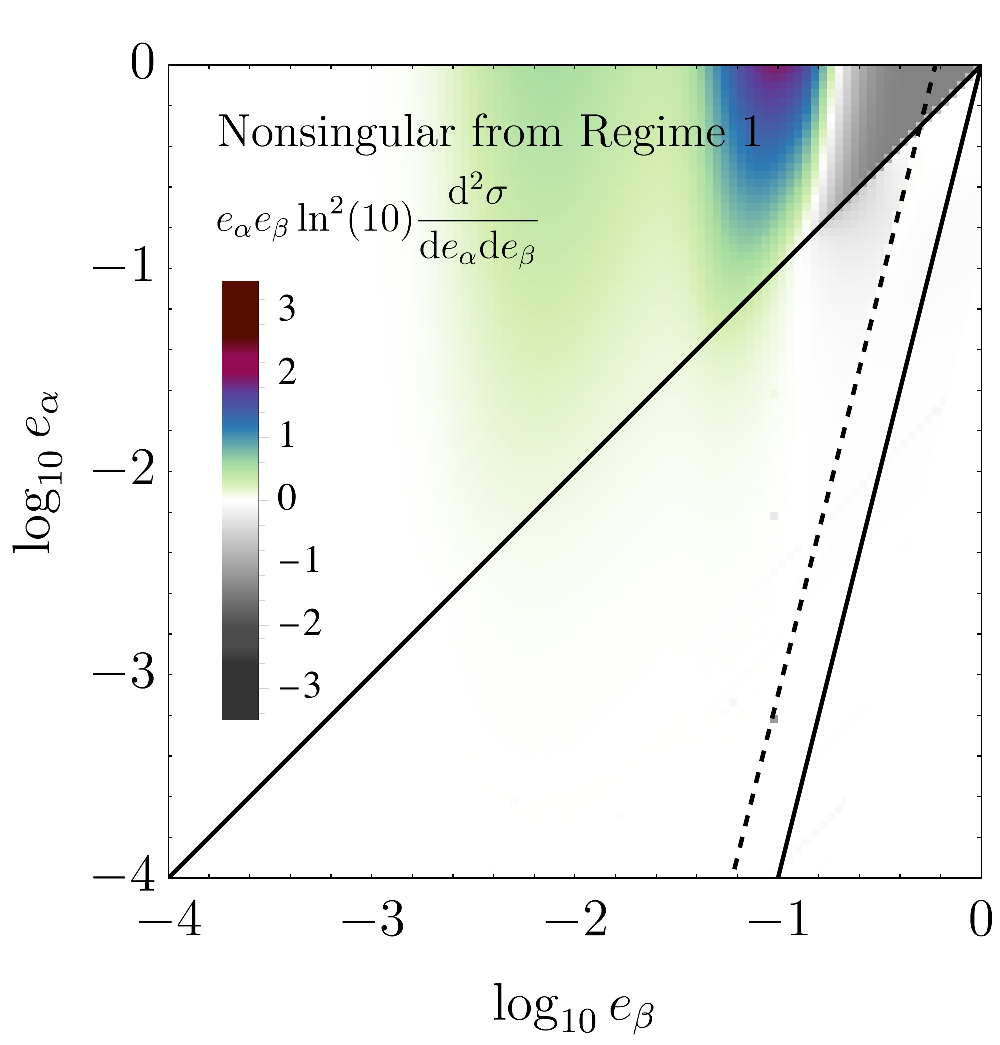}  
   \includegraphics[width=0.325\textwidth]{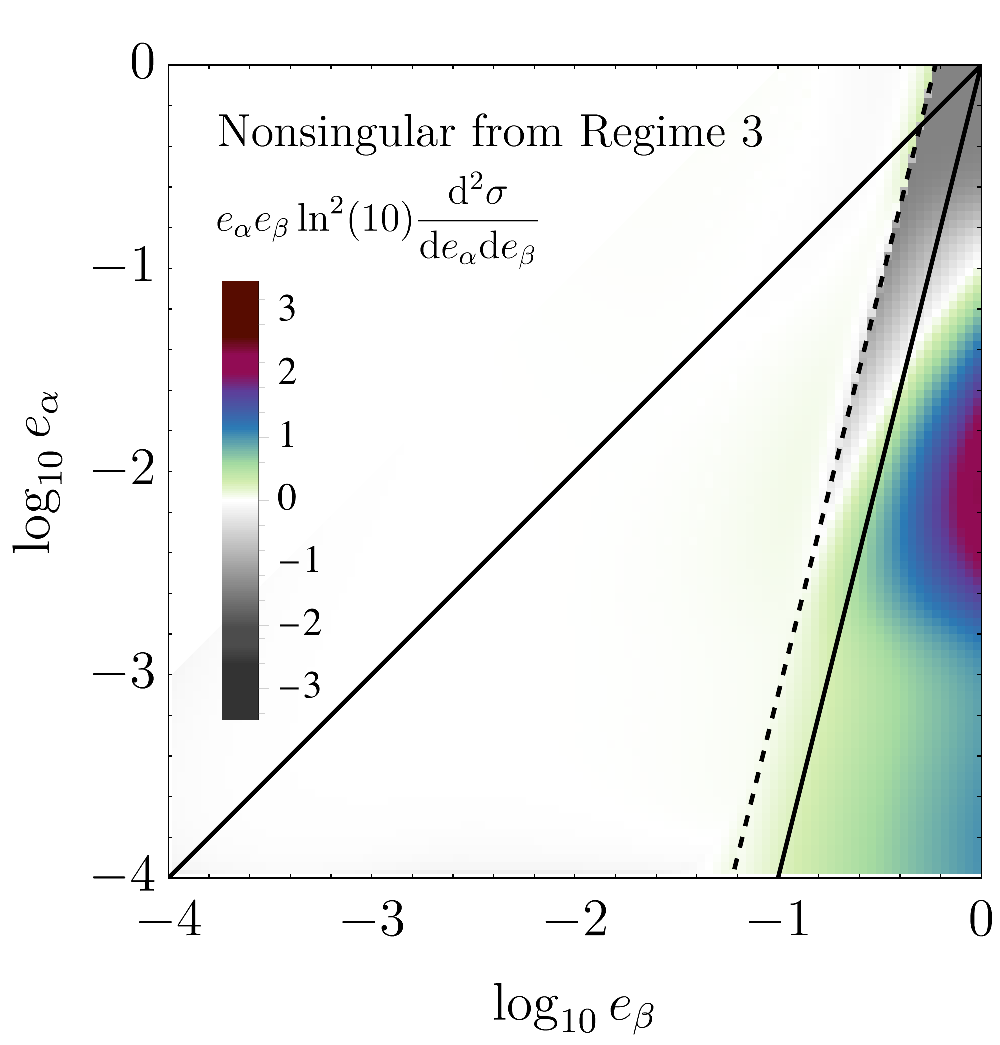}
   \caption{NNLL cross section in regime 2 (left) and nonsingular corrections from regimes 1 (center) and 3 (right), corresponding to the second and third line of \eq{matching}.}
   \label{fig:nnllsplitup}
  \end{figure}

In \fig{nnllsplitup} we show the contributions (besides the fixed-order nonsingular, already discussed in \sec{FOns}) which make up the total NNLL cross section.\footnote{The total NNLL cross section will be shown and discussed later, but can also already be seen in the left panel of \fig{nnllextraterms}. However, note the different color range.}
As shown in the left panel, in the bulk of the phase space the cross section is already captured by the regime 2 cross section. The nonsingulars from regimes 1 and 3 correct the regime 2 cross section close to the phase space boundaries and cause the cross section to vanish outside the boundaries.

%===============================================================================
\subsection{Nonperturbative effects}
\label{sec:nonperturbative}
%===============================================================================

We have also studied the effect of nonperturbative corrections, which we first discuss for single angularities before extending to the double angularity cross section.  We restrict to $e_\al$ with $\al>1$ because only in this case does the soft function capture the (dominant) nonperturbative corrections. We can factorize the soft function~\cite{Korchemsky:1999kt,Hoang:2007vb,Ligeti:2008ac}
%%%
\begin{align} \label{eq:S_factor}
  S(Qe_\al,\mu) = \int\! \df Q e_\al'\, S^{\rm pert}(Qe_\al - Qe_\al',\mu) F(Qe_\al')
\end{align}
%%%
into its perturbative contribution $S^{\rm pert}$ and nonperturbative contribution $F$. $F$  is dominated by momenta of the order $Qe_\al' \lesssim \lqcd$, and its integral must be one, since nonperturbative effects do not change the total cross section.
Expanding \eq{S_factor} for $Qe_\al \gg \lqcd$, 
%%%
\begin{align} \label{eq:S_exp}
  S(Qe_\al,\mu) = S^{\rm pert}(Qe_\al - \Omega_\al,\mu)\, \bigg[1 + \mathcal{O}\Big(\frac{\lqcd^2}{Q^2 e_\al^2}\Big)\bigg]
\,,
\qquad
 \Omega_\al = \int\! \df Q e_\al'\, Q e_\al'\, F(Qe_\al')
\,,\end{align}
%%%
where the leading nonperturbative correction is characterized by the parameter $\Omega_\al$, with a calculable dependence on $\al$~\cite{Dokshitzer:1995zt,Salam:2001bd,Lee:2006nr,Mateu:2012nk}\footnote{As is clear from our definition in \eq{ea_def}, hadron-mass effects are treated in the $E$-scheme. Eq.~\eqref{eq:omega} therefore still holds when accounting for hadron mass effects~\cite{Mateu:2012nk}.}
%%%
\begin{align} \label{eq:omega}
 \Omega_\al = \frac{2}{\al-1} \Omega
\,.\end{align}
%%%
We take $\Omega = 0.323$ GeV with 16\% uncertainty~\cite{Abbate:2010xh}.
Since a shift is rather crude, we implement nonperturbative effects in our analysis using the following functional form for the nonperturbative contribution $F$ in \eq{S_factor}\footnote{In the jet mass study of ref.~\cite{Stewart:2014nna}, this form captured the dominant features of the hadronization model of \Pythia rather well.}
%%%
\begin{align} \label{eq:shapefunction}
   F(Qe_\al) = \frac{4Q e_\al}{\Omega_\al^2}\, e^{-2Q e_\al/\Omega_\al}
\,,\end{align}
%%%
which is normalized and has the first moment required by \eq{S_exp}. 

 \begin{figure}
  \centering
   \includegraphics[width=0.49\textwidth]{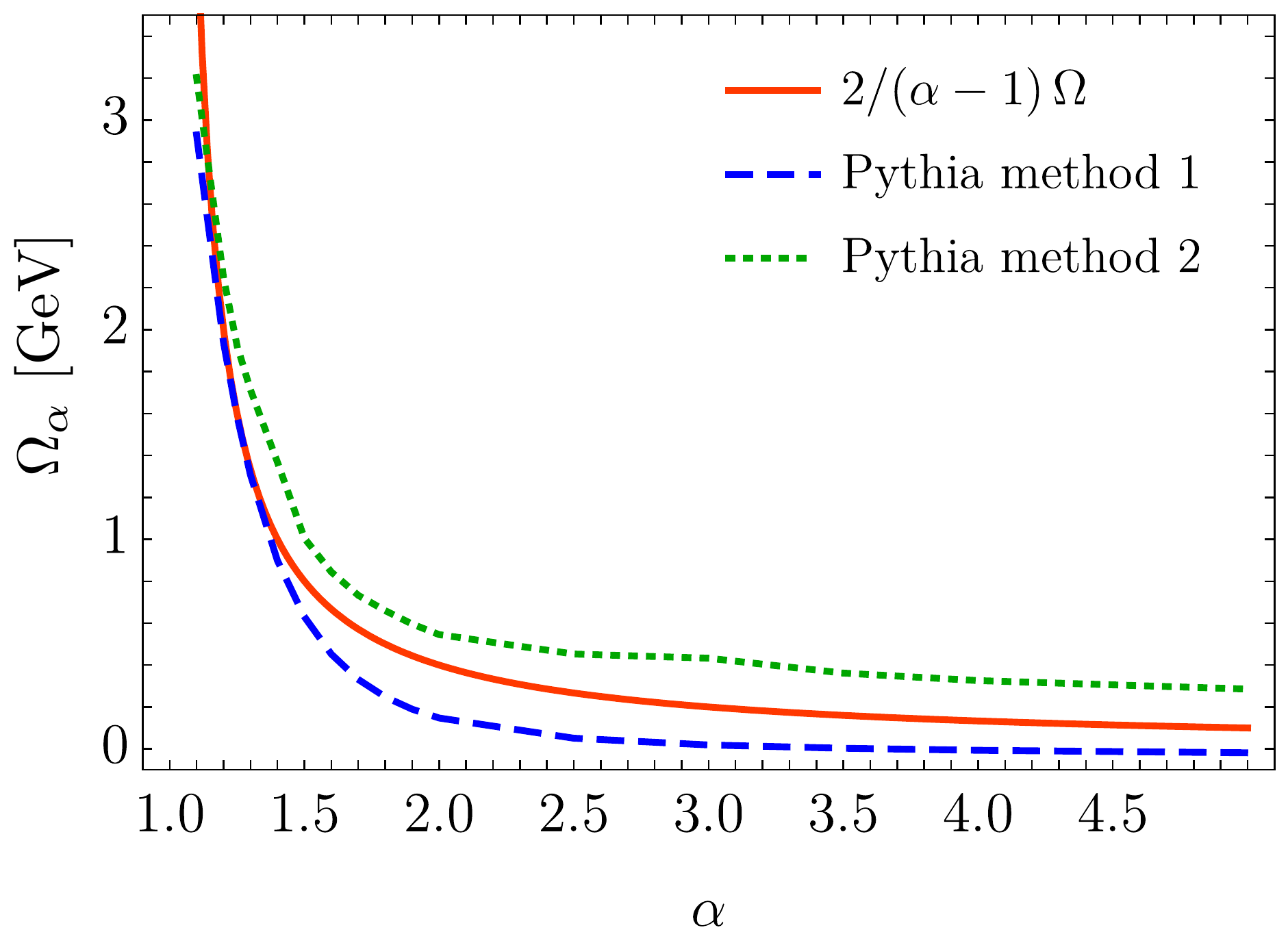}
   \caption{The leading nonperturbative parameter $\Omega_\al$ extracted from \Pythia.}
   \label{fig:omegaPythia}
  \end{figure}
We have tested \eq{omega} using \Pythia, applying two methods to extract $\Omega_\al$.\\
\underline{Method 1:} $\Omega_\al$ is obtained by taking the difference of the first moment of \Pythia cross sections at hadron- and parton-level, 
%%%
\begin{align}
  \Omega_\alpha \approx Q \int\df e_\al\, e_\al \Big(\frac{\df \si_{\rm hadr}}{\df e_\al} - \frac{\df \si_{\rm part}}{\df e_\al}\Big)~.
\end{align}
%%%
This follows directly from \eq{S_factor} and the definition of $\Omega_\al$ in \eq{S_exp}, but it assumes that the convolution in \eq{S_factor} is also valid for the nonsingular cross section. The resulting distribution for $\Omega_\al$, shown in \fig{omegaPythia}, approximately exhibits the $\al$-dependence of \eq{omega}. However, it clearly breaks down for large values of $\al$, where $\Omega_\al$ becomes negative. One possible explanation is that the above assumption on the nonsingular cross section is not justified. We have therefore attempted to extract $\Omega_\al$ in a second way. \\
\underline{Method 2:}
We performed the convolution of the parton-level \Pythia prediction with \eq{shapefunction} and determined $\Omega_\al$ by minimizing the distance between the resulting distribution and \Pythia's prediction at hadron level\footnote{As distance measure, we considered both the integral of the absolute difference and the integral of the difference squared, obtaining very similar results.}. This also approximately exhibits the $\al$-dependence in \eq{omega}, but now overshoots it for large values of $\al$. Note that this method also relies on the assumption that \eq{S_factor} extends to the nonsingular cross section.

Moving on to two angularities, in regime 2 and 3 the nonperturbative effects for $e_\al$ arise from the soft function $S(Qe_\al,\mu)$ discussed above. In regime 1, we encounter the double differential soft function, which can be factorized in a way similar to \eq{S_factor}
%%%
\begin{align}
  S(Qe_\al,Qe_\bt, \mu) = \int\! \df Q e_\al'\, \df Q e_\bt'\, S^{\rm pert}(Qe_\al - Qe_\al',Qe_\bt - Qe_\bt',\mu) F(Qe_\al',Qe_\bt')
\,.\end{align}
%%%
The leading nonperturbative corrections take on a particularly simple form
%%%
\begin{align} \label{eq:S_exp_2}
  S(Qe_\al,Qe_\bt, \mu) = S^{\rm pert}(Qe_\al - \Omega_\al,Qe_\bt - \Omega_\bt,\mu)\, \bigg[1 + \mathcal{O}\Big(\frac{\lqcd^2}{Q^2 e_\al^2},\frac{\lqcd^2}{Q^2 e_\bt^2}\Big)\bigg]
\,,\end{align}
%%%
since nonperturbative correlations vanish at this order. In our numerical analysis we set
%%%
\begin{align} \label{eq:F_num}
    F(Qe_\al,Qe_\bt) =  F(Qe_\al)\, F(Qe_\bt)\,F_{\rm cor}(Q e_\al, Qe_\bt)
\,,\end{align}
%%%
where the effect of nonperturbative correlations are encoded in
%%%
\begin{align}
  F_{\rm cor}(k_1, k_2) &= 1 + c \Big(k_1 k_2 - \frac{\Omega_2\,k_1^2}{3\Omega_1} - \frac{\Omega_1\,k_2^2}{3\Omega_2}\Big)
\,.\end{align}
%%%
$F_{\rm cor}$ was imposed to be a polynomial of degree 2 in $k_1$ and $k_2$ that introduces correlations such that $F$ in \eq{F_num} remains normalized and produces the first moments required by \eq{S_exp_2}. We explored correlations by varying the size of the correlation parameter $c \sim 1/\lqcd^2$. 

In regime 2 and 3 the nonperturbative effects involving $e_\bt$ are suppressed because $Qe_\bt \gg Q e_\al$. We will nevertheless use \eq{F_num} in these regimes as well. For the leading nonperturbative correction, this seems reasonable from the point of view of continuity.
The cross section for the ratio of angularities is particularly interesting, because nonperturbative corrections contribute to any value of the ratio, since this integrates over a line that goes through $(e_\al,e_\bt)=(0,0)$. 

%%%%%%%%%%%%%%%%%%%%%%%%%%%%%%%%%%%%%%%%%%%%%%%%%%%%%%%%%%%%%%%%%%%%%%%%%%%%%%%%
\section{Results}
\label{sec:results}
%%%%%%%%%%%%%%%%%%%%%%%%%%%%%%%%%%%%%%%%%%%%%%%%%%%%%%%%%%%%%%%%%%%%%%%%%%%%%%%%

%===============================================================================
\subsection{Single angularity}
%===============================================================================

%---------------------------------------
 \begin{figure}[t]
  \centering
   \includegraphics[width=0.49\textwidth]{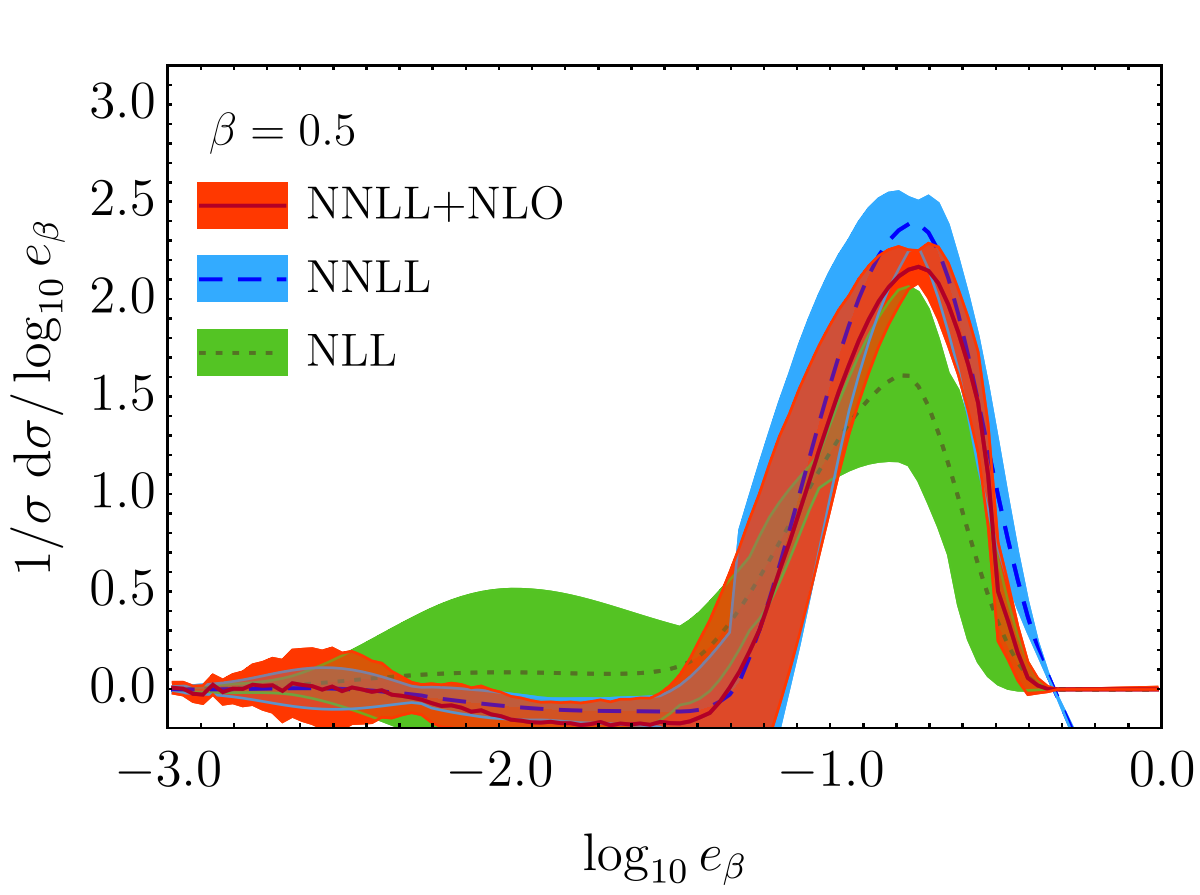}
   \includegraphics[width=0.49\textwidth]{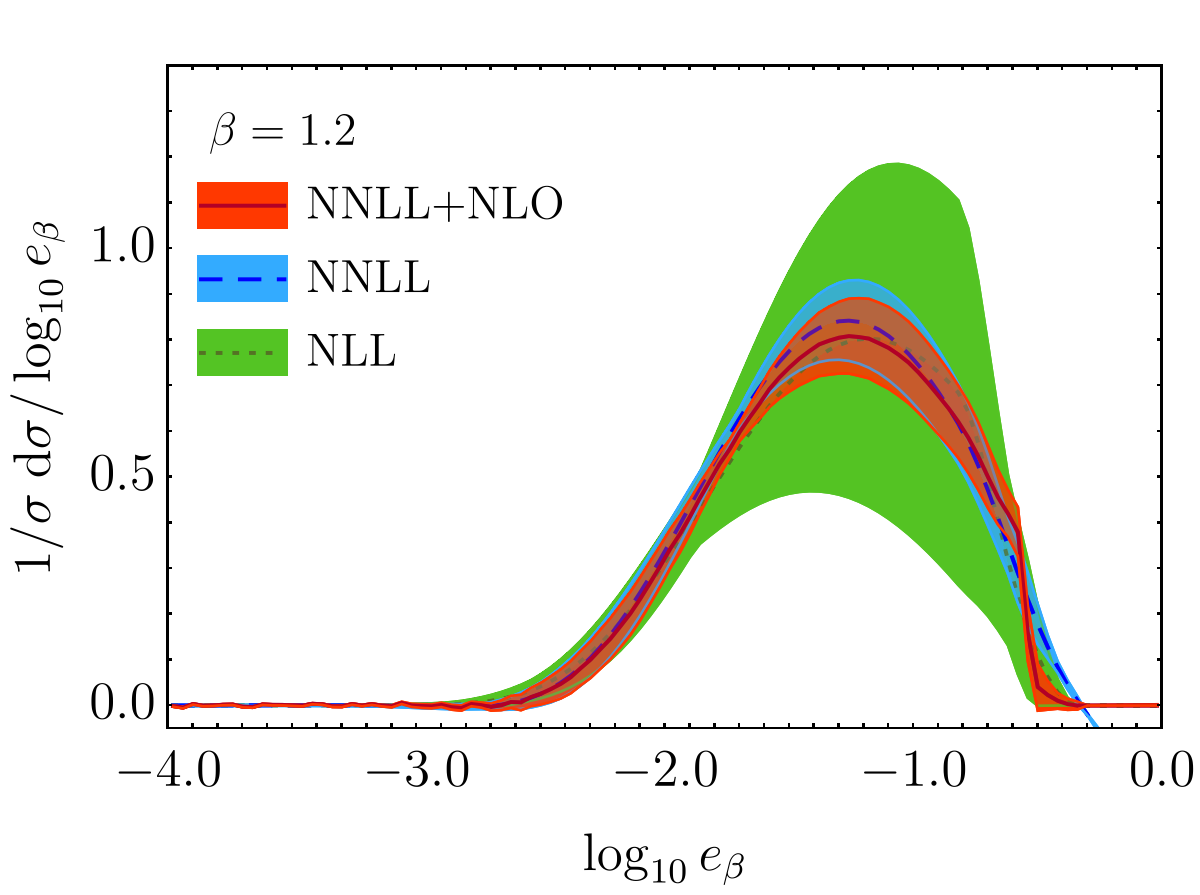}  \\
   \includegraphics[width=0.49\textwidth]{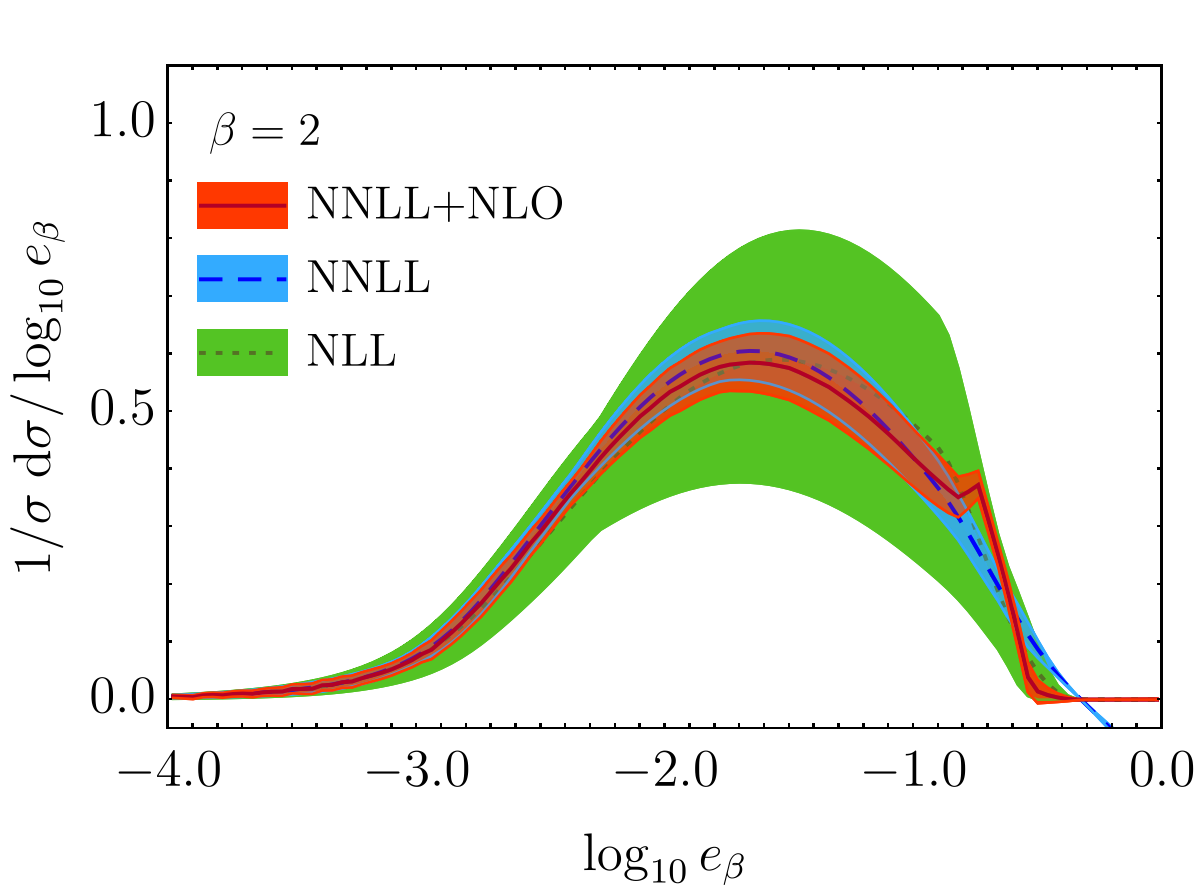}
   \includegraphics[width=0.49\textwidth]{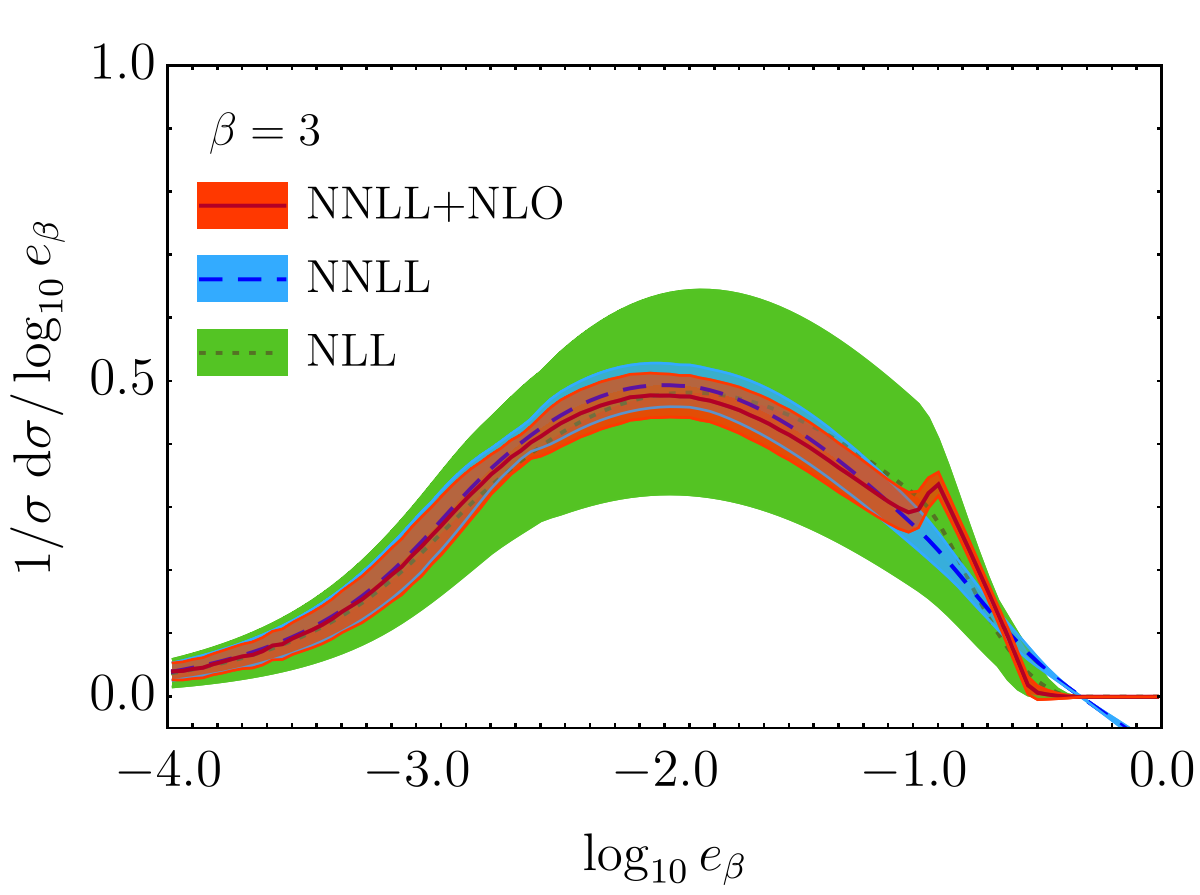}
   \caption{
   Results for the NLL, NNLL and NNLL+NLO cross sections with uncertainties for four angularities $\bt = 0.5, 1.2, 2$ and $3$ 
   (all normalized relative to the full NLO cross section).
   }
   \label{fig:1dresults}
  \end{figure}
%---------------------------------------
%---------------------------------------
 \begin{figure}[t]
  \centering
   \includegraphics[width=0.49\textwidth]{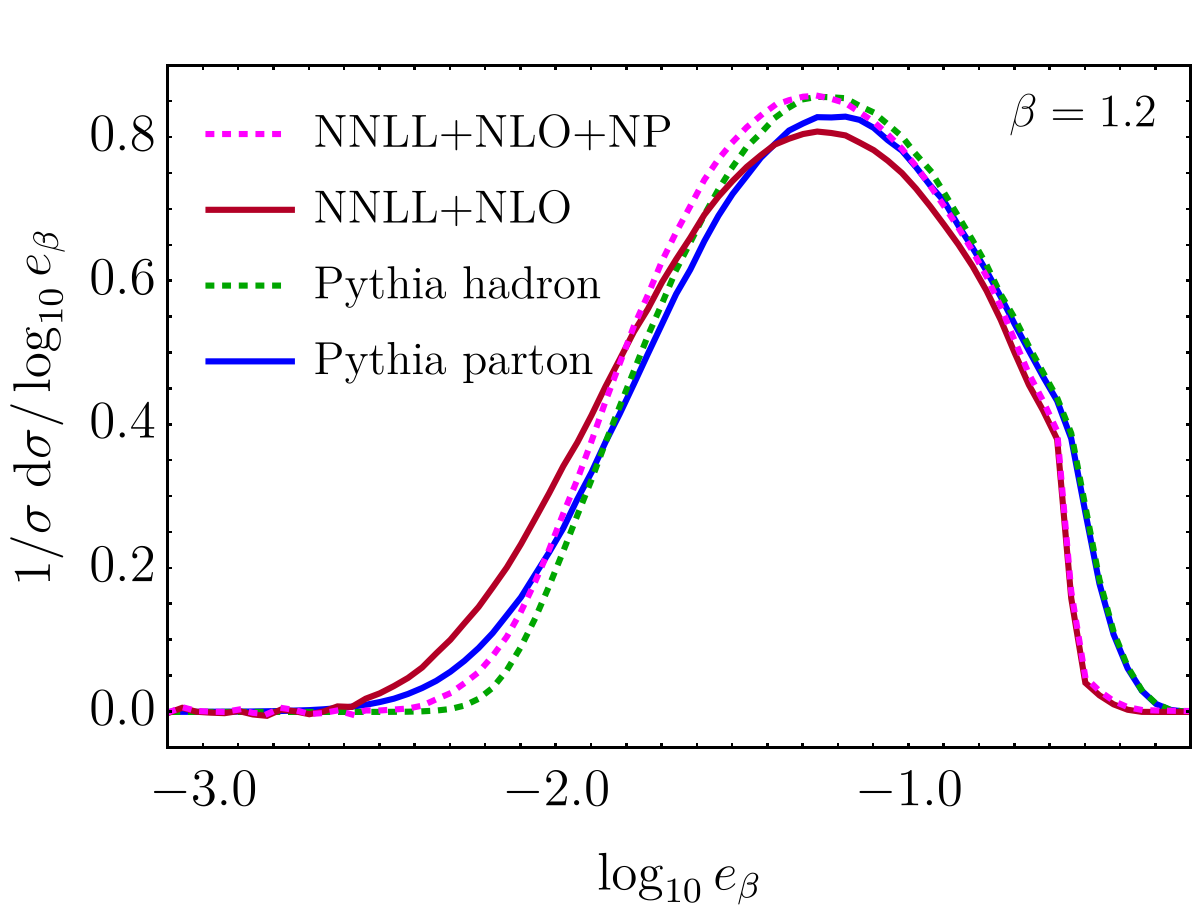}  
   \includegraphics[width=0.49\textwidth]{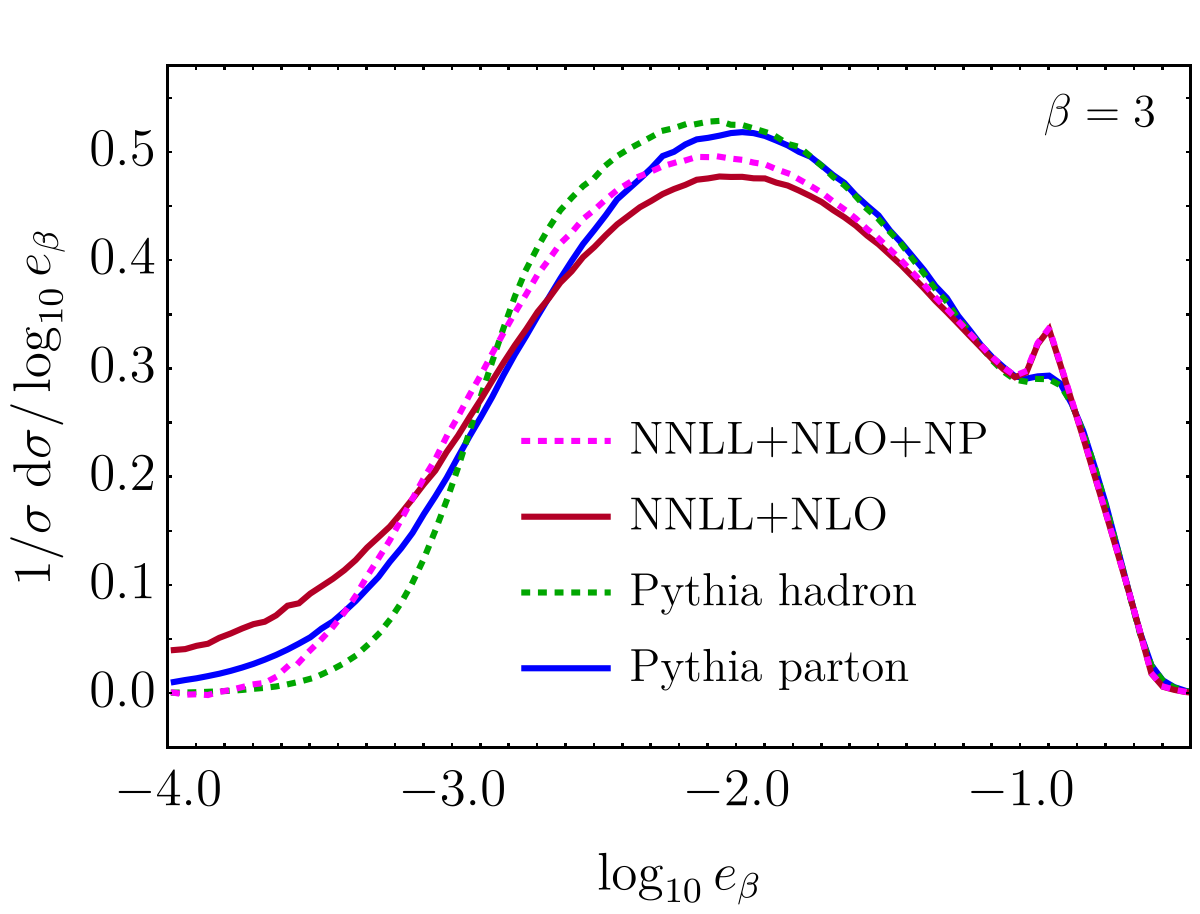} \\
   \includegraphics[width=0.49\textwidth]{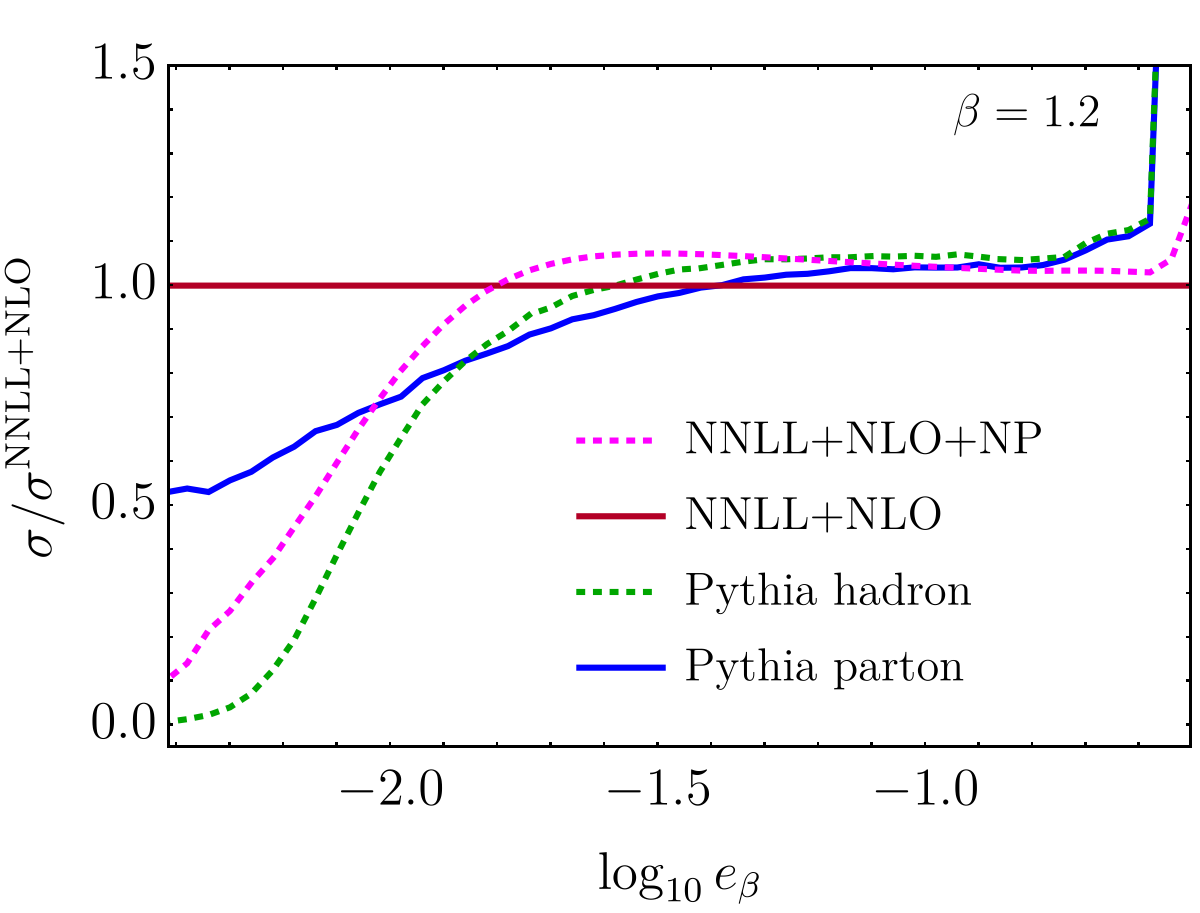}  
   \includegraphics[width=0.49\textwidth]{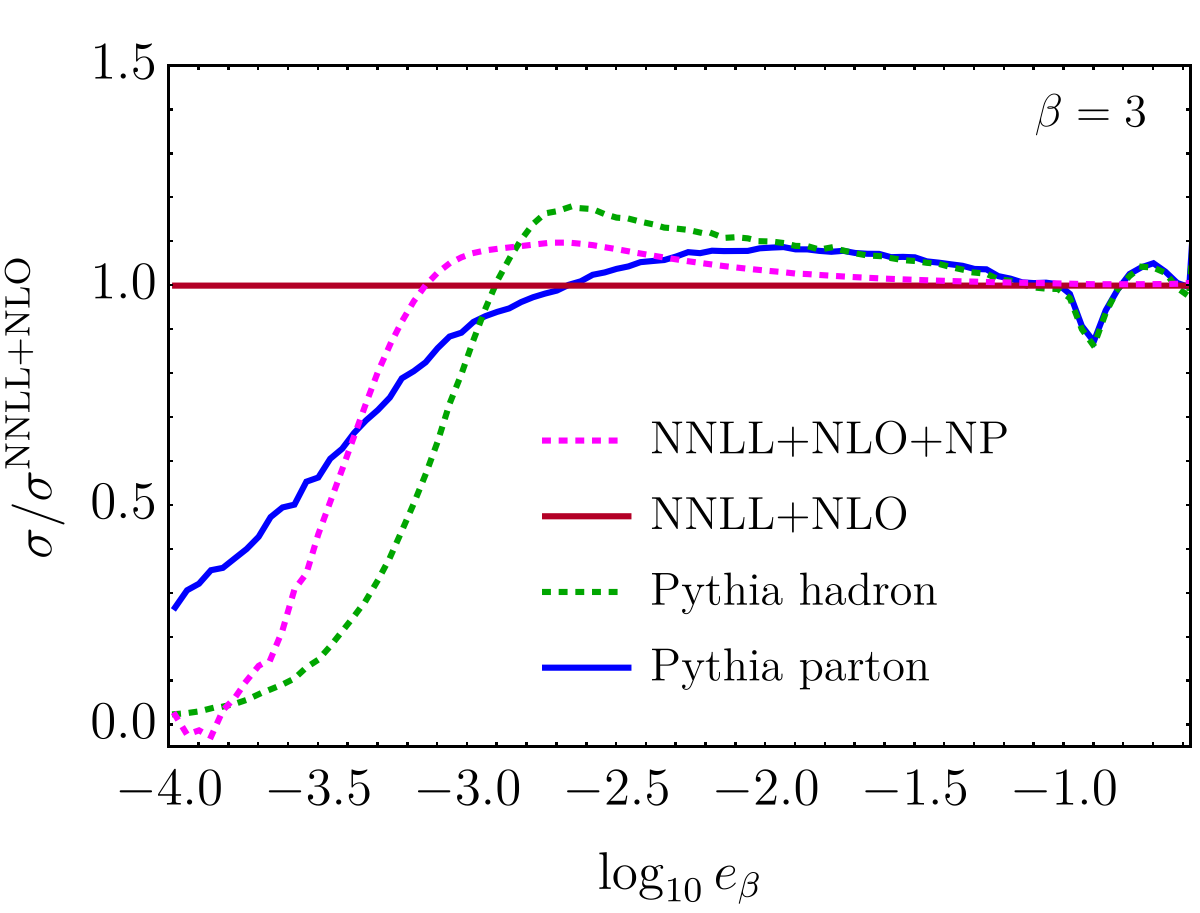}
   \caption{
   NNLL+NLO without and with nonperturbative effects, compared to \Pythia at parton and hadron level, for $\beta =1.2$ (left) and $\beta=3$ (right). The bottom row shows the ratio with the NNLL+NLO cross section.}
   \label{fig:1dresultsWithNP}
  \end{figure}
%---------------------------------------
%---------------------------------------
 \begin{figure}[t]
  \centering
   \includegraphics[width=0.49\textwidth]{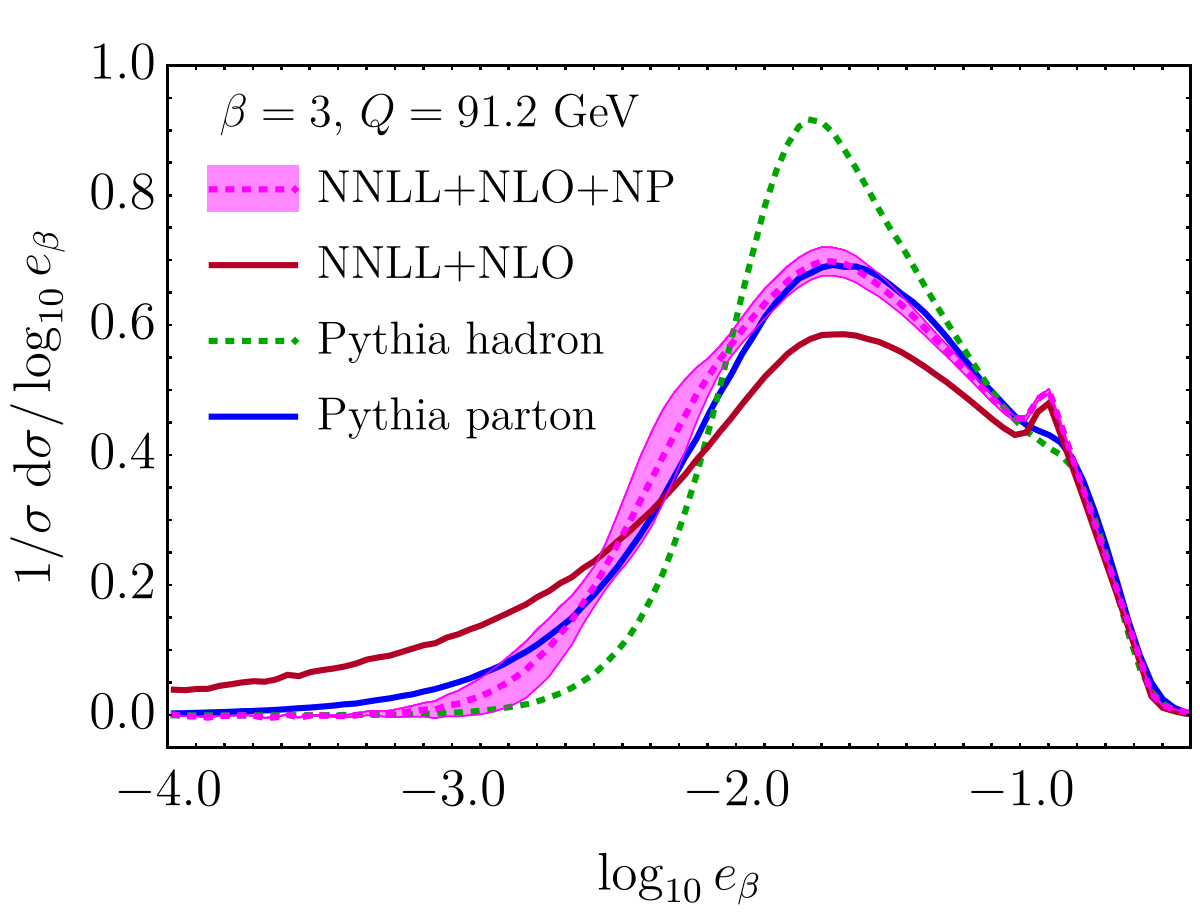} 
   \includegraphics[width=0.49\textwidth]{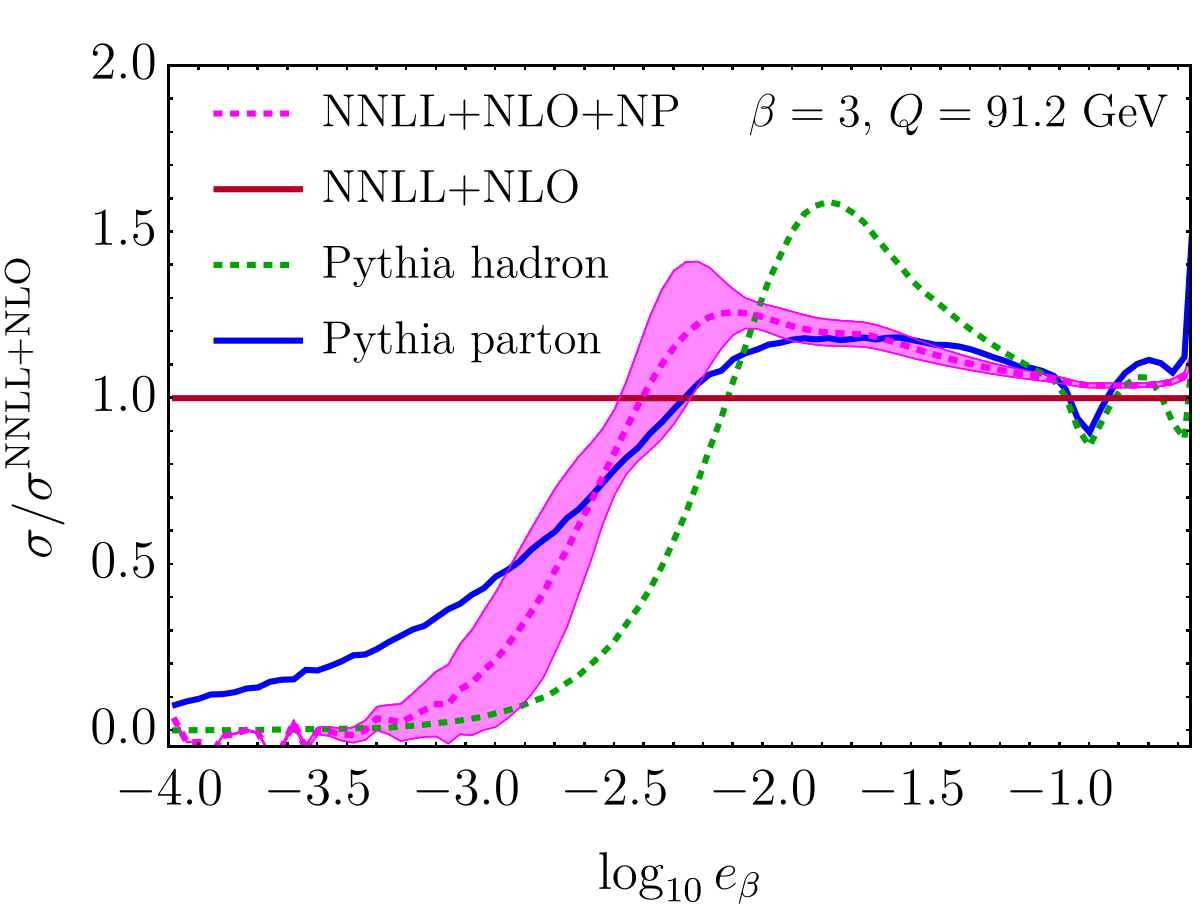}
   \caption{
   NNLL+NLO without and with nonperturbative effect, compared to \Pythia at parton and hadron level for $Q=91.2$ GeV and $\beta =3$. The band provides an estimate of the nonperturbative uncertainty, as described in the text, which is sizable due to the smaller value of $Q$.}
   \label{fig:1dresultsWithNPMZ}
  \end{figure}
%---------------------------------------

We start by presenting results for the cross section of a single angularity $e_\beta$ in \fig{1dresults}.
Shown are our predictions at NLL, NNLL and NNLL+NLO order (defined in table~\ref{tab:orders}) for angularity exponents $\beta=0.5$, 1.2, 2, 3 with $Q = 1000$ GeV. The bands show the perturbative uncertainty estimated by varying the profile scales, as described in \sec{prof_1D}. Our predictions for the central curves are normalized to 1.\footnote{For $\beta=0.5$ we normalize the region $\log_{10} e_\bt \geq -1.3$, to avoid a large effect from the negative cross section in the nonperturbative region.}
The variations are not normalized to 1, but rescaled by the same amount as the corresponding central curve. 
As expected, the uncertainty bands reduce at higher orders, and overlap between the different orders over most of the range. The one exception is the NNLL vs.~NNLL+NLO in the fixed-order region. This is not surprising, because in this region the matching with NLO cannot be neglected. 
The Sudakov shoulder~\cite{Catani:1997xc} that features in the spectrum at large values of $e_\bt$ comes from the matching with NLO and is due to our choice of using the WTA axis, as observed already in \sec{FOns}. 
We checked that the separation between the Sudakov shoulder and the peak of the distribution decreases for smaller values of $\beta$ but for $\beta=0.5$ and $Q=1\,{\rm TeV}$ it is sufficiently large to preserve the reliability of our matched result, in agreement with the discussion in ref.~\cite{Banfi:2018mcq}.
Interestingly, for smaller values $\bt$ the range of $e_\bt$ values gets squeezed, such that there is a fairly rapid transition from nonperturbative region at small $e_\bt$ to the fixed-order region at large $e_\bt$. 

In \fig{1dresultsWithNP} we show our NNLL+NLO result for $\beta = 1.2$ and $3$ with and without nonperturbative corrections, included using the procedure described in \sec{nonperturbative}. We compare these to \Pythia with and without hadronization. We also show the ratio with NNLL+NLO to make it easier to distinguish these curves. The effect of hadronization on our perturbative prediction is very similar to the difference between \Pythia at the parton and hadron level. The curves are not the same, but this difference is already present before including nonperturbative effects. 
Pythia smooths the Sudakov shoulder by taking into account additional resummation effects.

The corresponding plot for $Q=91.2$ GeV is shown in \fig{1dresultsWithNPMZ}. Here we added also a nonperturbative uncertainty band, which was obtained by varying $\Omega$ within its uncertainty \cite{Abbate:2010xh} and adding in quadrature the envelope of the variations obtained by considering 
%%%
\begin{align}
\tilde{F}(Qe_\al,a) = \frac{(Q e_\al)^a}{\Gamma(1+a)} \bigg(\frac{1+a}{\Omega_\al}\bigg)^{1+a} \, e^{-(1+a)Q e_\al/\Omega_\al}
\end{align}
%%%
with $a=1,2,3$ and $4$. $\tilde{F}(Qe_\al,1)$ coincides with $F(Qe_\al)$ in \eq{shapefunction}.
These alternative functional forms $\tilde F$ are all normalized and have the same leading nonperturbative correction, thus probing the effect of subleading nonperturbative effects. Indeed, the uncertainty band in \fig{1dresultsWithNPMZ} grows significantly at small values of the angularity, because of the sensitivity to the shape of $\tilde F$ and not just its first moment. For $Q=1000$ GeV this uncertainty is very small, which is why we do not show the corresponding plot.

%===============================================================================
\subsection{Single angularity distributions from \Eventtwo}
\label{sec:event2}
%===============================================================================

%---------------------------------------
 \begin{figure}[t!]
  \centering
\includegraphics[height=0.35\textwidth]{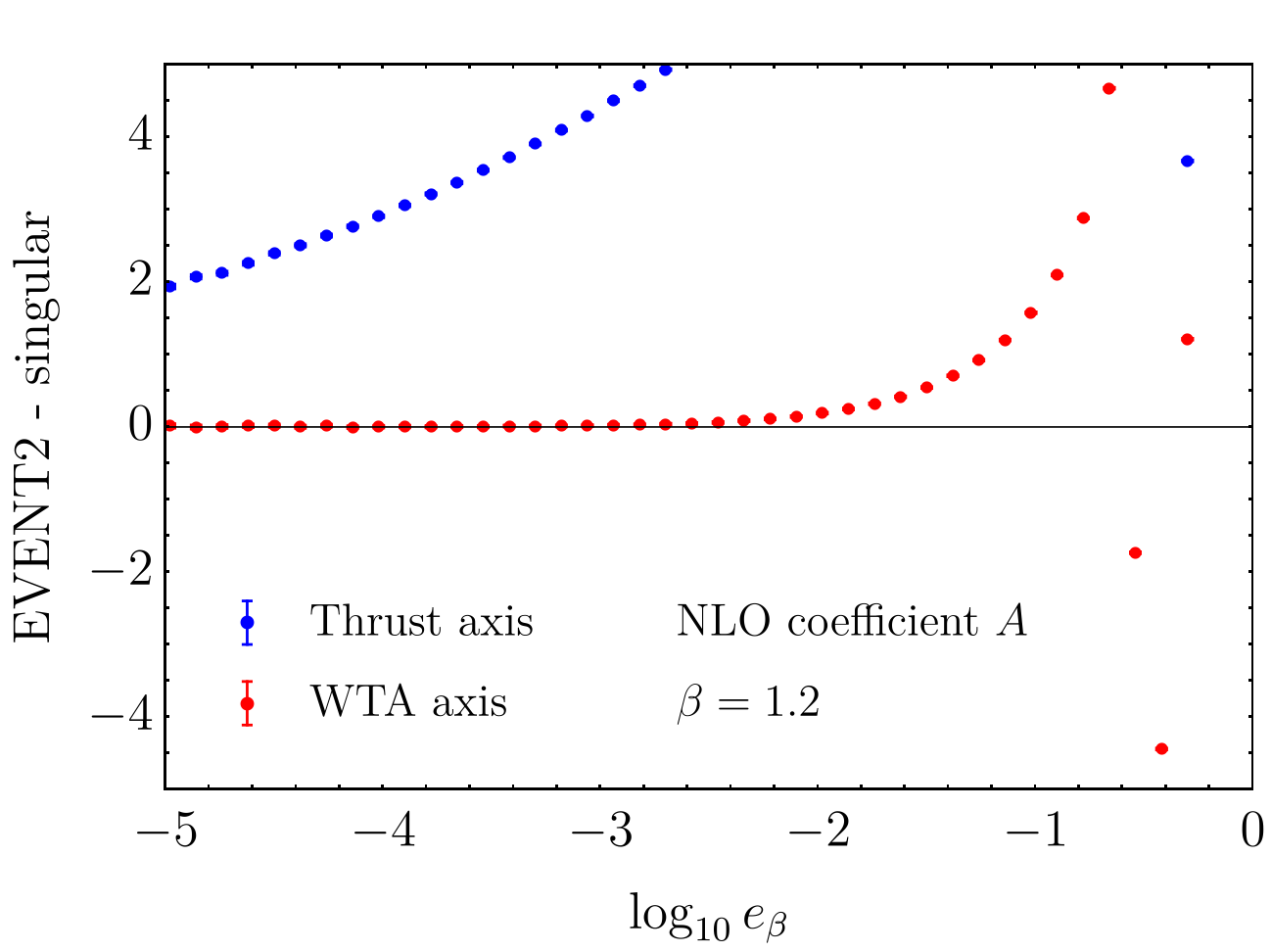}
\includegraphics[height=0.35\textwidth]{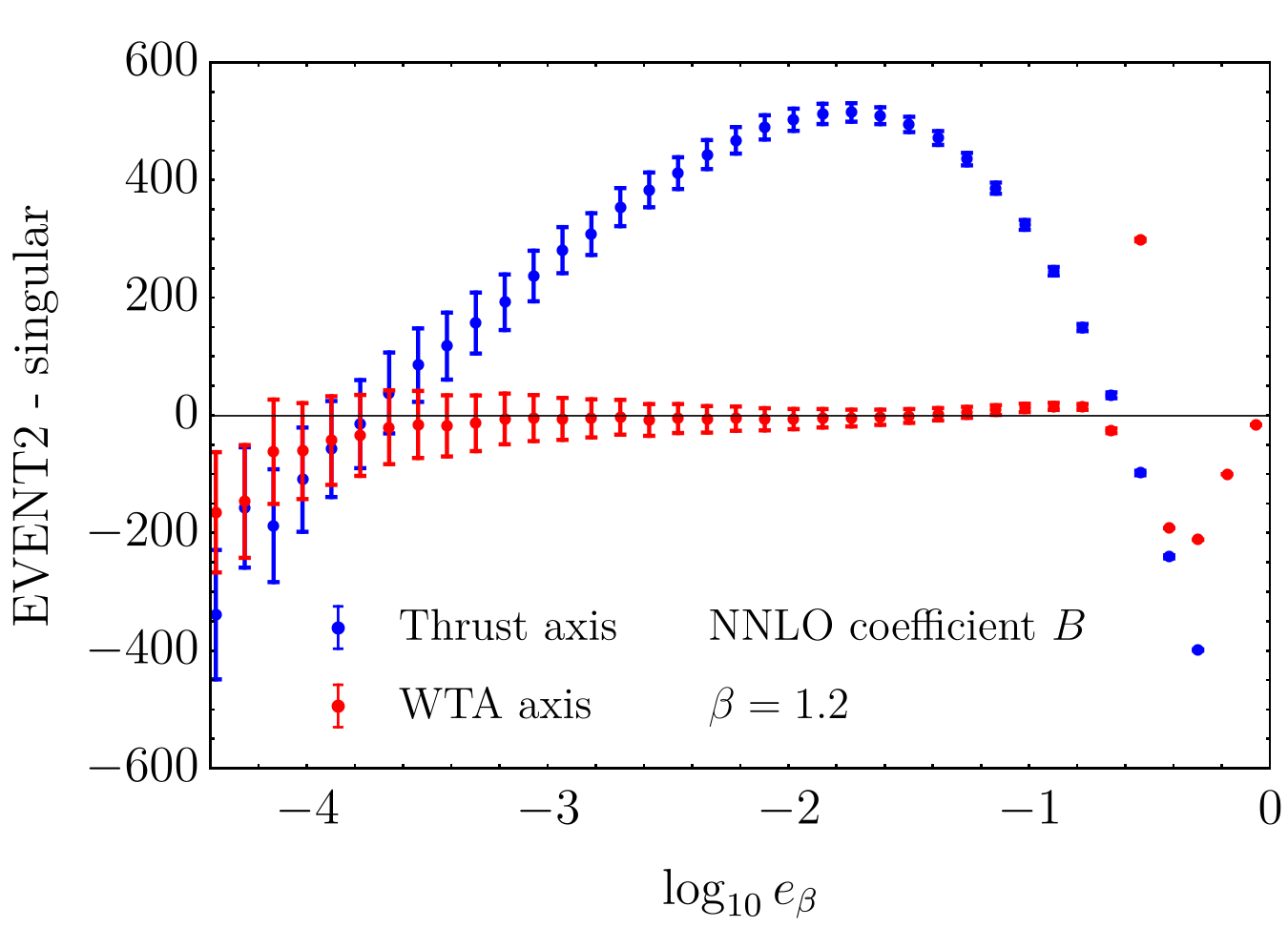} \\
\includegraphics[height=0.35\textwidth]{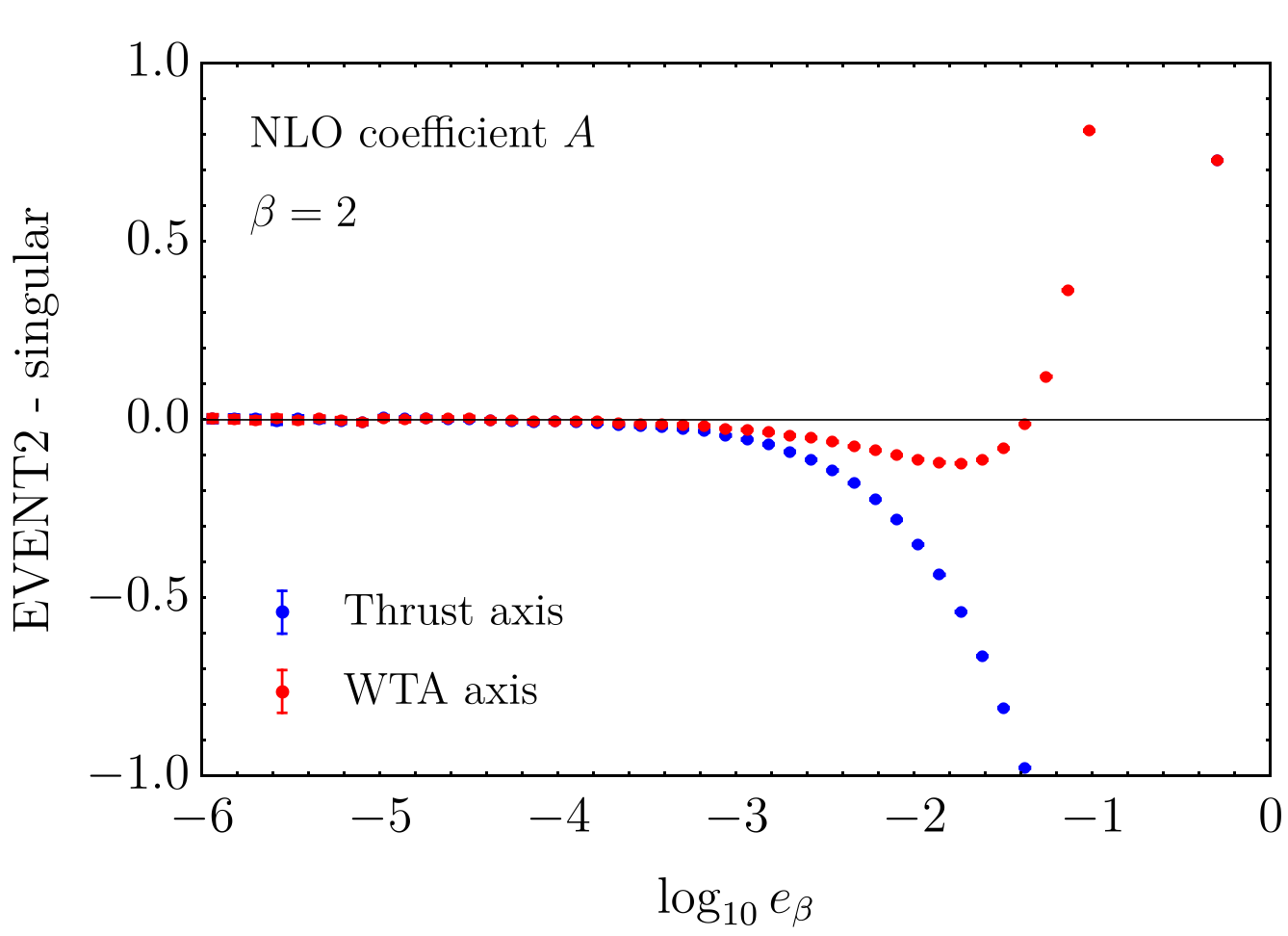}
\includegraphics[height=0.35\textwidth]{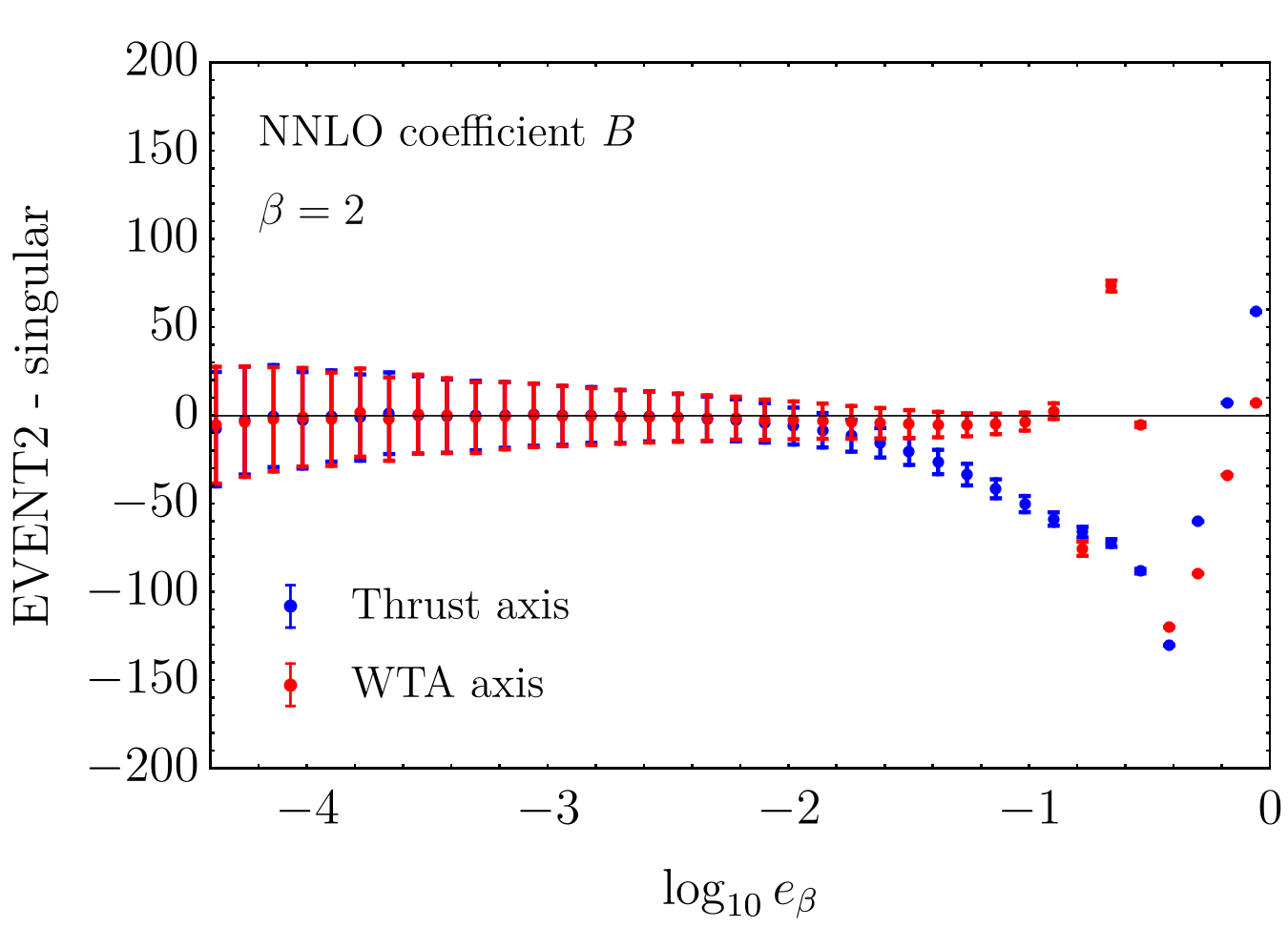} \\
   \caption{
Difference between the NLO and NNLO terms for the single angularity cross section calculated by \Eventtwo and our singular results. For $\beta = 1.2$ with the thrust axis, the absence of a plateau is due to recoil effects.}
  \label{fig:e2}
  \end{figure}
%---------------------------------------

To test the factorization framework, especially for the WTA axis choice, we compared the fixed-order expansions from our resummed single differential cross sections against numerical results from the \Eventtwo generator~\cite{Catani:1996vz}. For this purpose, we ran \Eventtwo with $n_f=5$ and an infrared cutoff $\rho = 10^{-10}$ and generated one trillion events.
To be explicit about what is compared here, we write the expansion of the cross section as 
%%%
\begin{align}
\frac{1}{\hat{\sigma}_0} \, \frac{\df \sigma}{\df \log_{10}{e_\bt}} = \frac{\al_s}{2 \pi} A(\log_{10}{e_\bt}) + \left(\frac{\al_s}{2 \pi} \right)^2 B(\log_{10}{e_\bt})  + \ord{\al_s^3}.
\end{align}
%%%
In \fig{e2}, we plot the difference between the \Eventtwo output and our singular contributions to the NLO and NNLO coefficients $A$ and $B$ for angularity exponents $\beta = 1.2$ and 2. Here we consider both the thrust axis and the WTA axis. Assuming that recoil effects can be ignored, the only difference at this order can be traced back to the constant in the (cumulative) one-loop jet function, which for the thrust axis was calculated in ref.~\cite{Hornig:2009vb}. In \app{sing} we collect our (N)NLO singular results for $A$ and $B$ for several angularity exponents and both axis choices. 
The (N)NLO coefficients can also be determined using the approach of ref.~\cite{Banfi:2014sua}, and agree with our results.\footnote{We are grateful to Pier Monni for providing this check before the publication of ref.~\cite{Banfi:2018mcq}.}
For the WTA case, the difference between \Eventtwo and our singular cross section goes clearly to zero at small values of the angularity (within statistical uncertainty) for both $\beta = 1.2$ and 2, at variance with the thrust axis case for $\beta = 1.2$ where power-suppressed terms become numerically large due to recoil effects. 
Interestingly, the turn off of the nonsingular contribution takes place substantially faster for the WTA axis.

At very small values of $e_\bt$, the comparison breaks down due to infrared cutoff effects in \Eventtwo. More specifically, \Eventtwo regulates infrared divergences by cutting on the invariant mass of pairs of partons, $(p_i + p_j)^2> \rho\, Q^2$. By applying this prescription to the SCET modes for the single angularity distribution,
%%%
\begin{align}
  (p_{\rm coll} + p_{\rm coll})^2 \sim e_\bt^{2/\bt} Q^2
  \,,\qquad  
  (p_{\rm coll} + p_{\rm soft})^2 \sim e_\bt Q^2
  \,, \qquad
  (p_{\rm soft} + p_{\rm soft})^2 \sim e_\bt^2 Q^2
\,,\end{align}
%%%
we conclude that \Eventtwo is expected to deliver reliable results for values of $e_\bt$ down to about $\rho^{\min(\bt/2,1)}$ at NLO, and about $\rho^{\min(\bt/2,1/2)}$ at NNLO. The further restriction at NNLO stems from the fact that at this order two soft emissions arise. We stress that this is simply an order-of-magnitude estimate, and judging from our numerical results, the true cutoff seems to be somewhat higher. 

%===============================================================================
\subsection{Two angularities}
%===============================================================================

%---------------------------------------
 \begin{figure}[t!]
  \centering
   \includegraphics[width=0.325\textwidth]{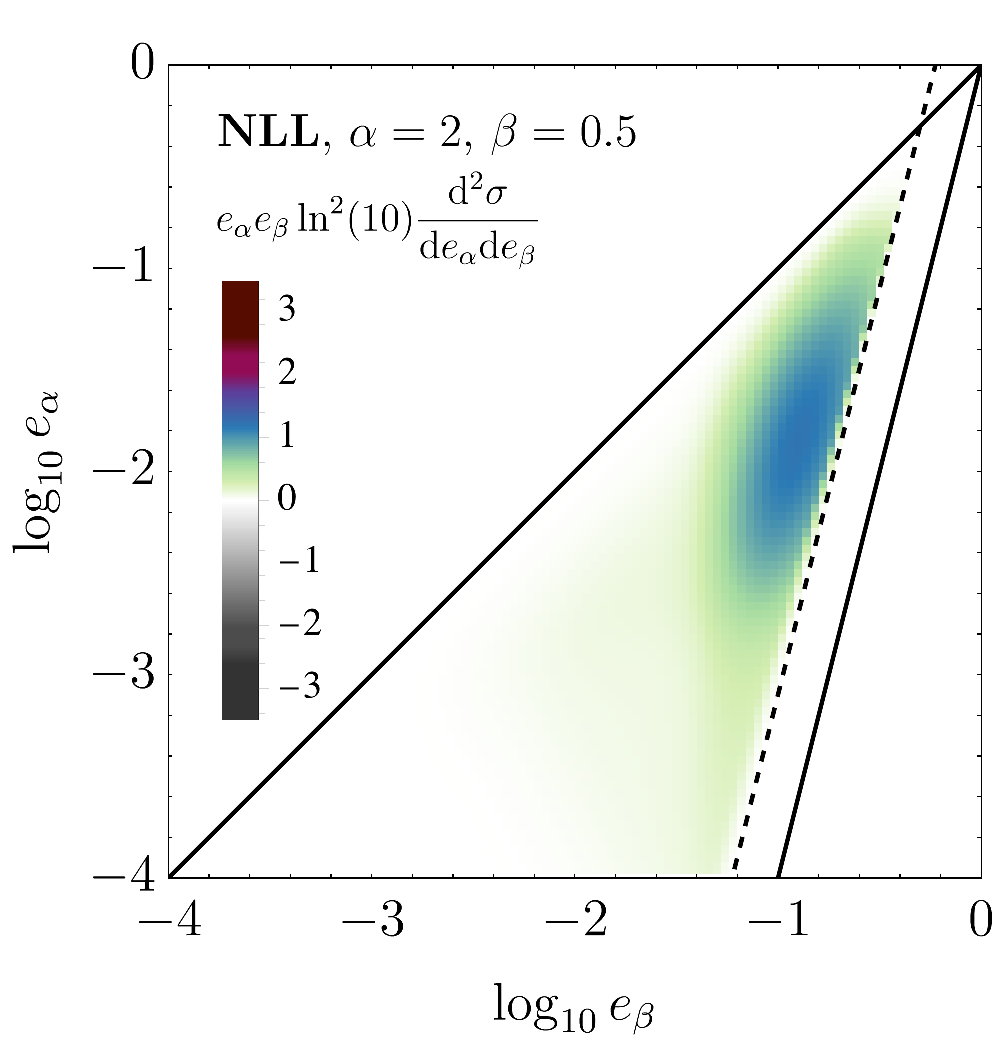}
   \includegraphics[width=0.325\textwidth]{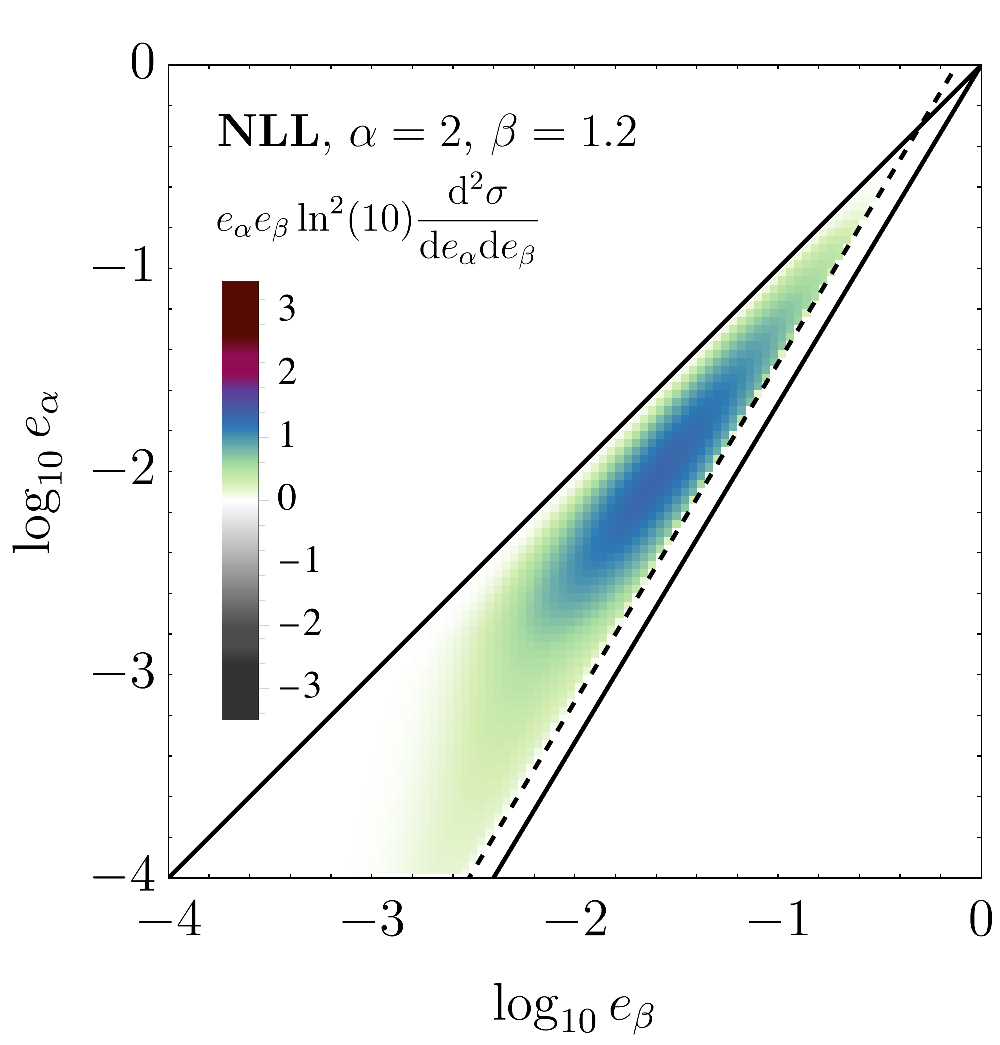}
   \includegraphics[width=0.325\textwidth]{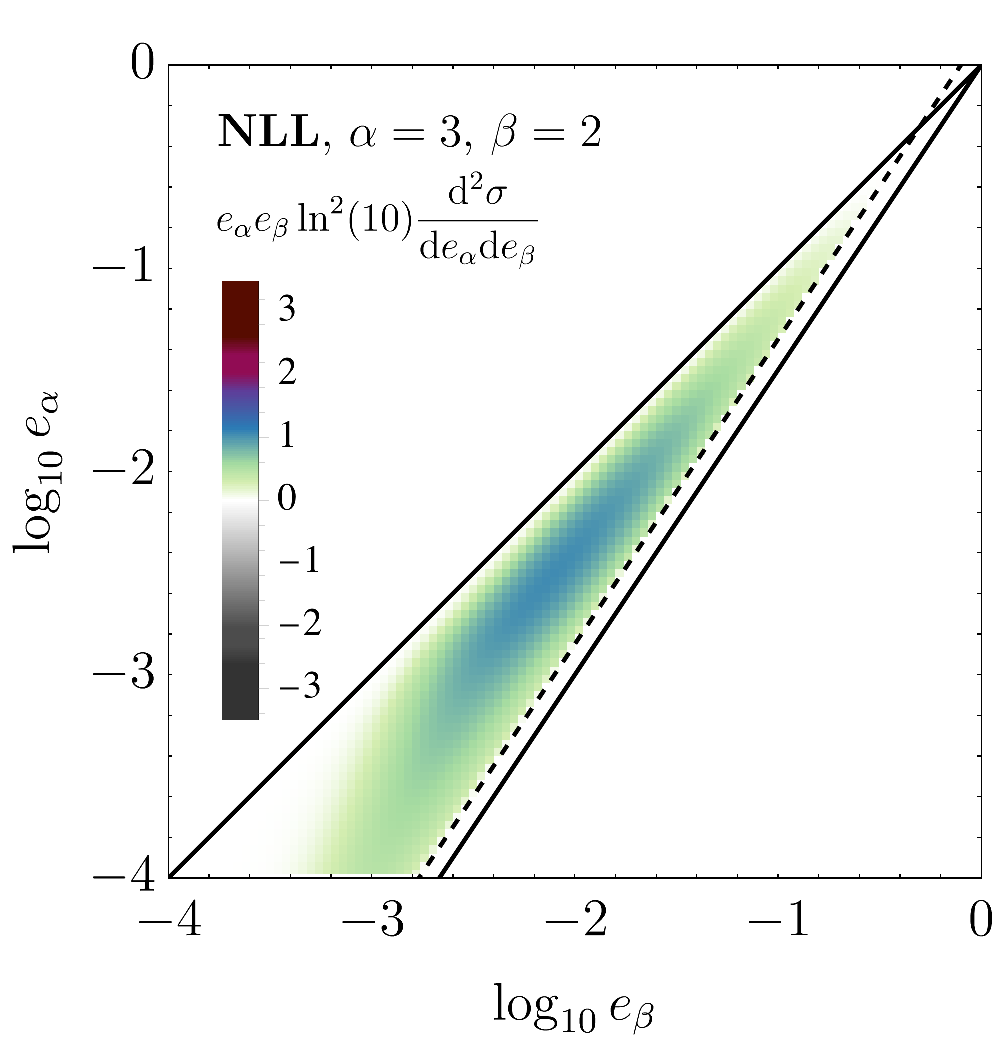}  \\
   \includegraphics[width=0.325\textwidth]{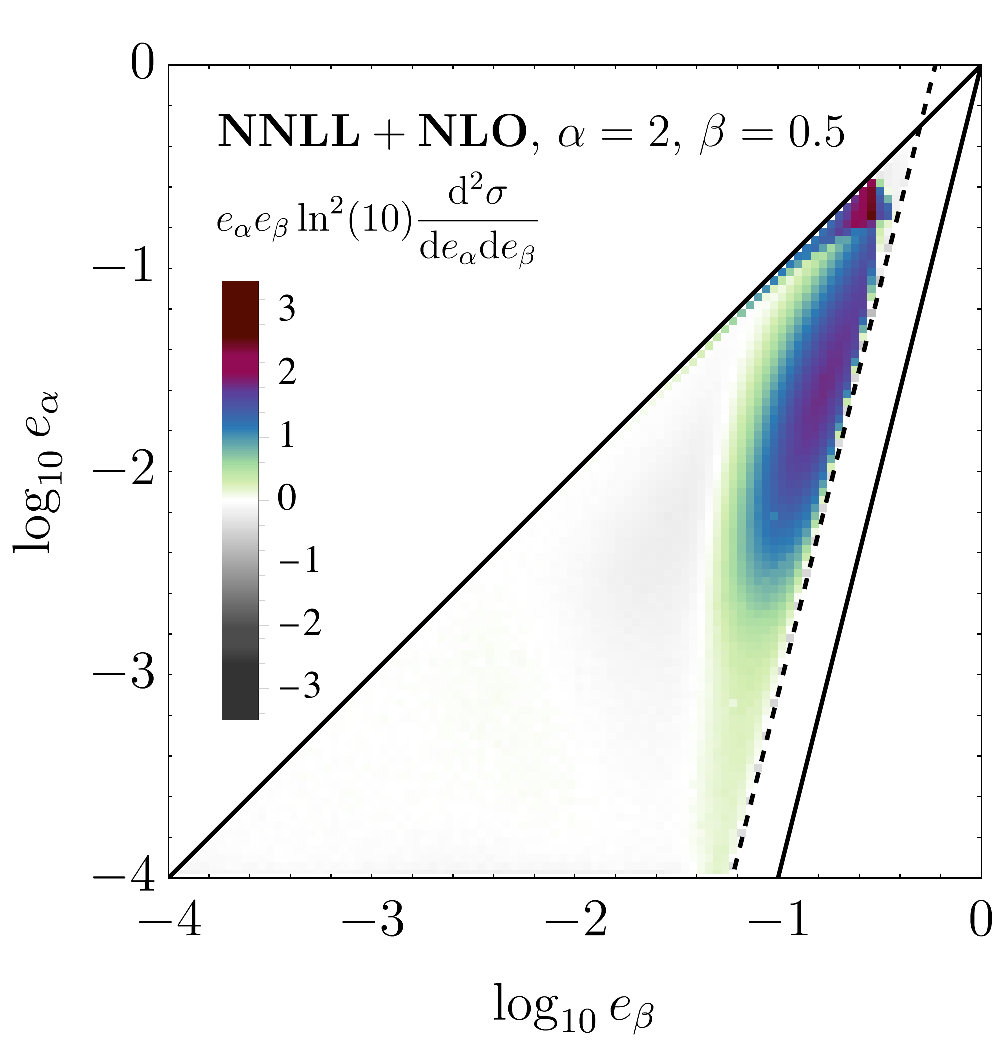}
   \includegraphics[width=0.325\textwidth]{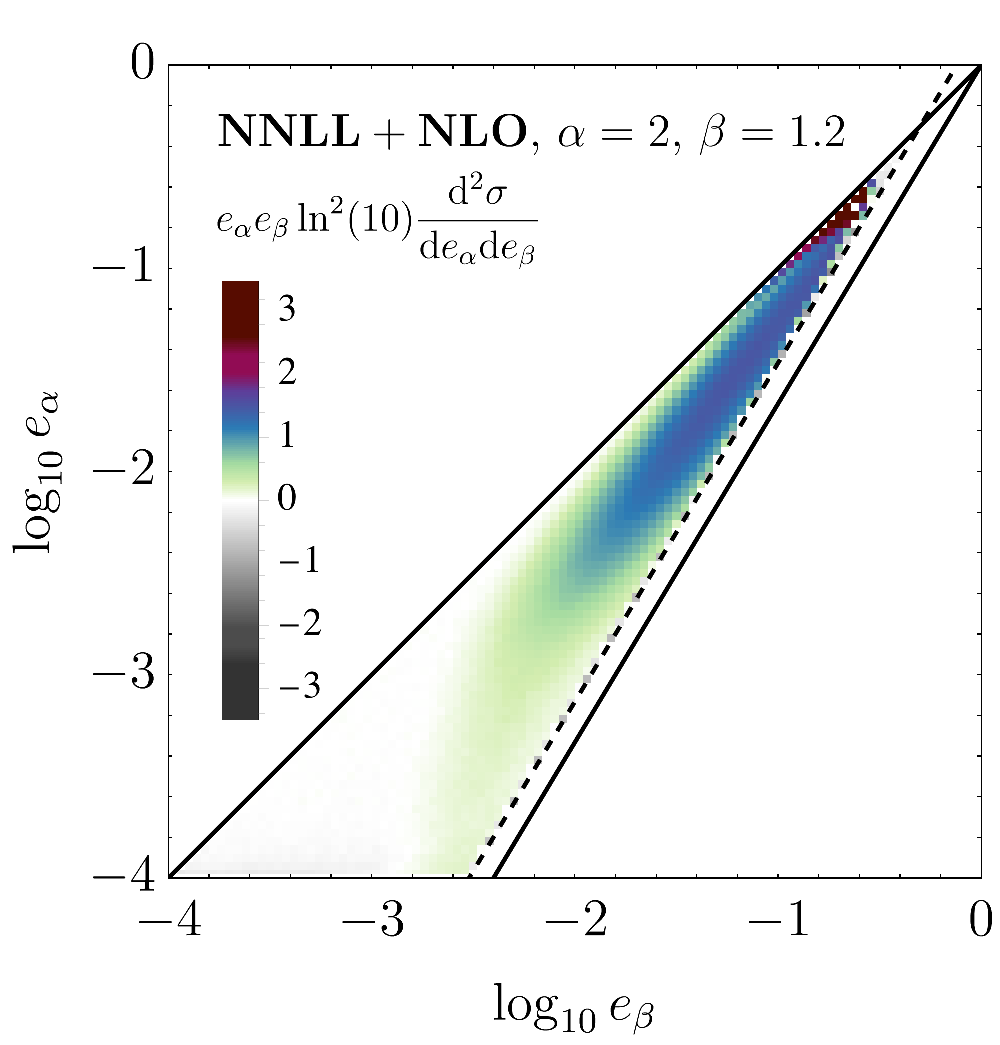}
   \includegraphics[width=0.325\textwidth]{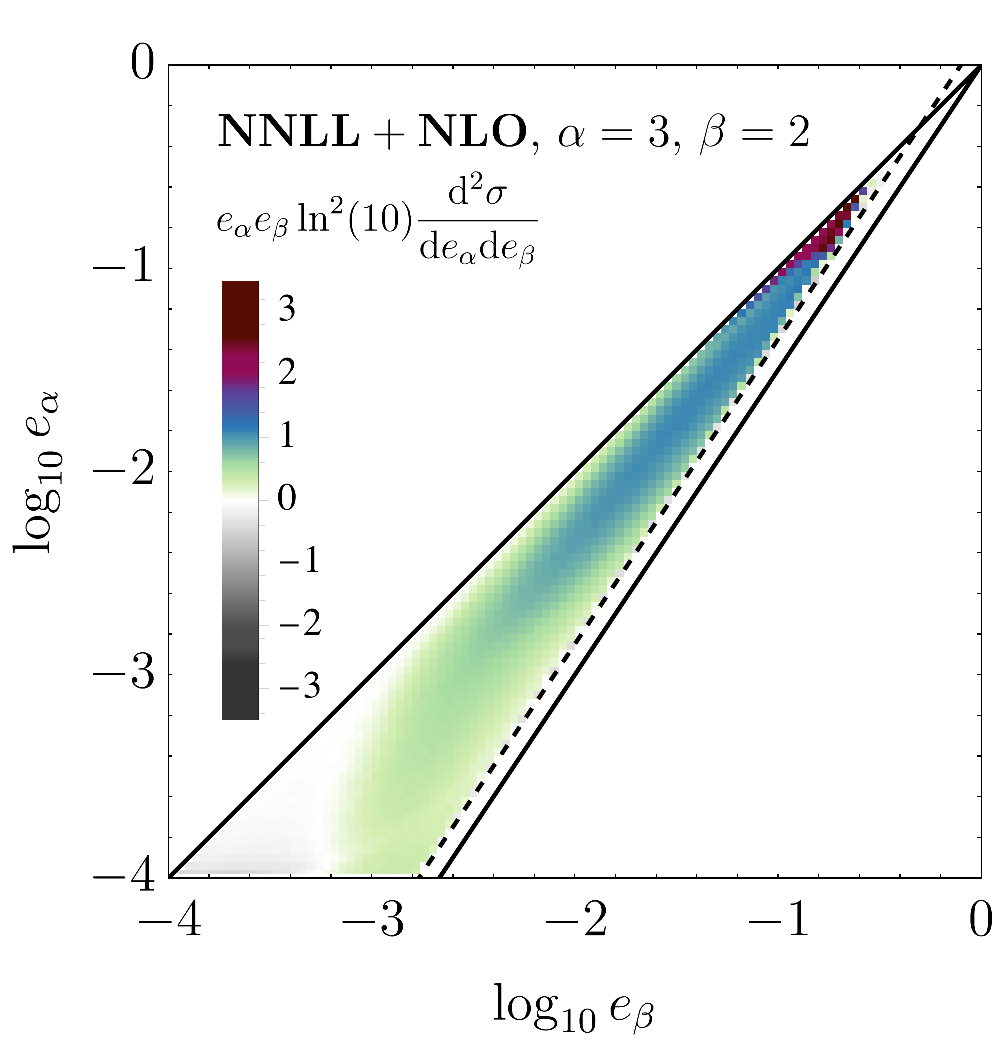}  \\
   \includegraphics[width=0.325\textwidth]{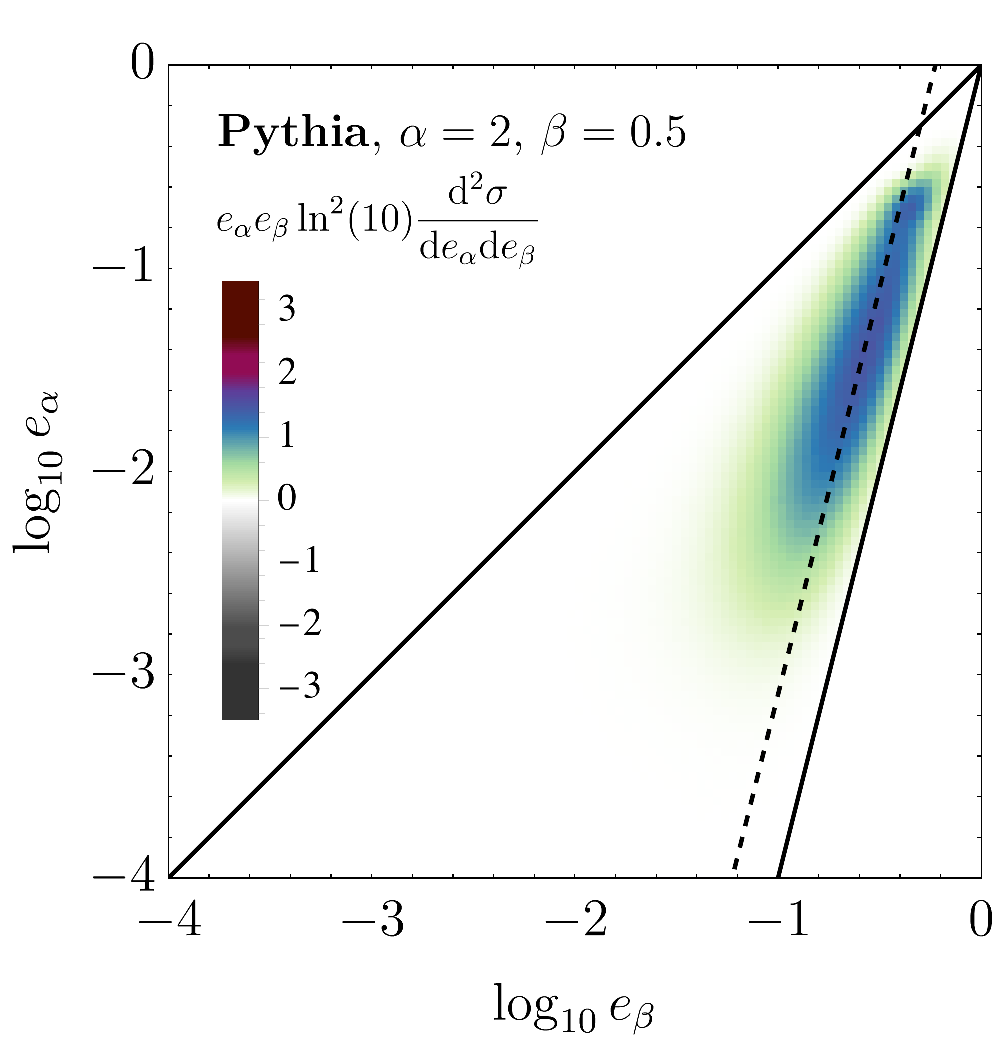}
   \includegraphics[width=0.325\textwidth]{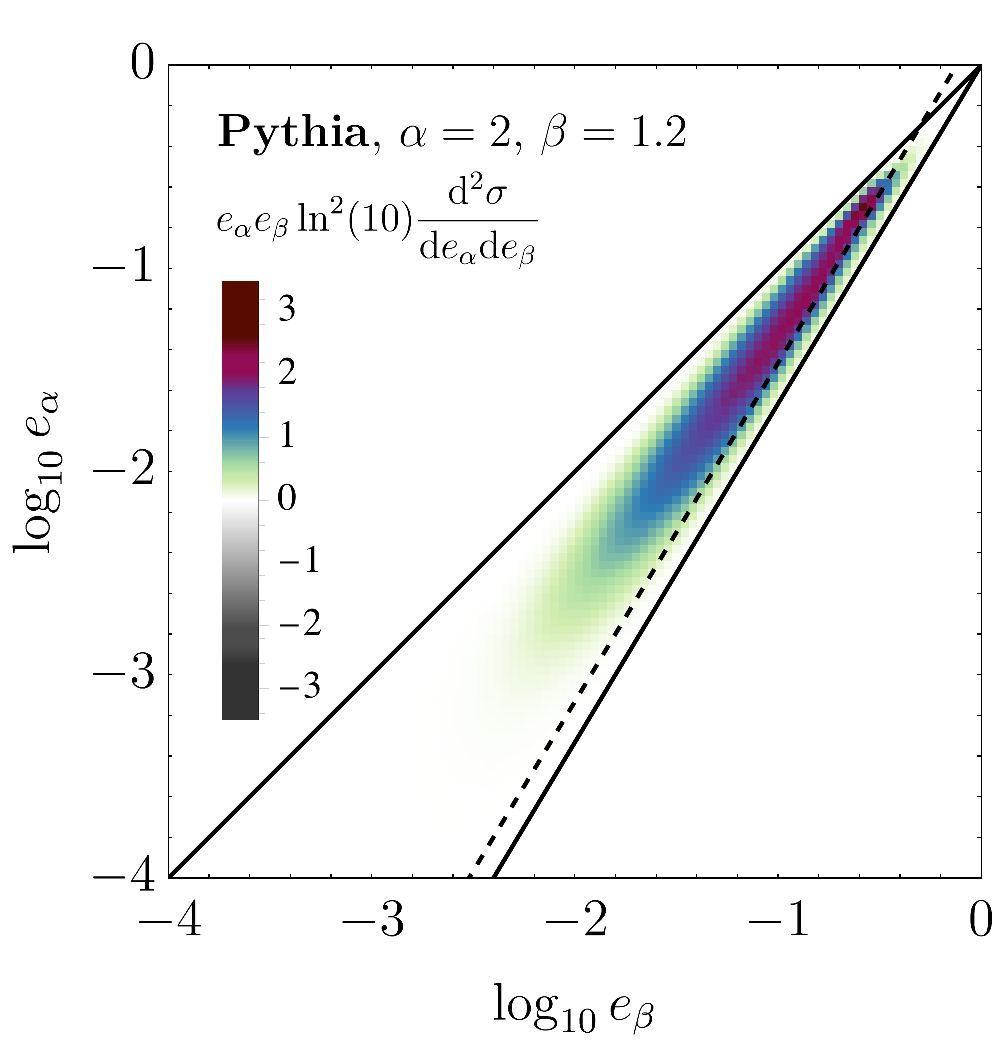}
   \includegraphics[width=0.325\textwidth]{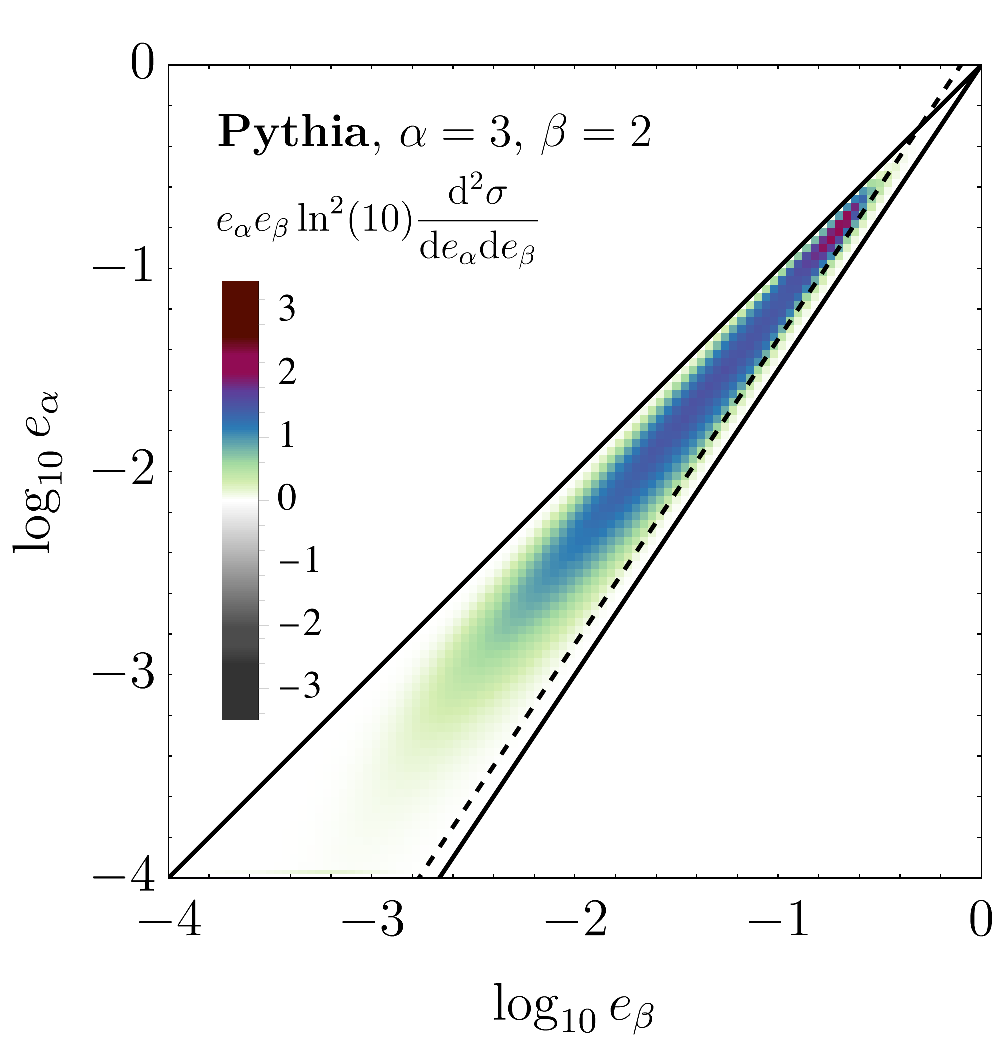}     
   \caption{The NLL (top), NNLL+NLO (middle) and \Pythia (bottom) cross section for three pairs of angularities
   $(\al,\bt) = (2, 0.5)$ (left), (2, 1.2) (middle) and (3,2) (right).
}
   \label{fig:2d}
  \end{figure}
%---------------------------------------

In \fig{2d} we show our results for the normalized cross section differential in the angularities $e_\al$ and $e_\bt$ at NLL and NNLL+NLO order, compared to \Pythia (parton level), where we again take $Q = 1000$ GeV. The difference between the NLL and NNLL+NLO is not very large, except in the fixed-order region. However, as is clear from our one dimensional plots, the uncertainties at NLL are pretty large. Indeed, the only reason that the cross section vanishes at NLL at the NLO phase-space boundaries is simply due to our choice of profile scales. If we would have turned off our profile scales at the canonical boundary instead, the peak region of the NLL cross section would be broader and extend (slightly) over the NLO phase-space boundary (dashed line). As discussed in \sec{FOns}, the sharp feature that the NNLL+NLO cross section exhibits in the fixed-order region is analogous to the bump of the single differential distributions, and is due to our choice of using the WTA axis. 
For $(\al,\bt) = (2,0.5)$ the peak of the distribution is close to the phase-space boundary corresponding to regime 3, while for the other angularity combinations it sits more in the middle between regime 1 and 3.

Comparing our results to \Pythia, we see that the \Pythia cross section is closer to our NNLL+NLO than NLL cross section. There are however notable differences: In \Pythia there is no sharp feature in the fixed-order region. Although it is expected that this would be somewhat washed out in \Pythia, it is surprising that there is no visible remnant (the corresponding bump for the one-dimensional distribution is still noticeable for \Pythia in \fig{1dresults}). The largest difference between the \Pythia and the NNLL+NLO distribution is for $(\al,\bt) = (2,0.5)$. 
In agreement with the discussion in ref.~\cite{Banfi:2018mcq}, \Pythia results, which take into account effects around the Sudakov shoulder, extend outside the NLO phase-space boundary in \eq{PS_NLO}.
That there are large differences in this case is not so surprising, because we have already seen that for $\beta=0.5$ the resummation region gets squeezed such that there is a quick transition between the fixed-order region and the nonperturbative region. 

%===============================================================================
\subsection{Ratio of angularities}
%===============================================================================
%---------------------------------------
 \begin{figure}[t]
  \centering
   \includegraphics[width=0.325\textwidth]{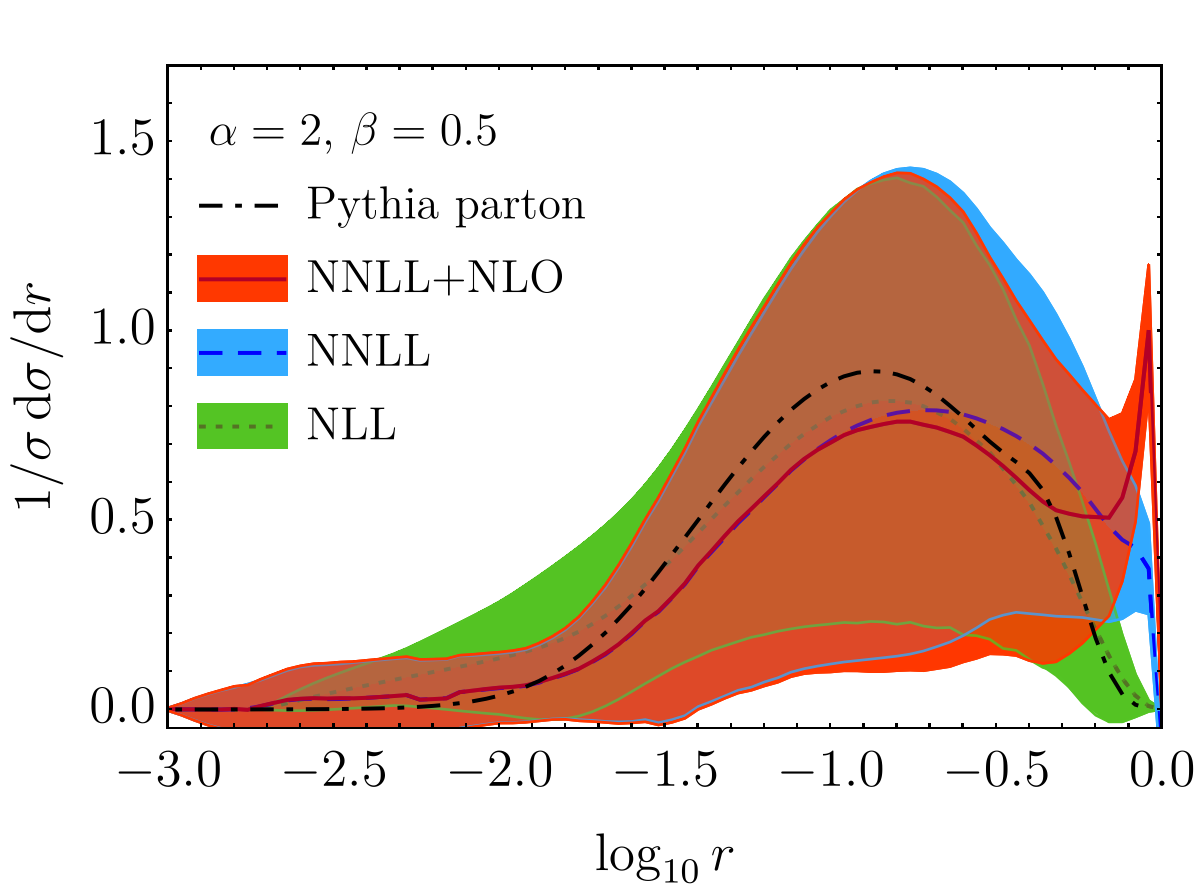}
   \includegraphics[width=0.325\textwidth]{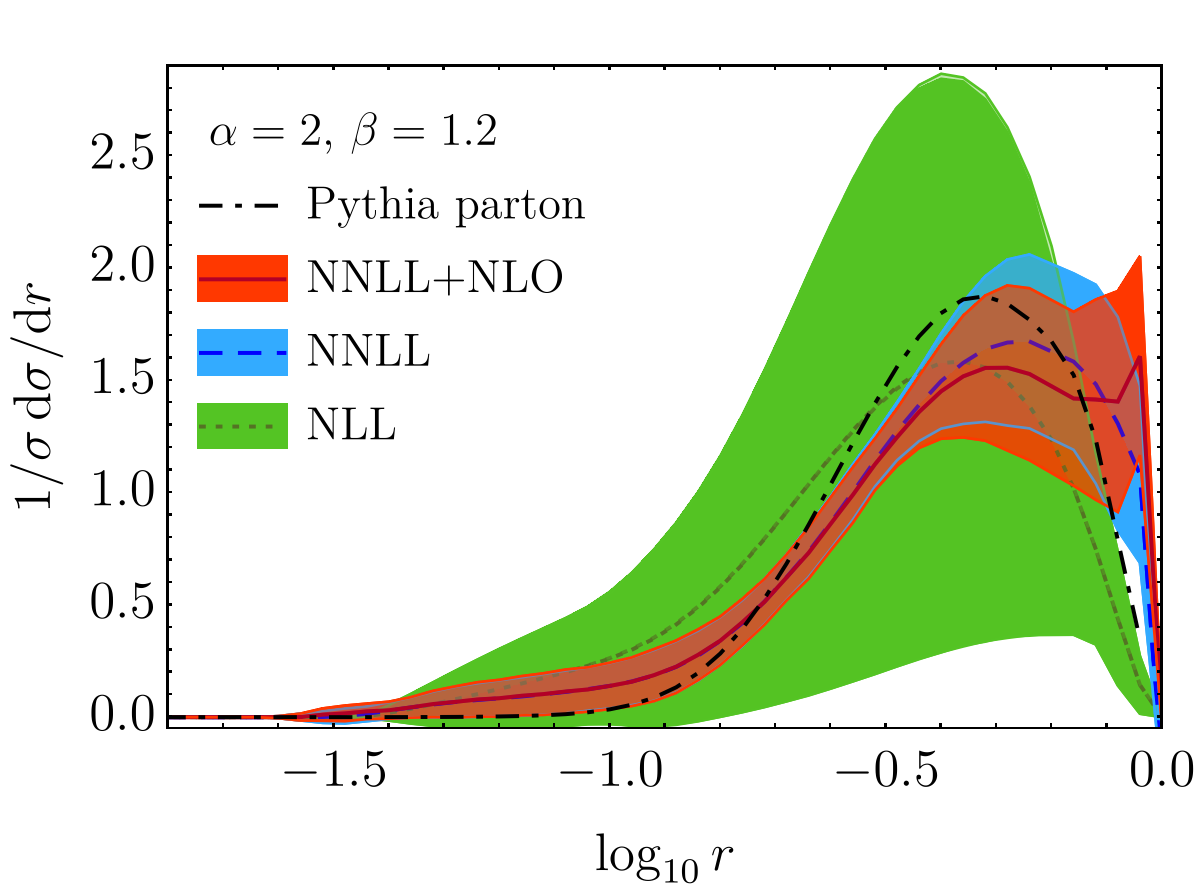}
   \includegraphics[width=0.325\textwidth]{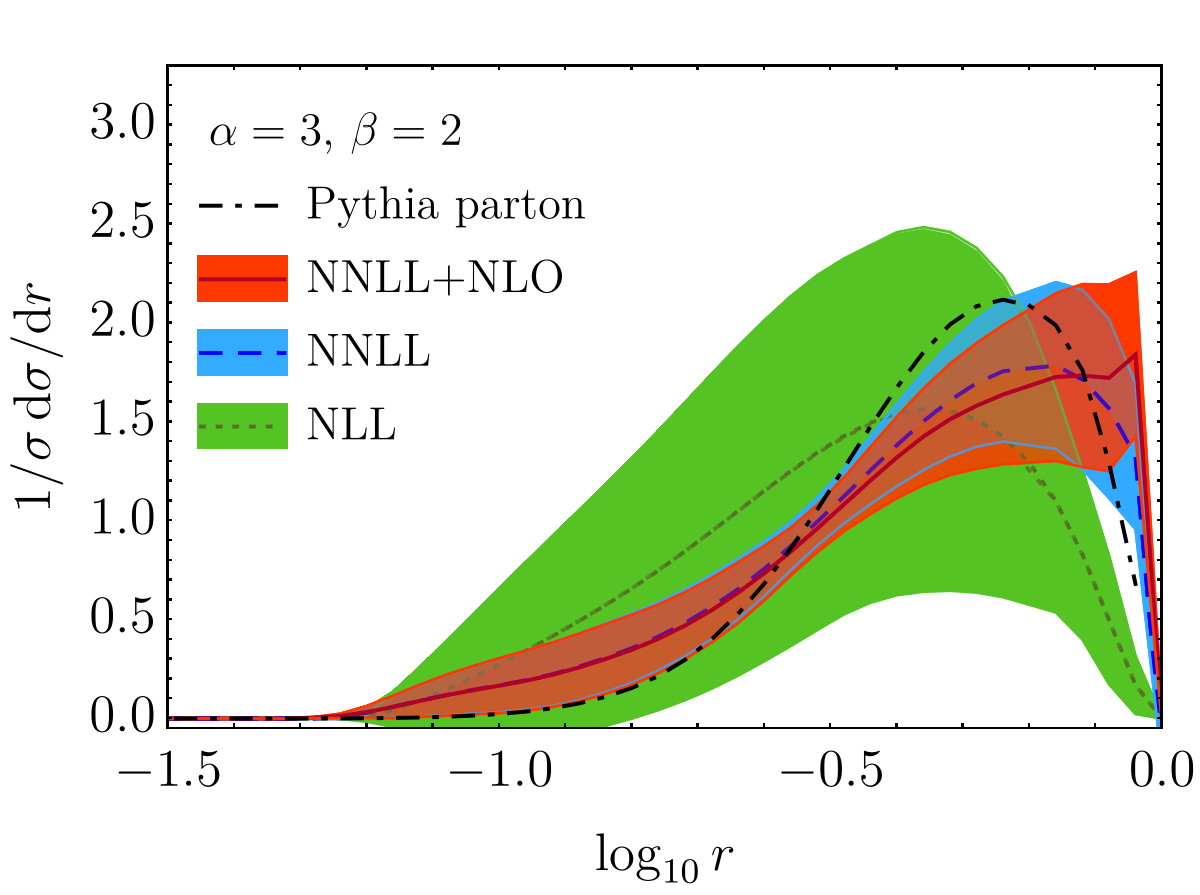}  \\
   \caption{The NLL, NNLL and NNLL+NLO cross section for the ratio of two angularities for $(\al,\bt) = (2, 0.5), (2, 1.2) \text{ and } (3,2)$
    (all normalized relative to the full NLO cross section), compared to parton-level predictions from \Pythia.
}
   \label{fig:rplots}
  \end{figure}
%---------------------------------------

Our results for the cross section differential in the ratio of two angularities $r=e_\al/e_\bt$ are shown in \fig{rplots} for angularity exponents $(\al,\bt) =$ (2,0.5), (2,1.2) and  (3,2) and $Q = 1000$ GeV. These are obtained from projecting the cross section differential in two angularities through\footnote{For the NNLL+NLO cross section, it is important to obtain the NLO nonsingular before performing the projection, since projecting the NLO first would yield a divergent result.}
%%%
\begin{align}
  \frac{\df \si}{\df r} = \int\! \df e_\al\, \df e_\bt\, \frac{\df \si}{\df e_\al\, \df e_\bt}\, \de\Big(r- \frac{e_\al}{e_\bt}\Big)
\,.\end{align}
%%%
The uncertainties are taken from the scale variations for the two-dimensional distributions using the procedure outlined in \sec{prof_2D}. As for the single angularity distributions, we have normalized the central curve, and rescaled the scale variations with the same factor. However, note that unlike for the single angularity case, the resummation region can contribute to all values of $r$. It is reassuring to see that the uncertainties decrease at higher orders, and the uncertainty bands overlap. The reason that the uncertainty is so large $(\al,\bt) = (2,0.5)$, is that the peak is close to the region 3 boundary, so almost all $r$ values are affected by large resummation uncertainties (see \fig{2d}). Given how close the central curves are, compared to the size of the uncertainty bands, our estimate is probably quite conservative. 

%---------------------------------------
 \begin{figure}[t]
  \centering
   \includegraphics[width=0.49\textwidth]{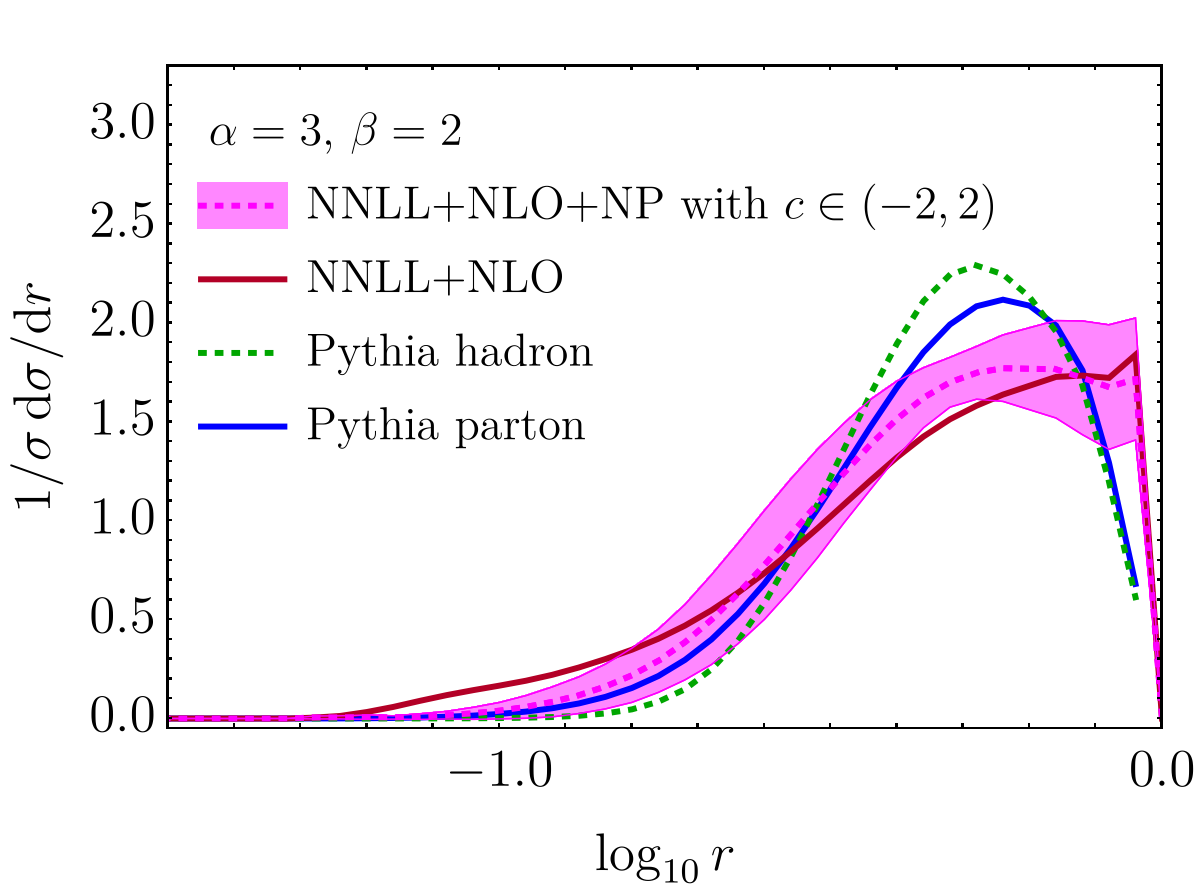}
   \includegraphics[width=0.49\textwidth]{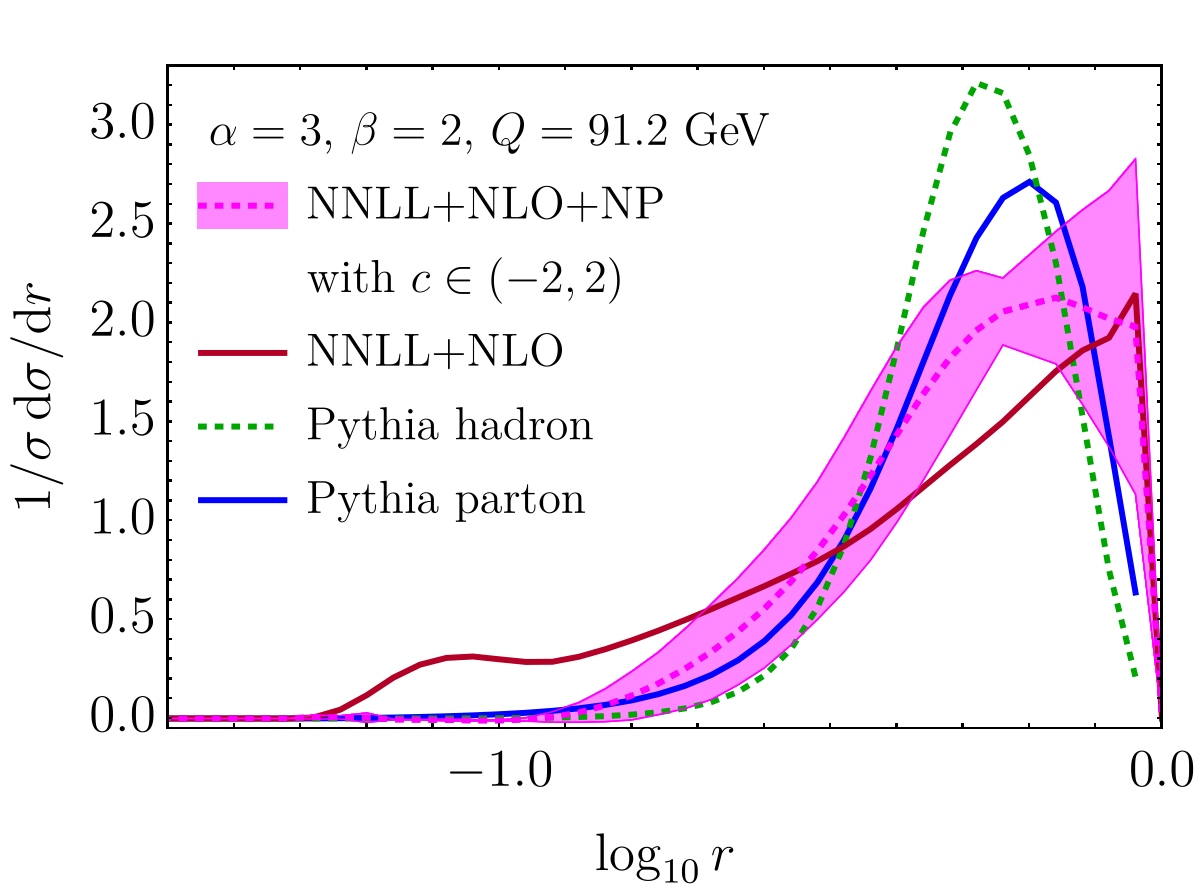}
   \caption{Distribution for the ratio of two angularities from the NNLL+NLO cross section with and without nonperturbative effects, for $(\al,\bt)=(3,2)$, compared to \Pythia at parton and hadron level for $Q=1000$ GeV (left) and $Q=91.2$ GeV (right). The band indicates the uncertainty from nonperturbative effects, as described in the text.}
   \label{fig:rplotsNP}
  \end{figure}
%---------------------------------------
The nonperturbative effect on the cross section differential in $r$ is shown in \fig{rplotsNP} for $(\al,\bt)=(3,2)$. 
Referring to \sec{nonperturbative} for the notation, the uncertainty band here includes both the $c$-variation (within -2 to 2),\footnote{We determined a reasonable range in $c$ by applying our procedure for including nonperturbative effects on \Pythia parton level predictions and comparing to \Pythia at hadron level.} the variation of $\Omega$ within its uncertainty of $16 \%$ and of $\tilde F$ (with $a=1$,\,2,\,3,\,4) as described for the single angularity case. We constructed a separate envelope for each of them and added these three uncertainties in quadrature. For $Q= 1000$ GeV, the correlations probed by $c$ dominate the uncertainty. For $Q=91.2$ GeV, the subleading nonperturbative corrections estimated by varying $a$ are the largest instead, for $\log_{10} r<-0.5$.
Compared to the single angularity distribution in \fig{1dresultsWithNPMZ}, the band is sizable over the whole plot range, because nonperturbative effects contribute to all values of the ratio.

%%%%%%%%%%%%%%%%%%%%%%%%%%%%%%%%%%%%%%%%%%%%%%%%%%%%%%%%%%%%%%%%%%%%%%%%%%%%%%%%
\section{Conclusions}
\label{sec:conc}
%%%%%%%%%%%%%%%%%%%%%%%%%%%%%%%%%%%%%%%%%%%%%%%%%%%%%%%%%%%%%%%%%%%%%%%%%%%%%%%%

In this paper we presented our calculation of the cross section for $e^+e^- \to$ hadrons differential in two angularities. We simultaneously resummed the logarithms of each angularity, employing the \SCETab framework we developed in ref.~\cite{Procura:2014cba}. The resummation was performed at NNLL accuracy and matched  to NLO, thereby obtaining a prediction that is valid throughout the phase space. By using exclusive $k_T$ clustering with the WTA recombination scheme, we could ignore the issue of recoil. We performed a detailed numerical study, assessed the perturbative uncertainties through variations of each of the various scales entering factorization, and studied the impact of the leading nonperturbative corrections. 

The one-loop matching with the full QCD calculation shows that our \SCETab factorization  correctly captures the singular limit at this order. We extended this check for the factorization theorem of a single angularity to $\ord{\al_s^2}$ by using \Eventtwo, and found agreement. We also showed that the effect of recoil can not be ignored for the angularity exponent $\beta=1.2$, highlighting the advantage of the WTA axis. In the fixed-order region the cross section has a Sudakov shoulder. This arises because the position of the WTA axis can change abruptly depending on the precise momentum configuration, as we are no longer in the dijet region.

We have tested the perturbative convergence of our resummed calculation, finding that the uncertainty bands at higher orders become smaller and (mostly) overlap with those at lower orders. For the double angularity distribution, \Pythia seems closer to our NNLL+NLO prediction than our NLL prediction, though the Sudakov shoulder in our predictions that arises in the fixed-order region is washed out. Of course a benefit of our calculation is that it provides an estimate of the perturbative uncertainty, and is systematically improvable. 
We point out that reaching NNLL+NNLO accuracy for the double differential cross section to match the precision for the single-angularity case, would require the calculation of the two-loop double differential jet and soft function, which we expect to be quite intricate, based on the complexity of the two-loop double-differential beam function calculation~\cite{Gaunt:2014xxa}. 

We also considered the cross section differential in the ratio of two angularities, which is not infrared safe but still Sudakov safe. This is interesting to investigate because many jet substructure observables are also Sudakov-safe ratio observables, for which calculations have typically been restricted to NLL accuracy (with the exception of the ratio $\tau_{2,1}^{(2)}$ of 2- to 1-subjettiness with angular exponent 2 for signal events~\cite{Feige:2012vc}). 
Since the resummation region contributes to most of the plot range for the angularity ratio, the uncertainty on the cross section is larger than for the single angularity measurement, but still reasonable. As may be expected, nonperturbative corrections similarly play a more important role. We expect these features to carry over to other Sudakov-safe ratio observables. 

In this paper we restricted ourselves to two-jet production in $e^+e^-$ collisions to have a clean theoretical setup, but it is our goal to extend this analysis to the measurement of (multiple) angularities of jets in LHC collisions at NNLL+NLO. One concrete application is the simultaneous extraction of $\alpha_s$ and the quark/gluon fraction performed in ref.~\cite{Bendavid:2018nar}. Also, only very few jet substructure observables have been calculated at this accuracy so far.
Our framework allows us to reliably account for correlations between jet observables, and demonstrates the feasibility of performing higher-order resummation for more differential measurements.

%%%%%%%%%%%%%%%%%%%%%%%%%%%%%%%%%%%%%%%%%%%%%%%%%%%%%%%%%%%%%%%%%%%%%%%%%%%%%%%%
\begin{acknowledgments}
We thank C.~Bauer, G.~Bell, Z.~Ligeti, P.~Monni, I.~Moult and  I.~Stewart for discussions.
M.P. acknowledges support by a Marie Curie Intra-European Fellowship of the European Community's 7th Framework Program under contract number PIEF-GA-2013-622527. 
W.W.~is supported by the ERC grant ERC-STG-2015-677323 and the D-ITP consortium, a program of the Netherlands Organization for Scientific Research (NWO) that is funded by the Dutch Ministry of Education, Culture and Science (OCW). 
L.Z.~is supported by NWO through a Veni grant (project number 680-47-448), and thanks the LBNL theory group for hospitality and support.
This article is based upon work from COST Action CA16201 PARTICLEFACE, supported by COST (European Cooperation in Science and Technology).
This research was supported by the Munich Institute for Astro- and Particle Physics (MIAPP) of the DFG cluster of excellence ``Origin and Structure of the Universe''.
\end{acknowledgments}
%%%%%%%%%%%%%%%%%%%%%%%%%%%%%%%%%%%%%%%%%%%%%%%%%%%%%%%%%%%%%%%%%%%%%%%%%%%%%%%%

\appendix

%%%%%%%%%%%%%%%%%%%%%%%%%%%%%%%%%%%%%%%%%%%%%%%%%%%%%%%%%%%%%%%%%%%%%%%%%%%%%%%%
\section{Renormalization Group Evolution}
\label{app:RGE}
%%%%%%%%%%%%%%%%%%%%%%%%%%%%%%%%%%%%%%%%%%%%%%%%%%%%%%%%%%%%%%%%%%%%%%%%%%%%%%%%

The integrals $K_\Gamma$, $\eta_\Gamma$ and $K_{\ga_F}$ that enter in the evolution kernels, and were defined in \eq{Keta_def}, can be performed analytically in a perturbative expansion. Up to NNLL order their expressions are given by
%%%
\begin{align} \label{eq:Keta}
K_\Gamma(\mu, \mu_0) &= -\frac{\Gamma_0}{4\beta_0^2}\,
\biggl\{ \frac{4\pi}{\alpha_s(\mu_0)}\, \Bigl(1 - \frac{1}{r} - \ln r\Bigr)
   + \biggl(\frac{\Gamma_1 }{\Gamma_0 } - \frac{\beta_1}{\beta_0}\biggr) (1-r+\ln r)
   + \frac{\beta_1}{2\beta_0} \ln^2 r
\nn\\ & \hspace{10ex}
+ \frac{\alpha_s(\mu_0)}{4\pi}\, \biggl[
  \biggl(\frac{\beta_1^2}{\beta_0^2} - \frac{\beta_2}{\beta_0} \biggr) \Bigl(\frac{1 - r^2}{2} + \ln r\Bigr)
  + \biggl(\frac{\beta_1\Gamma_1 }{\beta_0 \Gamma_0 } - \frac{\beta_1^2}{\beta_0^2} \biggr) (1- r+ r\ln r)
\nn\\ & \hspace{10ex}
  - \biggl(\frac{\Gamma_2 }{\Gamma_0} - \frac{\beta_1\Gamma_1}{\beta_0\Gamma_0} \biggr) \frac{(1- r)^2}{2}
     \biggr] \biggr\}
\,, \nn\\
\eta_\Gamma(\mu, \mu_0) &=
 - \frac{\Gamma_0}{2\beta_0}\, \biggl[ \ln r
 + \frac{\alpha_s(\mu_0)}{4\pi}\, \biggl(\frac{\Gamma_1 }{\Gamma_0 }
 - \frac{\beta_1}{\beta_0}\biggr)(r\!-\!1)
 + \frac{\alpha_s^2(\mu_0)}{16\pi^2} \biggl(
    \frac{\Gamma_2 }{\Gamma_0 } - \frac{\beta_1\Gamma_1 }{\beta_0 \Gamma_0 }
      + \frac{\beta_1^2}{\beta_0^2} -\frac{\beta_2}{\beta_0} \biggr) \frac{r^2\!-\!1}{2}
    \biggr]
\,, \nn\\
K_{\gamma_F}(\mu, \mu_0) &=
 - \frac{\gamma_{F,0}}{2\beta_0}\, \biggl[ \ln r
 + \frac{\alpha_s(\mu_0)}{4\pi}\, \biggl(\frac{\gamma_{F,1} }{\gamma_{F,0} }
 - \frac{\beta_1}{\beta_0}\biggr)(r-1) \biggr]
\,,\end{align}
%%%
where  $r = \alpha_s(\mu)/\alpha_s(\mu_0)$. The running coupling is given by the three-loop expression
%%%
\begin{equation} \label{eq:alphas}
\frac{1}{\alpha_s(\mu)} = \frac{X}{\alpha_s(\mu_0)}
  +\frac{\beta_1}{4\pi\beta_0}  \ln X
  + \frac{\alpha_s(\mu_0)}{16\pi^2} \biggr[
  \frac{\beta_2}{\beta_0} \Bigl(1-\frac{1}{X}\Bigr)
  + \frac{\beta_1^2}{\beta_0^2} \Bigl( \frac{\ln X}{X} +\frac{1}{X} -1\Bigr) \biggl]
\,,\end{equation}
%%%
with $X = 1+\alpha_s(\mu_0)\beta_0 \ln(\mu/\mu_0)/(2\pi)$. 

The coefficients of the cusp anomalous dimension that enter in \eq{Keta} are~\cite{Moch:2004pa}
%%%
\begin{align}
\Gamma_0 &= 4 C_F
\,,\nn\\
\Gamma_1 &= 4 C_F \Bigl[\Bigl( \frac{67}{9} -\frac{\pi^2}{3} \Bigr)\,C_A  -
   \frac{20}{9}\,T_F\, n_f \Bigr]
\,,\nn\\
\Gamma_2 &= 4 C_F \Bigl[
\Bigl(\frac{245}{6} -\frac{134 \pi^2}{27} + \frac{11 \pi ^4}{45}
  + \frac{22 \zeta_3}{3}\Bigr)C_A^2
  + \Bigl(- \frac{418}{27} + \frac{40 \pi^2}{27}  - \frac{56 \zeta_3}{3} \Bigr)C_A\, T_F\,n_f
\nn\\* & \hspace{8ex}
  + \Bigl(- \frac{55}{3} + 16 \zeta_3 \Bigr) C_F\, T_F\,n_f
  - \frac{16}{27}\,T_F^2\, n_f^2 \Bigr]
\,, 
\end{align}
%%%
and for the $\beta$ function they are given by~\cite{Tarasov:1980au,Larin:1993tp}
%%%
\begin{align}
\beta_0 &= \frac{11}{3}\,C_A -\frac{4}{3}\,T_F\,n_f
\,,\nn\\
\beta_1 &= \frac{34}{3}\,C_A^2  - \Bigl(\frac{20}{3}\,C_A\, + 4 C_F\Bigr)\, T_F\,n_f
\,, \nn\\
\beta_2 &=
\frac{2857}{54}\,C_A^3 + \Bigl(C_F^2 - \frac{205}{18}\,C_F C_A
 - \frac{1415}{54}\,C_A^2 \Bigr)\, 2T_F\,n_f
 + \Bigl(\frac{11}{9}\, C_F + \frac{79}{54}\, C_A \Bigr)\, 4T_F^2\,n_f^2
 \,.
\end{align}
%%%
The coefficients for the non-cusp anomalous dimension for the hard function are~\cite{Manohar:2003vb,Bauer:2003di} 
%%%
\begin{align}
\gamma_{H,0} &= -12 C_F
\,, \nn \\
\gamma_{H,1} & = - 2 C_F \Bigl[
  \Bigl(\frac{82}{9} - 52 \zeta_3\Bigr) C_A
+ (3 - 4 \pi^2 + 48 \zeta_3) C_F
+ \Bigl(\frac{65}{9} + \pi^2 \Bigr) \beta_0 \Bigr]
\,,\end{align}
%%%
and for the soft function~\cite{Bell:2018vaa,Bell:2017wvi,TalbertSCET} 
%%%
\begin{align}
\gamma_{S,0}^q &= 0
\,, \nn \\
\gamma_{S,1}^q & = \frac{2}{\al-1} C_F C_A \bigg[  - \frac{808}{27} + \frac{11 \pi^2}{9} + 28 \zeta_3 - \int_0^1 \df x \int_0^1 \df y \ln \Big( \frac{(x^{2-\al} + xy)(x+x^{2-\al} y)}{x^{2-\al} (1+xy)(x+y)}\Big)
\nn \\ & \quad \times
\frac{32 x^2 (1+xy+y^2)\big(x(1+y^2)+(x+y)(1+xy)\big)}{y(1-x^2)(x+y)^2(1+xy)^2}  \bigg]
+ \frac{2}{\al-1} C_F T_F n_f \bigg[ \frac{224}{27} - \frac{4 \pi^2}{9} 
\nn \\ & \quad 
 - \int_0^1 \df x \int_0^1 \df y \ln \Big( \frac{(x^{2-\al}  + xy)(x+x^{2-\al}  y)}{x^{2-\al}  (1+xy)(x+y)}\Big)
\frac{64 x^2 (1+y^2)}{(1-x^2)(x+y)^2 (1+xy)^2} \bigg]
\,.\end{align}
%%%
The other non-cusp anomalous dimensions follow from \eq{noncusp_rel}.

%%%%%%%%%%%%%%%%%%%%%%%%%%%%%%%%%%%%%%%%%%%%%%%%%%%%%%%%%%%%%%%%%%%%%%%%%%%%%%%%
\section{NLO and NNLO singular terms in the single angularity distribution}
\label{app:sing}
%%%%%%%%%%%%%%%%%%%%%%%%%%%%%%%%%%%%%%%%%%%%%%%%%%%%%%%%%%%%%%%%%%%%%%%%%%%%%%%%
The fixed-order single angularity distribution can be written as
%%%
\begin{align}
\frac{1}{\hat{\sigma}_0} \, \frac{\df \sigma}{\df \Lb} = \frac{\al_s}{2 \pi} A(L_\bt) + \left(\frac{\al_s}{2 \pi} \right)^2 B(L_\bt)  + \ord{\al_s^3}
\end{align}
%%% 
for $\Lb \equiv \log_{10} (e_\bt)$. Our resummed results allow us to derive the singular contributions to the $A(L_\bt)$ and $B(L_\bt)$ coefficients for angularities with respect to the WTA axis. In particular, for the angularity exponents considered in our plots, 
%%%
\begin{align}
A_{\rm sing}(L_3)&= C_F\,(-4.60517 - 14.1384\, L_3) \nn \\
B_{\rm sing}(L_3)&= C_A C_F\,(k_3 - 16.4093 \,L_3 + 79.5785 \, L_3^2) \nn \\
&\quad+ C_F^2 \,(m_3 + m_3^L \,L_3 + 97.6646 \,L_3^2 + 99.9471 \,L_3^3) \nn \\
&\quad+ C_F \,T_F \,n_f \,(n_3 + 10.9965 \,L_3 - 28.9377 \, L_3^2) \nn \\
A_{\rm sing}(L_2)&= C_F\,(-6.90776 - 21.2076\, L_2) \nn \\
B_{\rm sing}(L_2)&= C_A C_F\,(k_2 - 14.8938 \,L_2 + 134.289 \,L_2^2) \nn \\
&\quad+ C_F^2 \,(m_2 + m_2^L \,L_2 + 219.745 \,L_2^2 + 224.881\, L_2^3) \nn \\
&\quad+ C_F \,T_F \,n_f \,(n_2 + 12.9602 \,L_2 - 48.8323 \,L_2^2)  \nn \\
A_{\rm sing}(L_{1.2})&= C_F\,(-11.5129 - 35.346 \,L_{1.2}) \nn \\
B_{\rm sing}(L_{1.2})&= C_A C_F\,(k_{1.2} + 7.57742 \,L_{1.2} + 273.551 \,L_{1.2}^2) \nn \\
&\quad+ C_F^2 \,(m_{1.2} +m_{1.2}^L \,L_{1.2} + 610.404 \,L_{1.2}^2 + 624.669 \,L_{1.2}^3) \nn \\
&\quad+ C_F \,T_F \,n_f \,(n_{1.2} + 9.81833 \,L_{1.2} - 99.4732 \,L_{1.2}^2) \nn \\
A_{\rm sing}(L_{0.5})&= C_F\,(-27.631 - 84.8304 \,L_{0.5}) \nn \\
B_{\rm sing}(L_{0.5})&= C_A C_F\,(k_{0.5} + 290.35 \,L_{0.5} + 1074.31 \,L_{0.5}^2) \nn \\
&\quad+ C_F^2 \,(m_{0.5} +m_{0.5}^L \,L_{0.5} + 3515.92 \,L_{0.5}^2 + 3598.1 \,L_{0.5}^3) \nn \\
&\quad+ C_F \,T_F \,n_f \,(n_{0.5} - 75.4048 \,L_{0.5} - 390.658 \,L_{0.5}^2)
\,,
\end{align}
%%%%
where the remaining coefficients for the three color structures at NNLO are
%%%
\begin{align}
k_3 &= -0.152521\,, & m_3 &=  -11.043\,,&m_3^L &=  - 48.1211\,, & n_3&=7.72881\,, \nn \\
k_2 &= -6.20299\,,  & m_2 &=  -12.8324\,, & m_2^L &=  - 90.1385\,, & n_2&=7.8759\,, \nn \\
k_{1.2} &= -6.01464\,, & m_{1.2} &=  -8.94722\,, & m_{1.2}^L &=   - 210.087\,, & n_{1.2}&=8.5379\,, \nn \\
k_{0.5} &= 58.1583\,,  & m_{0.5} &=  83.024\,,& m_{0.5}^L &=  -1007.0\,, & n_{0.5}&=-4.34318\,.
\end{align}
%%%%
In our comparison against \Eventtwo, we also analyzed angularities with respect to the thrust axis, with exponents $\bt=1.2, \,2,\, 3$. The corresponding NLO coefficients $A^{\rm thr} _{\rm sing}(L_\bt)$ coincide with the ones calculated with respect to the WTA axis. Differences first appear at NNLO and are due to the non-logarithmic terms in the one-loop cumulative jet function, which for the thrust axis can be obtained from ref.~\cite{Hornig:2009vb}. Thus in the NNLO coefficients $B^{\rm thr}_{\rm sing}(L_\bt)$ the only changes are
%%%
\begin{align}
k_3^{\rm thr} &= -11.4406\,,  & m_3^{\rm thr} &= -29.5143\,, & m_3^{L,\,{\rm thr} }&= -104.83\,, & n_3^{\rm thr}&=11.8336 \,, \nn \\
k_2^{\rm thr} &=-16.2048 \,,  & m_2^{\rm thr} &= -29.199\,, & m_2^{L,\,{\rm thr} }&= -140.386\,, & n_2^{\rm thr}&=11.5129\,, \nn \\
k_{1.2}^{\rm thr} &= 133.46\,,  & m_{1.2}^{\rm thr} &= 219.284\,, & m_{1.2}^{L,\,{\rm thr} }&= 490.607\,, & n_{1.2}^{\rm thr}&=-42.18
\,.
\end{align}
%%%%
For the thrust case, the coefficients $A^{\rm thr} _{\rm sing}(L_2)$ and $B^{\rm thr} _{\rm sing}(L_2)$ agree with the well-known results from the literature (see {\it e.g.}~ref.~\cite{Becher:2008cf}).

\phantomsection
\bibliographystyle{jhep}
\bibliography{dang}

\end{document}